\documentclass[nofootinbib,amsmath,prd,aps,superscriptaddress,tightenlines,12pt]{revtex4}
\usepackage{graphicx}
\usepackage{epstopdf}
\usepackage{bm}
\usepackage{epsfig}
\usepackage{graphics}
\usepackage{xspace}
\usepackage{subfigure}
\usepackage{amsmath}
\usepackage{xcolor}
\usepackage{url}
\usepackage{graphicx}

\usepackage{amssymb}

\usepackage{comment}

\usepackage[normalem]{ulem}

\def\OMIT#1{}

\newcommand{\nn}{\nonumber}

\newcommand{\bea}{\begin{eqnarray}}
\newcommand{\eea}{\end{eqnarray}}

\newcommand{\gsim}{\mathrel{\rlap{\lower4pt\hbox{\hskip1pt$\sim$}}\raise1pt\hbox{$>$}}}

\newcommand{\be}{\begin{equation}}
\newcommand{\ee}{\end{equation}}

\allowdisplaybreaks

\begin{document}



\title{\bf Hadronization effects and electroweak contributions \\ in DIS one-jettiness }

\author{Radja Boughezal}
\email{radja.boughezal@gmail.com}
\affiliation{Northwestern University, Evanston, IL, USA}

\author{Haotian Cao}
\email{haotiao.cao@northwestern.edu}
\affiliation{Northwestern University, Evanston, IL, USA}
\affiliation{Center for Frontiers in Nuclear Science, Stony Brook University, Stony Brook, NY 11794, USA}

\author{Zhong-Bo Kang}
\email{zkang@physics.ucla.edu}
\affiliation{Department of Physics and Astronomy, University of California, Los Angeles, CA 90095, USA}
\affiliation{Mani L. Bhaumik Institute for Theoretical Physics,
University of California, Los Angeles, CA 90095, USA}
\affiliation{Center for Frontiers in Nuclear Science, Stony Brook University, Stony Brook, NY 11794, USA}
                   
\author{Xiaohui Liu}
\email{xiliu@bnu.edu.cn}
\affiliation{School of Physics and Astronomy, Beijing Normal University,
and Key Laboratory of Multiscale Spin Physics (Beijing Normal University), Ministry of Education, Beijing 100875, China}
 \affiliation{Southern Center for Nuclear Science Theory (SCNT),
Institute of Modern Physics, Chinese Academy of Science, Huizhou 516000, China}

\author{Sonny Mantry}
\email{sonny.mantry@ung.edu}
\affiliation{Department of Physics and Astronomy, 
                   University of North Georgia,
                   Dahlonega, GA 30597, USA}
\author{Frank Petriello}
\email{f-petriello@northwestern.edu}
\affiliation{Northwestern University, Evanston, IL, USA}   



\newpage


\begin{abstract}
  \vspace*{0.3cm}

We study the structure of leading hadronization effects and the contributions of massive electroweak gauge boson exchange in the $\tau_1$ and $\tau_{1a}$ 1-Jettiness global event shapes  for  deep inelastic scattering (DIS). The leading hadronization effects in $\tau_1$  acquire a non-trivial  dependence on the hard scattering kinematics. This kinematic dependence is explicitly calculable, so that the leading hadronization effects in $\tau_1$, $\tau_{1a}$, and DIS thrust are universal and described by the same underlying shape function. The additional calculable kinematic dependence in $\tau_1$ provides an independent lever arm for simultaneously constraining hadronization effects in these observables. We  present corresponding results at the N$^3$LL+${\cal O}(\alpha_s^2)$ level of accuracy. We extend previous results by including the contributions mediated by the exchange of the massive $Z$ and $W$ electroweak gauge bosons for neutral current (NC) and charged current (CC) DIS, respectively, including ${\cal O}(\alpha_s)$ QCD corrections to
obtain N$^2$LL+${\cal O}(\alpha_s)$ results. We compare theoretical predictions to Pythia simulation and demonstrate the universality of leading hadronization effects  across NC and CC DIS processes and a wide range of kinematics relevant to HERA and the EIC.

\end{abstract}

\maketitle

\newpage
\tableofcontents

\section{Introduction}

Global event shape variables~\cite{Dasgupta:2003iq,Banfi:2010xy,Stagnitto:2025air} have long been used as powerful tools to probe the structure of Quantum Chromodynamics (QCD). They were first used to study gluon bremsstrahlung~\cite{ellis:1977,Farhi:1977sg,Georgi:1977sf,Fox:1978vu,Basham:1978,Basham:1979} in $e^+e^-\to$ hadrons, leading to confirmation of the gluon as the vector gauge boson~\cite{Ellis:1996mzs} of QCD. Event shapes are infrared and collinear safe observables that characterize the geometry of the final state hadronic energy flow through the value of a continuous event shape parameter. They describe whether the overall shape of final state radiation is spherical or pencil-like. Event shapes group final state particles into beam or jet regions using a set of null reference axes. For small values of the event shape variable, the directions of the reference axes are  correlated with the direction of dominant energy flow emerging from the underlying hard scattering process. For large values of the event shape variable, the energy flow is more spherical and uncorrelated with the directions of the reference axes. 

Events with small event shape values are characterized by final state radiation that is either soft compared to the hard scale or collinear to one of the reference axes.  These soft and collinear emissions are logarithmically enhanced in perturbation theory and require resummation. Events with large values of the event shape, on the order of the hard scale, are not logarithmically enhanced and can be described by fixed-order perturbative QCD. While the infrared and collinear safety of event shapes allow for well-defined calculations in perturbation theory, they are subject to significant non-perturbative effects associated with the evolution of quarks and gluons into bound states of hadrons in the final state, especially in the region of small event shape values. Factorization and resummation frameworks provide rigorous field-theoretic definitions of the leading hadronization effects in terms of nonperturbative vacuum matrix elements of soft Wilson lines that describe soft eikonal emissions along the reference axes.  In  Monte Carlo (MC) event generators~\cite{Corcella:2000bw,Sjostrand:2019zhc,bierlich2022comprehensive,Sherpa:2024mfk,Putschke:2019yrg,Chang:2022hkt}, hadronization effects are implemented through  hadronization models~\cite{Andersson:1983momentum,Webber:1983ps,Cheng:1985zz,Sjostrand:2019zhc,Corcella:2000bw}, with parameters tuned to data. Comparing high precision theoretical predictions of global event shape observables with simulation and real data can help probe and constrain  our understanding of the dynamics of hadronization.

Examples of global event shapes include thrust~\cite{Brandt:1964sa,Farhi:1977sg}, broadening~\cite{Rakow:1981qn}, the C-parameter~\cite{Parisi:1978eg}, heavy jet mass~\cite{Catani:1991bd}, and  N-Jettiness~\cite{Stewart:2010tn}. They are also used in Deep Inelastic Scattering (DIS) of electrons off nuclei~\cite{Antonelli:1999kx,Dasgupta:2001sh,Dasgupta:2001eq,Dasgupta:2002bw,Dasgupta:2002dc}, including the more recent 1-Jettiness class of event shapes~\cite{Kang:2012zr,Kang:2013wca,Kang:2013nha,Kang:2013lga,Kang:2014qba,Chu:2022jgs,Cao:2024ota,Ee:2025scz}. Event shapes can be computed to very high accuracy and have been used for precision extractions of the strong coupling constant~\cite{Becher:2008cf, Abbate:2010xh,Abbate:2012jh,dEnterria:2022hzv,Bell:2023dqs,Benitez:2024nav,Benitez:2025vsp,Nason:2025qbx} and the top quark mass~\cite{Fleming:2007qr,Fleming:2007xt,Bachu:2020nqn,Dehnadi:2023msm}. In Ref.~\cite{Chien:2025rbp}, the ``1-Jettiness jet charge" observable was introduced, combining the jet charge~\cite{Field:1977fa,Krohn:2012fg,Waalewijn:2012sv,Kang:2021ryr} measurement with the 1-Jettiness global event shape into a unified factorization framework to probe quark flavor dynamics in the initial state proton and in final state hadronization. Ref.~\cite{Dotson:2026ttc} introduced  Centauric 1-Jettiness with adjustable beam and jet reference vectors, and correspondingly adjustable beam and jet regions, and derived universality relations of leading hadronization effects.  Recently, there have been considerable developments on the energy-energy correlator (EEC) class of event shapes~\cite{Basham:1979gh,Basham:1978zq,Basham:1978,Basham:1977iq} as a probe of hadronization. Refs.~\cite{Neill:2022lqx,Moult:2025nhu} give detailed and comprehensive reviews of EECs, including recent developments~\cite{Li:2020bub,Ali:2020ksn,Li:2021txc,Liu:2022wop,Liu:2023aqb,Cao:2023oef,Devereaux:2023vjz,Andres:2023xwr,Kang:2023big,Kang:2023oqj,Cao:2023qat,Fu:2025hpc,Cao:2025icu,Guo:2025qnz,Kang:2026pro}. Recently a new class of event shapes called Lund-Tree Shapes (LTS)~\cite{Dasgupta:2020fwr,vanBeekveld:2022ukn,vanBeekveld:2023chs,vanBeekveld:2025zjh} were introduced. They probe the detailed fine-grained geometry of radiation in multijet processes with the property that the integrated cross sections correspond to jet production rates. Event shapes have also been proposed as probes of TMDs through transverse momentum measurements relative to the event shape reference axes~\cite{Kang:2020yqw,Fernandez:2026nya}. 

With the proposed Electron-Ion Collider (EIC)~\cite{AbdulKhalek:2021gbh,DOE_LRP,DPAPreport2022} on the horizon, DIS global event shapes have gained renewed interest~\cite{Antonelli:1999kx,Dasgupta:2002dc,Kang:2012zr,Kang:2013wca,Kang:2013nha,Kang:2013lga,Kang:2014qba,Chu:2022jgs,Cao:2024ota,Ee:2025scz,Chien:2025rbp}. The EIC is expected to be built at the Brookhaven National Laboratory over the next decade. It will become one of the world's leading facilities for nuclear physics. Its mission is to explore fundamental questions about QCD. It will investigate the origin of nucleon mass and spin, how partons are distributed within the nucleon, the structure of heavier nuclei and its effects on parton distributions, and the evolution of partons into bound states of hadrons. These questions will be studied by analyzing collisions of electron and ion beams over a wide range of center of mass energies, $\sqrt{s}\sim 20-140$ GeV, and at high luminosity, ${\cal L}\sim 10^{33}-10^{34}$ cm$^{-2}$ s$^{-1}$. The EIC will use a variety of ion beams, ranging from protons to heavy nuclei such as uranium, and will use a hermetic detector for almost full solid angle coverage. The combination of variable center of mass energy and a hermetic detector will provide access to a wide range of electron-ion collision kinematics. The EIC will also have the capability to achieve $70-80\%$ polarization for the electron and proton beams.  These unique experimental design capabilities will allow for studies of nuclear structure and hadronization in unprecedented detail.

The success of the EIC program requires a wide array of high precision observables that are independent, complementary, and overconstraining.  DIS global event shapes, characterizing the pattern of final state radiation in electron-nucleus collisions, is one such class of observables that can be computed with high accuracy to facilitate precision studies of nuclear structure, hadronization dynamics, and a precision extraction of the strong coupling constant. DIS global event shapes were first studied~\cite{Antonelli:1999kx,Dasgupta:2001sh,Dasgupta:2001eq,Dasgupta:2002bw} and measured at HERA by the H1\cite{Adloff:1997gq,Aktas:2005tz,Adloff:1999gn} and ZEUS\cite{Breitweg:1997ug,Chekanov:2002xk,Chekanov:2006hv} collaborations more than two decades ago. 

The jet-based  1-Jettiness DIS  global event shape, $\tau_1$, was first introduced~\cite{Kang:2012zr,Kang:2013wca} over a decade ago, along with a factorization theorem derived in the Soft-Collinear Effective Theory (SCET)~\cite{Bauer:2000ew,Bauer:2000yr,Bauer:2001ct,Bauer:2001yt,Bauer:2002nz,Beneke:2002ph}. The $\tau_{1a}$ global event shape, a dimensionless variant of $\tau_1$, was introduced in Ref.~\cite{Kang:2013nha},  which also introduced two other 1-jettiness event shapes, denoted as $\tau_{1b}$ and $\tau_{1c}$.  The $\tau_{1b}$ event shape is equivalent to the DIS thrust event shape~\cite{Antonelli:1999kx}.  Since the introduction of this class of 1-Jettiness DIS event shapes, there have been many theoretical developments. These include the 1-Jettiness spectrum having been computed at the next-to-leading-logarithm (NLL)~\cite{Kang:2012zr},  N$^2$LL~\cite{Kang:2013wca,Kang:2013nha}, N$^2$LL+${\cal O}(\alpha_s)$~\cite{Kang:2013lga,Kang:2014qba,Chu:2022jgs}, and N$^3$LL+${\cal O}(\alpha_s^2)$~\cite{Cao:2024ota,Ee:2025scz}
 level of accuracies. Furthermore, the $\tau_{1b}$ distribution was measured using HERA data~\cite{Hessler:2021usr} and compared to ${\cal O}(\alpha_s^2)$ predictions generated by the  NNLOJET~\cite{Gehrmann-DeRidder:2016cdi,Currie:2016ytq,Currie:2017tpe,Gehrmann:2019hwf} program. The $\tau_{1b}$ event shape has also been studied~\cite{Knobbe:2023ehi} with soft drop grooming~\cite{Larkoski:2014wba,Frye:2016aiz,Baron:2020xoi}.

The focus of our study is on the $\tau_1$ and $\tau_{1a}$ DIS event shapes. We will often denote them collectively by the symbol $\xi$, defined as 
\bea
\label{eq:xi}
\tau_1: \>\>\>\>\xi &=&\tau_1, \nn \\
\tau_{1a}: \>\>\>\> \xi &=& Q\tau_{1a}, 
\eea
where $Q=\sqrt{Q^2}$ is the momentum transfer delivered to the proton. We will also generically denote the hard scattering scale  as $Q_H$. The region of small 1-Jettiness, $\xi\ll Q_H$, corresponds to events with the topology of a single jet recoiling against the final state lepton. This region is dominated by final state radiation that is either soft or energetic but collinear with either the beam or jet directions. Restricting  the 1-Jettiness event shape to small values, $\xi\ll Q_H$, effectively acts as a veto on additional jets and gives rise to large Sudakov logarithms of the form $\alpha_s^n\ln ^{m}(\xi/Q_H)$ for $m\leq 2n$. The  resummation of these Sudakov logarithms is achieved using an SCET factorization formula which involves the hard,  beam, jet, and soft functions that describe the physics of the hard scattering, initial state radiation collinear to the beam,  radiation collinear  to the jet direction, and soft radiation throughout the event, respectively. The beam functions are further matched onto the parton distribution functions (PDFs), factoring out the dynamics of the perturbative initial state radiation from the non-perturbative physics of nucleon structure.  Predictions obtained at N$^3$LL+${\cal O}(\alpha_s^2)$~\cite{Cao:2024ota} level of accuracy  require knowing the
${\cal O}(\alpha_s)$~\cite{Manohar:2003vb} and ${\cal O}(\alpha_s^2)$~\cite{Idilbi_2006,Becher:2006mr} hard function, the ${\cal O}(\alpha_s)$~\cite{Bosch:2004th} and ${\cal O}(\alpha_s^2)$~\cite{Becher:2006qw,Becher:2010pd} jet function, the ${\cal O}(\alpha_s)$~\cite{Stewart:2009yx, Stewart:2010qs, Mantry:2009qz,Berger:2010xi} and ${\cal O}(\alpha_s^2)$~\cite{Gaunt:2014xga,Gaunt:2014cfa} beam functions, the ${\cal O}(\alpha_s)$~\cite{Jouttenus:2011wh} and ${\cal O}(\alpha_s^2)$~\cite{Boughezal:2015eha} soft function, and  the four loop cusp anomalous dimension~\cite{Henn:2019swt,Moult:2022xzt}.  The jet function~\cite{Bruser:2018rad}, including the jet function for massive top quarks~\cite{Clavero:2024yav}, is now known to ${\cal O}(\alpha_s^3)$. The ${\cal O}(\alpha_s^2)$ fixed order perturbative QCD calculation, which includes up to three final state colored partons, was implemented numerically using the NLOJET++~\cite{Nagy:2005gn} program. Recently~\cite{Buonocore:2026yai}, a new method was proposed for efficiently computing the N-Jettiness soft functions for arbitrary in perturbative QCD.

In this work, we focus on advancing the theoretical framework for the $\tau_1$ and $\tau_{1a}$ 1-Jettiness global event shapes in two key ways. First, we study the structure of hadronization effects
in the small 1-Jettiness limit, $\xi \ll Q_H$, providing more detailed derivations and results in support of our companion paper~\cite{Boughezal:2026dvu}. In this limit, the leading hadronization effects are described by vacuum matrix elements of soft Wilson lines that describe soft eikonal emissions from the beam and jet directions. In the peak region, $\xi\sim \Lambda_{\rm QCD}\ll Q_H$, these vacuum matrix elements become nonperturbative and sensitive to the dynamics of hadronization of the soft partons. We show that the leading hadronization effects can acquire a non-trivial but calculable kinematic dependence on the underlying hard scattering kinematics, while still retaining properties of universality. This kinematic dependence arises from the dependence on the relative angle between the directions of the beam and jet reference axes that dynamically varies event-by-event, correlated with the hard scattering kinematics. Correspondingly, the impact of  hadronization effects can vary across  kinematic bins. This can be contrasted with DIS thrust ($\tau_{1b}$) \cite{Kang:2013nha,Ee:2025scz} in the Breit frame where the beam and current jet axes are defined to have a fixed relative angle, always being back-to-back along a thrust axis. In this case, the leading hadronization effects are constant across different kinematic bins. 

Using boost transformation properties~\cite{Kang:2013nha} of the 1-Jettiness soft functions, we show that the leading hadronization effects in $\tau_1$, $\tau_{1a}$, and DIS thrust ($\tau_{1b}$) are still described by a single underlying universal shape function.  The additional hard kinematic dependence for $\tau_1$ is explicitly calculable, and we derive the corresponding explicit expressions. We show that for $\tau_{1a}$ the kinematic dependence cancels in a specific manner so that its leading hadronization effects are described by the same shape function as for DIS thrust. As part of this universality, we also show that the required renormalon subtractions for $\tau_1$ are related to the renormalon  subtractions in DIS thrust up to calculable kinematic factors. The renormalon subtractions for $\tau_{1a}$ are identical to DIS thrust.  One can also explore new definitions of 1-Jettiness that give rise to different kinematic behavior of the leading hadronization effects. These results suggest performing combined global analyses of DIS thrust and jet-based 1-Jettiness global event shapes, making use of the hard kinematic dependence as a lever arm to better constrain hadronization dynamics and extract the strong coupling constant with high precision. Such global analyses could also include  the 1-Jettiness jet charge~\cite{Chien:2025rbp} observable for quark flavor separation of hadronization effects and proton structure.

In the tail region of the 1-Jettiness spectrum, $\Lambda_{\rm QCD}\ll \xi\ll Q_H$, one can perform an operator product expansion (OPE) in powers of $\Lambda_{\rm QCD}/\xi$. The OPE implies that hadronization effects can be captured in the tail region through a simple shift~\cite{Korchemsky:1994is,Dokshitzer:1995zt,Dokshitzer:1995qm,Dokshitzer:1997ew,Korchemsky:1999kt,Berger:2003pk,Salam:2001bd,Belitsky:2001ij,Berger:2004xf,Bauer:2003di,Lee:2006fn,Lee:2006nr}  in the partonic (part.) cross section as
\bea
\label{eq:OPEshift}
\frac{d\sigma}{d\{O_i\} \>d\xi } (\xi) \to \frac{d\sigma^{\rm part.}}{d\{O_i\} \>d\xi} (\xi - 2 \Omega_1),  \qquad \Lambda_{\rm QCD} \ll \xi \ll Q_H,
\eea
where $\{O_i\}$  denotes any other set of hard kinematic variables such as the momentum transfer, the Bjorken-$x$ variable, the 1-Jettiness jet transverse momentum or rapidity, or the final lepton transverse momentum or rapidity. This non-perturbative shift in the tail region, determined  by the first moment of the shape function, $2\Omega_1$, is now a well-known~\cite{Korchemsky:1994is,Dokshitzer:1995zt,Dokshitzer:1995qm,Dokshitzer:1997ew,Korchemsky:1999kt,Berger:2003pk,Salam:2001bd,Belitsky:2001ij,Berger:2004xf,Bauer:2003di,Lee:2006fn,Lee:2006nr} result for many global event shapes. The new feature here is that the first moment, $2\Omega_1$, inherits any hard kinematic dependence of the shape function, as in the case of $\tau_1$. As a result, the non-perturbative shift through $2\Omega_1$  in general can be a function of  the hard kinematic variables $\{O_i\}$ 
\bea
\label{eq:OmegaOi}
\Omega_1 = \Omega_1 (\{O_i\}).
\eea
This implies that  the non-perturbative shift can in general vary across different kinematic bins.  However,  this dependence can be explicitly calculated for any choice of the underlying shape function model parameters. We present results at the N$^3$LL+${\cal O}(\alpha_s^2)$ level of accuracy for the single photon exchange contributions, extending previous results~\cite{Cao:2024ota} by incorporating the above-discussed kinematic dependence, renormalon subtractions, and universality properties of hadronization effects. 

The second key result of this work is that we include the contributions from the exchange of the massive electroweak gauge bosons, $Z$ and $W$, for  neutral current (NC) and charged current (CC) DIS, respectively, along with the ${\cal O}(\alpha_s)$ QCD corrections.
This allows us to achieve the N$^2$LL+${\cal O}(\alpha_s)$ level of accuracy for the 1-Jettiness distribution for both NC and CC DIS.  All previous studies of the $\tau_1$ and $\tau_{1a}$ distributions have been restricted to NC DIS. For the first time we present the theoretical framework and numerical results for CC DIS. The NC and CC DIS processes are complementary to each other, probing different quark flavor combinations of nucleon structure functions.  At leading order in the electroweak couplings, the NC DIS  process is mediated by the exchange of a single virtual photon or $Z$-boson. The full cross section requires both of these contributions, including their interference.  In the resummation region of small 1-Jettiness, adding the $Z$-boson contribution simply modifies~\cite{Kang:2013nha,Cao:2024ota} the overall normalization in the SCET factorization and resummation formula. However,  in the fixed order region of large 1-Jettiness, adding the required $Z$-boson contribution is more non-trivial and has not yet been included in the   ${\cal O}(\alpha_s)$ and ${\cal O}(\alpha_s^2)$ perturbative QCD calculations for the 1-Jettiness event shape.  For this reason, all previous predictions for the 1-Jettiness event shape have been restricted to the region of low momentum transfer, $Q^2\ll M_Z^2$, where the $Z$-boson contribution is suppressed relative to the photon exchange contribution. Given the wide kinematic range of the EIC and HERA, it is essential to include the $Q^2\sim M_Z^2$ kinematic region where the $Z$-boson contribution is no longer suppressed.

For CC DIS, the entire process is mediated by the exchange of the massive $W$-boson, converting the initial state electron into a final state neutrino that goes undetected.  Since the final state neutrino escapes detection, the event kinematics must be reconstructed from the observed hadronic final state using the Jacquet-Blondel (JB) method~\cite{Amaldi:1979yh,Jacquet:1979,Akhundov:1993wvk}.  The  resummation region, $\xi \ll Q_H$, can be described by an SCET factorization and resummation formula with the same underlying structure as for the case of NC DIS. However, as in the case of NC DIS,  in the fixed order region of large 1-Jettiness, $\xi \sim Q_H$, the required perturbative QCD corrections have not yet been studied. In order to achieve the N$^2$LL+${\cal O}(\alpha_s)$ level of accuracy, we use the SCET factorization framework for the resummation region and use the DISTRESS~\cite{Abelof:2016pby} code for the ${\cal O}(\alpha_s)$ contribution in the fixed order region. We modify the DISTRESS code, which originally only included the single photon exchange contribution, to include the $Z$ and $W$ boson exchange contributions at ${\cal O}(\alpha_s)$. We provide results for a wide range of kinematic settings relevant to HERA and the EIC for both NC and CC DIS 1-Jettiness event shapes which further demonstrate the kinematic dependence and universality structure of leading hadronization effects.

In section~\ref{sec:KO} we define the various 1-Jettiness observables and the relevant kinematics for both NC and CC DIS. Section~\ref{sec:TF} describes the theoretical framework of factorization, resummation, the soft function, renormalon subtractions, and universality relations of leading hadronization effects. We give numerical results for a wide range of observables and kinematics relevant to HERA and the EIC in section~\ref{sec:NR}. Finally, we make concluding remarks in section~\ref{sec:Conc}.

\section{Kinematics and Observables}
\label{sec:KO}

\subsection{DIS Kinematics}
We consider both the NC and CC DIS processes:
\bea
\label{process}
\text{NC:} \>&& e^-(k) + p(P) \to e^-(k') + J+  X, \nn \\
\text{CC:} \>&& e^-(k) + p(P) \to \nu_e(k') + J + X, 
\eea
where  $k^\mu, k'^\mu,$ and $P^\mu$ denote the four-momenta of the initial lepton, the final lepton, and the initial proton, respectively, and $J$ denotes the final state jet that is defined within the 1-Jettiness framework.  In the single boson exchange approximation, the NC DIS  and CC DIS processes are mediated by the exchange of a photon or $Z$-boson and a $W$-boson, respectively.  We work in the center of the mass frame, using benchmark center of mass energy values of $\sqrt{s}=$90.0 GeV, 140 GeV, and 319 GeV covering typical kinematics at the EIC and HERA. The initial electron and proton momenta are given by
\bea
\label{eq:Pmukmu}
P^\mu &=& \frac{\sqrt{s}}{2}n_B^\mu, \qquad n_B^\mu = (1,0,0,1), \nn \\
k^\mu &=& \frac{\sqrt{s}}{2}\bar{n}_B^\mu, \qquad \bar{n}_B^\mu = (1,0,0,-1),
\eea
where we have ignored the electron and proton masses. The standard DIS kinematic variables are defined as:
\bea
s=(k+P)^2 , &&\qquad  q= k -k', \nn \\
Q^2 = -q^2, \qquad x &=& \frac{Q^2}{2P\cdot q}, \qquad y = \frac{P\cdot q}{P\cdot k}, 
\eea
where $Q^2= x y s$ when the proton mass is ignored.  

\subsection{1-Jettiness Observables}

\begin{figure}
\centering
\includegraphics[scale=0.12]{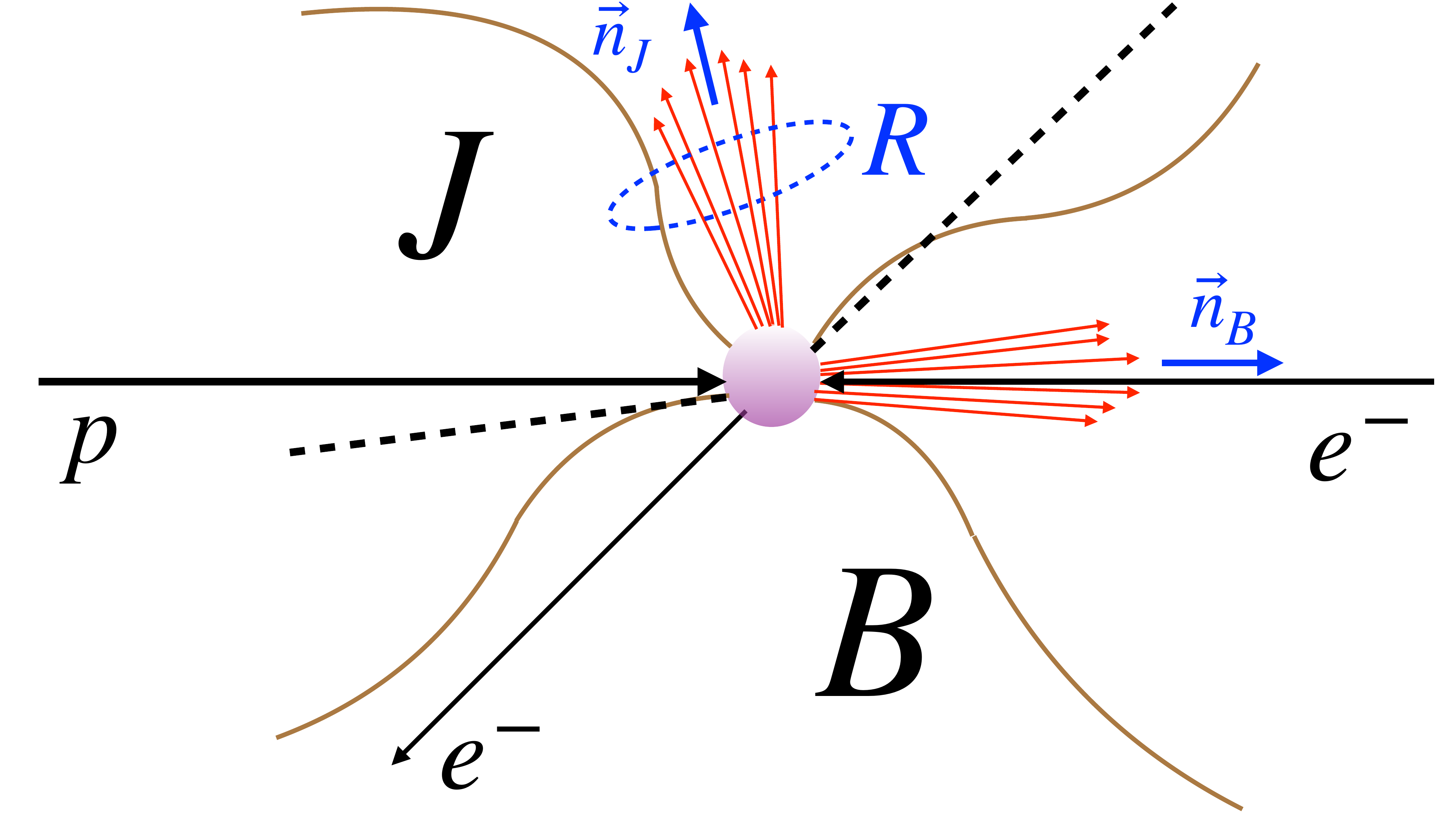}
\includegraphics[scale=0.12]{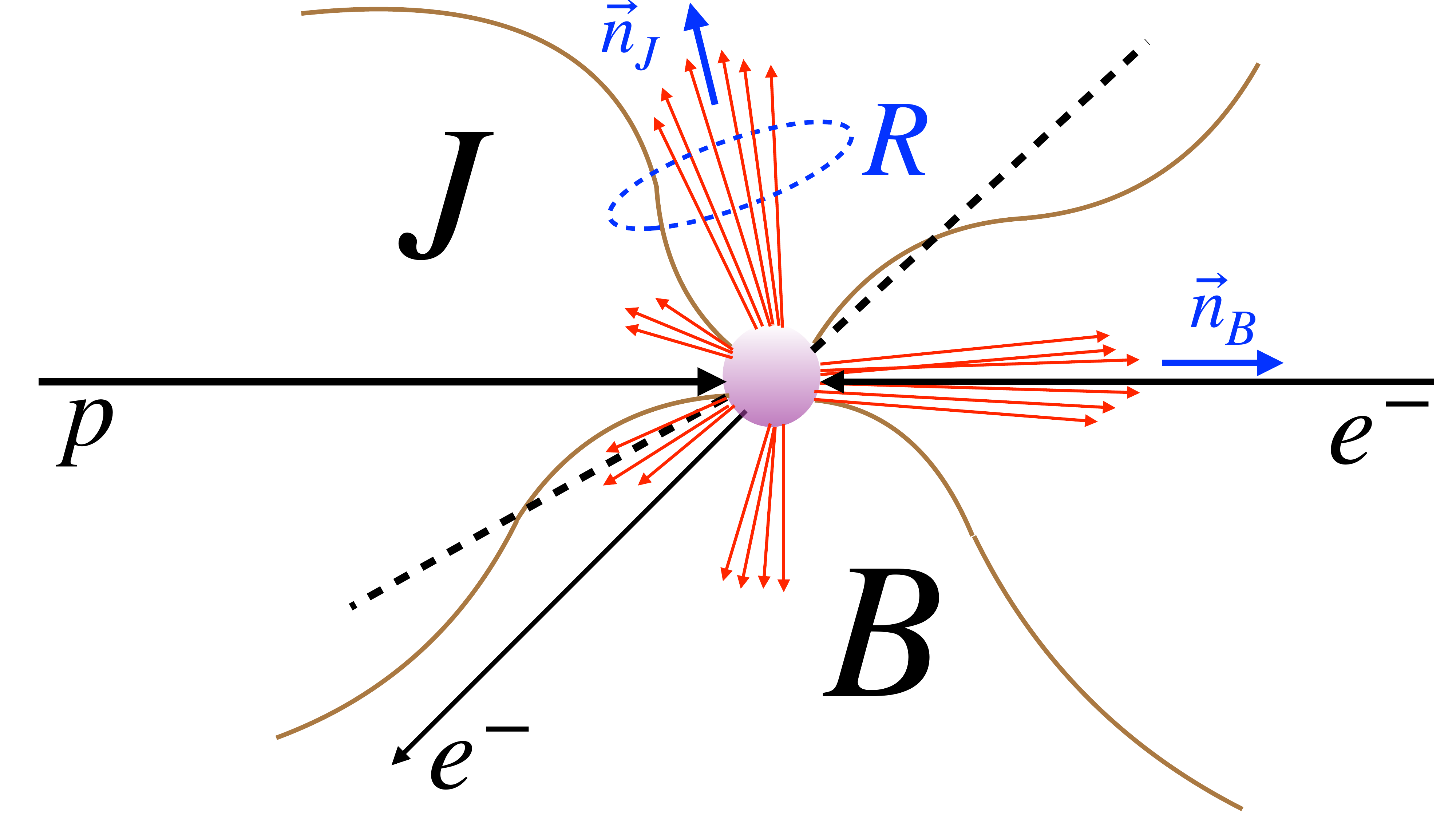}
\caption{Typical DIS events in the small (left panel) and large (right panel) 1-Jettiness limits, corresponding to events with the topology of a single jet and multiple jets. The red arrows and brown wavy lines denote  energetic and soft hadrons, respectively. The unit vectors $\vec{n}_B$ and $\vec{n}_J$ point along the proton beam axis and the leading jet direction. $R$ denotes the jet radius of the jet algorithm used to determine the jet reference vector $q_J^\mu=\omega_J n_J^\mu/2$, used in the definition of 1-Jettiness. The jet and beam regions, $J$ and $B$, are defined according to the minimization condition in the definition of $\tau_1$ or $\tau_{1a}$ in Eq.~(\ref{tau1andtau1a}).}
\label{fig:jettiness}
\end{figure}

We define a general class of jet-based 1-Jettiness observables, $\xi$, as:
\bea
\label{tau1andtau1a}
\xi &=& \sum_k \text{min} \Big \{ \frac{2q_B\cdot p_k}{Q_B}, \frac{2q_J\cdot p_k }{Q_J}\Big \}.
\eea
The sum is over all final state particles except the final state lepton.  The $Q_B$ and $Q_J$ constants in the definition of $\xi$ are associated with the beam and jet sectors, respectively, and are of the order of the hard scale in the process. Different choices for these constants correspond to  different definitions of $\xi$. The $q_B^\mu$ and $q_J^\mu$ null four vectors denote the beam and jet reference vectors, respectively. They can be parameterized as
\bea
\label{eq:qBqJ}
q_B^\mu = \omega_B \frac{n_B^\mu}{2}, \qquad q_J^\mu = \omega_J \frac{n_J^\mu}{2},
\eea
where $\omega_B$ and $\omega_J$ are chosen to be of the order of the hard scale. The null vectors are of the form
\bea
\label{eq:nBnJ}
n_B=(1,\vec{n}_B) , \qquad n_J=(1,\vec{n}_J)
\eea
so that the choice of the beam and jet reference vectors amounts to choice of the directions of the unit vectors $\vec{n}_B$ and $\vec{n}_J$. The direction of  $\vec{n}_B=(0,0,1)$ is usually chosen to be fixed along the proton beam direction. The direction of $\vec{n}_J$ is fixed to be in the direction of the leading jet determined by  a standard jet algorithm. In our analysis, we use the anti-$k_T$ jet algorithm with a jet radius $R=0.5$. The $\vec{n}_B$ and $\vec{n}_J$ vectors are illustrated in Fig.~\ref{fig:jettiness} for $\xi \ll Q_H$ (left panel) and $\xi \sim Q_H$ (right panel), corresponding to events characterized by the topology of a single jet events or multi-jet events, respectively.

We work with the canonical choice for the beam and jet reference vectors:
\bea
\label{eq:ref_vectors}
q_B^\mu = x P^\mu, \qquad q_J^\mu= (K_{J_T}\cosh y_K, \vec{K}_{J_T},K_{J_T}\sinh y_K),
\eea
where the leading jet momentum  is $K_J^\mu$.  The transverse momentum $K_{J_T}=|\vec{K}_{J_T}|$ and rapidity $y_K$ of the leading jet are  used to construct the null jet reference vector $q_J^\mu$. The choice of reference vectors in Eq.~(\ref{eq:ref_vectors}) corresponds to $\omega_B$ and $\omega_J$ in Eq.~(\ref{eq:qBqJ}) taking on the values
\bea
\label{eq:wBwJ}
\omega_B=x\sqrt{s}, \qquad \omega_J=2K_{J_T}\cosh y_K .
\eea

As seen in Fig.~\ref{fig:jettiness}, the beam and jet reference vectors divide the final state into beam and jet regions such that  each final state particle with momentum, $p_k$, is grouped either with the beam or jet region according to the minimization condition in Eq.~(\ref{tau1andtau1a}), and contributes accordingly to $\xi$. The largest contributions  come from energetic final state particles at large angles relative to both the beam and jet reference vectors. The contribution of soft particles or energetic particles collinear with the beam or jet reference vectors is suppressed. In this manner, the $\xi$ global event shapes can quantify the pattern of final state radiation in NC and CC DIS.

The $\tau_1$ and $\tau_{1a}$ event shapes are given in terms of $\xi$ as in Eq.~(\ref{eq:xi}) and correspond to choosing $Q_B$ and $Q_J$ in Eq.~(\ref{tau1andtau1a}) as
\bea
\label{eq:beam_ref_choices}
&&\xi=\tau_1: \qquad \>\>\>\>\>Q_B=\omega_B, \qquad  Q_J = \omega_J, \nn \\
&&\xi=Q\tau_{1a}: \qquad Q_B=Q_J = Q. 
\eea
Using Eqs.~(\ref{eq:qBqJ}) and (\ref{eq:beam_ref_choices}) in Eq.~(\ref{tau1andtau1a}), allows us to write the $\tau_1$ and $\tau_{1a}$ 1-Jettiness event shapes in the form
\bea
\label{eq:tau1tau1a2}
\tau_1 &=& \sum_k \text{min} \Big \{ n_B\cdot p_k,  n_J\cdot p_k \Big \}, \qquad \tau_{1a} = \sum_k \text{min} \Big \{ \frac{\omega_B n_B\cdot p_k}{Q^2}, \frac{\omega_Jn_J\cdot p_k }{Q^2}\Big \}.
\eea
 We note that from this form of $\tau_1$ that one can  determine the $n_B$ and $n_J$ directions through a minimization procedure that  minimizes the value of $\tau_1$. If we fix $\vec{n}_B=(0,0,1)$ in the direction of the proton beam as in Eq.~(\ref{eq:Pmukmu}), in practice it only remains to determine the direction of $\vec{n}_J$ through the minimization procedure. It is then possible to define $\tau_1$ independent of any jet algorithm. However, in our analysis we use the anti-$k_T$ jet algorithm  for simplicity. 
 
The 1-jettiness jet momentum for $\xi$ is defined as
\bea
\label{eq:pjet}
P_J^\mu &=& \sum_k p_k^\mu \>\theta \left(\frac{2q_B\cdot p_k}{Q_B} - \frac{2q_J\cdot p_k}{Q_J} \right), 
\eea
corresponding to the sum of the momenta of all final state particles grouped with the jet sector. The jet transverse momentum, $P_{J_T}$, and rapidity, $y_J$, are constructed from this 1-jettiness jet momentum $P_J^\mu$, defined in the center of mass frame.
In the resummation region, $ \xi \ll Q_H$, the difference between jet transverse momentum and rapidity determined by the jet algorithm, $(K_{J_T},y_K)$, and determined by the 1-Jettiness  jet momentum, $(P_{J_T},y_J)$, is power suppressed. This corresponds with the fact that the resummation region describes events with the topology of a single jet in addition to the beam jet with soft radiation in between. Any difference arises in how the soft radiation is clustered which gives a power suppressed contribution to the jet transverse momentum and rapidity.

In this work, we consider four types of observables for $\tau_1$ and $\tau_{1a}$, each. The first two:
\bea
\label{obs-1}
d\sigma \left [\tau_{1(a)}, P_{J_T}, y_J \right ] &\equiv& \frac{d^3\sigma (e^- + p \to  J + X)}{dy_J\> dP_{J_T}\>d\tau_{1(a)}},  \nn \\
d\sigma \left [\tau_{1(a)}, p_{T_e}, y_e \right ]  &\equiv& \frac{d^3\sigma (e^- + p \to e^- + J + X)}{dy_e\> dp_{T_e}\>d\tau_{1(a)}}, 
\eea
are differential in $(P_{J_T}$, $y_J)$  and  $(p_{T_e}$, $y_e)$ where the latter corresponds to the  transverse momentum and rapidity of the final state electron, respectively. The observable $d\sigma \left [\tau_{1(a)}, p_{T_e}, y_e \right ]$  is only relevant for NC DIS since it requires detection and measurement of the final state lepton momentum. The next two types of observables we consider are:
\bea
\label{obs-2}
d\sigma \left [\tau_{1(a)}, Q^2, x \right ] &\equiv& \frac{d^3\sigma (e^- + p \to e^- + J + X)}{dx\> dQ^2\>d\tau_{1(a)}} , \nn  \\
d\sigma \left [\tau_{1(a)}, Q^2, y \right ]  &\equiv& \frac{d^3\sigma (e^- + p \to e^- + J + X)}{dy\> dQ^2\>d\tau_{1(a)}}, 
\eea
being differential in the standard DIS kinematic variables $(Q^2,x)$ and $(Q^2,y)$. Due to the relation $Q^2=xys$,
they are related to each other by a replacement of variables and an overall Jacobian factor:
\bea
\label{eq:tau1a_x_vs_y}
d\sigma \left [ \tau_{1(a)}, Q^2, y \right ] &=& \frac{Q^2}{y^2 s}\> d\sigma \left [\tau_{1(a)}, Q^2, x=\frac{Q^2}{y s} \right ],  \nn \\
 d\sigma \left [ \tau_{1(a)}, Q^2, x \right ] &=& \frac{Q^2}{x^2 s}\> d\sigma \left [\tau_{1(a)}, Q^2, y=\frac{Q^2}{x s} \right ].
\eea
The observables  $d\sigma \left [\tau_{1(a)}, p_{J_T}, y_J \right ]$ and $d\sigma \left [\tau_{1(a)}, p_{T_e}, y_e \right ]$ only involve measurements over the final  particles and do not require reconstruction of intermediate kinematic variables such as $(Q^2,x,y)$, making them less susceptible to uncertainties associated with such reconstruction. This is especially important when taking into account collision-induced photon radiation~\cite{Kripfganz:1991,BLUMLEIN2003242,Afanasev:2004,Liu:2020rvc}  from the initial or final electrons, which complicate the relationship of the $(Q^2,x,y)$ variables to the initial and final state lepton momenta. These effects can be treated in the recently proposed joint QED and QCD factorization approach~\cite{Liu:2020rvc, Cammarota:2025jyr,Qiu:2026fed}. Ignoring QED radiation from the initial and final state leptons, so that the initial and final state lepton momenta can be used to reconstruct $(Q^2,x,y)$, the NC DIS observable $d\sigma \left [\tau_{1(a)}, p_{T_e}, y_e \right ]$ is related to  $d\sigma \left [\tau_{1(a)}, Q^2, y \right ]$ by a replacement of variables and an overall jacobian factor as:
\bea
\label{eq:Q2_y_vs_pTe_ye}
d\sigma \left [\tau_{1(a)}, p_{T_e}, y_e \right ] &=& 2 p_{T_e}  \>d\sigma \left [\tau_{1(a)}, Q^2 = \sqrt{s} \>p_{T_e} e^{y_e}, y= 1-  \frac{p_{T_e}}{\sqrt{s}}\> e^{-y_e}\right ]. 
\eea
In the rest of the paper, we will often use the notation
\bea
d\sigma \left [\xi, \{ O_i \} \right ],
\eea
to collectively denote the cross section that is differential in $\xi=\tau_1$ or $\xi=Q\tau_{1a}$ and the set $\{O_i\}$ which could denote any one of  the set of observables below:
\bea
\label{eq:Oi}
\{O_i\}= \{(P_{J_T},y_J ), (p_{T_e},y),(Q^2,x), (Q^2,y)\}. 
 \eea
For the observables in Eqs.~(\ref{obs-1}) and (\ref{obs-2}), the hard scale $Q_H$ will have the characteristic scaling
\bea
Q_H\sim \{P_{J_T}, p_{T_e}, Q\}.
\eea

\subsection{Neutral Current DIS}
The  electromagnetic and weak neutral currents for the electron and quarks  are given by
\bea
\label{eq:current_ops}
J_{f, \gamma}^\mu = Q_f \bar{\psi}_f \gamma^\mu \psi_f, \>\> J_{f, Z}^\mu = \bar{\psi}_f (v_f \gamma^\mu - a_f \gamma^\mu \gamma_5 )\psi_f , 
\eea
where $Q_f, v_f, a_f$ denote the electric charge,  weak vector charge, and  weak axial-vector charge, respectively, of the fermion $f$ in units of the proton charge $e$. The $Z$-boson vector and axial-vector couplings are:
\bea
v_f =  \frac{T_f^3 - 2 Q_f \sin^2\theta_w}{\sin 2\theta_w}, \qquad a_f =  \frac{T_f^3}{\sin 2\theta_w}.
\eea
At the tree level, ignoring hadronization effects, the final state is just a single quark or anti-quark recoiling against the final state electron. In this case, the 1-jettiness event shape vanishes so that the resulting tree level partonic 1-jettiness distribution is proportional to $\delta(\xi)$. The corresponding tree level result for $d\sigma[\xi, Q^2,x] $ is given by
\bea
\label{eq:tree_obs1}
d\sigma_0[\xi, Q^2,x] = \delta(\xi) \>\sigma_0^b  \> \Big  [ \sum_q L_q f_q(x,\mu) + \sum_{\bar{q}} L_{\bar{q}} f_{\bar{q}}(x,\mu)  \Big ],
\eea
where $f_q$ and $f_{\bar{q}}$ denote the quark and anti-quark PDFs, and  $\sigma_0^b$ is given by
\bea
\sigma_0^b = \frac{2\pi \alpha_{em}^2}{Q^4} \left [ 1+ \left (1-\frac{Q^2}{x s} \right )^2  \right ] .
\eea
Following the notation of Ref.~\cite{Kang:2013nha}, the $L_q$ and $L_{\bar{q}}$ factors are  given by
\bea
L_{q,\bar{q}} &=& Q_q^2 - \frac{2 Q_q v_q v_e}{1+m_Z^2/Q^2} + \frac{(v_q^2 + a_q^2)(v_e^2+a_e^2)}{(1+m_Z^2/Q^2)^2} \nn \\ 
&&\mp \frac{2y(2-y)}{(1-y)^2+1} \frac{a_qa_e[Q_q(1+m_Z^2/Q^2)-2v_qv_e]}{(1+m_Z^2/Q^2)^2},
\eea
where $m_Z$ denotes the $Z$-boson mass. The tree level result for $d\sigma_0[\xi,P_{J_T},y_J]$ is given by
\bea
\label{eq:tree_obs0}
d\sigma_0[\xi,P_{J_T},y_J] = \delta(\xi)\>\sigma_0 \> \Big  [ \sum_q L_q f_q(x_*,\mu) + \sum_{\bar{q}} L_{\bar{q}} f_{\bar{q}}(x_*,\mu)  \Big ] ,
\eea
where we have defined $\sigma_0$ and $x_*$ as
\bea
\label{eq:sig0_xstar}
\sigma_0 = 4\pi\alpha_{em}^2\frac{e^{y_J}}{\sqrt{s}\>  P_{J_T}^2} \left [1 + \left(1-\frac{P_{J_T}}{\sqrt{s}} e^{-y_J} \right )^2 \right ], \qquad x_* = \frac{\frac{P_{J_T}}{\sqrt{s}}\>e^{y_J}}{1 - \frac{P_{J_T}}{\sqrt{s}} \>e^{-y_J}}.
\eea
The corresponding tree level cross sections for $d\sigma \left [\xi, Q^2, y \right ]$ and $d\sigma \left [\xi, p_{T_e}, y_e \right ] $ can be obtained from Eqs.~(\ref{eq:tau1a_x_vs_y}) and (\ref{eq:Q2_y_vs_pTe_ye}), respectively. 

\subsection{Charged Current DIS}
For CC DIS, the lepton and quark currents are given by
\bea
\label{eq:current_ops}
J_{W_{ij}}^\mu = C_{ij} \>\bar{\psi}_i  (v_f \gamma^\mu - a_f \gamma^\mu \gamma_5 ) \psi_j , 
\eea
corresponding to the $\bar{f}_if_jW$ coupling, where the flavor coupling $C_{ij}$ is given by
\bea
C_{ij} = \delta_{ij}, \qquad C_{ij} = V_{ij},
\eea
for leptons and quarks, respectively, and $V$ denotes the CKM matrix. The vector and axial-vector couplings are given by
\bea
v_f =  a_f =  \frac{1}{2\sqrt{2} \sin 2\theta_w} .
\eea
The CC DIS tree level cross sections $d\sigma_0[\xi, \{ O_i\}] $  are obtained from the previous expressions for NC DIS by replacing the $L_q$ and $L_{\bar{q}}$ functions with the following expressions:
\bea
L_{q=u,c} &=& \frac{1}{1+(1-y)^2} \left (\frac{Q^2}{Q^2+m_W^2} \right )^2 \frac{G_F^2m_W^4}{4\pi^2\alpha_{em}^2}, \nn \\
L_{\bar{q}=\bar{d},\bar{s},\bar{b}} &=& \frac{(1-y)^2}{1+(1-y)^2} \left (\frac{Q^2}{Q^2+m_W^2} \right )^2 \frac{G_F^2m_W^4}{4\pi^2\alpha_{em}^2} \sum_{\bar{q}'=\bar{u},\bar{c}} |V_{\bar{q}'\bar{q}}|^2 , \nn \\
L_{d}&=&L_{s}=L_{b}=L_{\bar{u}}=L_{\bar{c}}=0
\eea
where in the first line we have used the unitarity of the CKM matrix, $\sum_{q'=d,s,b} |V_{q'q}|^2 =1$. The vanishing $L_q$ are the result of charge conservation. Thus, charge conservation in the partonic scattering only allows the up-type quarks and the down-type anti-quarks in the initial state. Note that the top quark flavor has been ignored since its PDF is highly suppressed and it cannot be produced in the final state for the chosen kinematics.

\section{Theoretical Framework}
\label{sec:TF}

The 1-Jettiness spectrum is characterized by three different regions: the peak region, the tail region, and the fixed order region, as shown in Table~\ref{tab:1-jettiness-regions}.
\begin{table}[h]
    \centering
    \begin{tabular}{|c|c|c|}
    \hline
      Regions & $\tau_1$ &  $\tau_{1a}$\\
       \hline
       \hline
      Peak Region &$\tau_1 \sim \Lambda_{{\rm QCD}}$ &  $\tau_{1a} \sim \>\>\Lambda_{\rm QCD}/Q_H$ \\
       (nonperturbative soft radiation) &  &\\
       \hline
        Tail Region   & $\>\>\Lambda_{\rm QCD} \ll \tau_1  \ll Q_H\>\>$ & $\>\>\Lambda_{\rm QCD}/Q_H \ll  \tau_{1a}  \ll 1\>\>$ \\
         (perturbative soft radiation) & & \\
         \hline
        Fixed Order Region   & $\tau_1 \sim Q_H$ & $\tau_{1a} \lesssim 1$ \\
       \hline
    \end{tabular}
    \caption{The peak, tail, and fixed order  regions of the $\tau_1$ and $\tau_{1a}$ 1-jettiness spectra. }
    \label{tab:1-jettiness-regions}
\end{table}

The resummation region $\xi\ll Q_H$, corresponding to the left panel in Fig.~\ref{fig:jettiness}, restricts final state radiation such that the energetic particles ($E\sim Q_H$) are closely aligned either with the beam or jet reference vectors and only soft radiation ($E\sim \xi\ll Q_H$) is allowed in other directions. Within the resummation region, the peak and tail regions correspond to  $\xi\sim \Lambda_{\rm QCD}$ and $\Lambda_{\rm QCD} \ll \xi \ll Q_H$, respectively. In the peak region,  the soft radiation becomes non-perturbative ($E\sim \Lambda_{\rm QCD}$) and must be treated with a  shape function model. The fixed order region, $\xi \sim Q_H$, corresponding to the right panel of Fig.~\ref{fig:jettiness}, is characterized by energetic radiation at wide angles from the beam or jet directions and corresponds to a looser veto on additional jets. There are no large Sudakov logarithms in this region, allowing for a treatment in perturbative QCD. These distinct regions of the 1-Jettiness spectrum can be smoothly combined by the schematic formula
\bea
\label{eq:spectrum_xsec}
d\sigma = [d\sigma_{\rm resum} - d\sigma_{\rm resum}^{\rm FO}] + d\sigma^{\rm FO},
\eea
where $d\sigma_{\rm resum}$ denotes the  resummed cross section, $d\sigma_{\rm resum}^{\rm FO}$ denotes the partonic resummed cross section expanded to fixed order in perturbation theory, and $d\sigma^{\rm FO}$ denotes the full partonic cross section at the same fixed order in perturbation theory. The $d\sigma_{\rm resum}^{\rm FO}$ cross section differs from $d\sigma^{\rm FO}$ by terms that are non-singular in the limit $\xi\to 0$. In the resummation region where the singular terms dominate, the cross section is dominated by $d\sigma_{\rm resum}$ due to the cancellation between $d\sigma_{\rm resum}^{\rm FO}$ and $d\sigma^{\rm FO}$, up to the relatively suppressed non-singular terms in $d\sigma^{\rm FO}$. Likewise, in the fixed order region where the Sudakov logarithms become small, the cross section is dominated by $d\sigma^{\rm FO}$ due to the cancellation between  $d\sigma_{\rm resum}$ and  $d\sigma_{\rm resum}^{\rm FO}$, up to relatively suppressed terms of higher order in $\alpha_s$ in $d\sigma_{\rm resum}$. In this manner, the schematic formula in Eq.~(\ref{eq:spectrum_xsec}) smoothly interpolates between the resummation and fixed order regions, providing a complete description of the  1-Jettiness spectrum.

\subsection{Factorization and Resummation}

In this section, we focus on the factorization  formula in the resummation region, $\xi\ll Q_H$. In particular, we examine  the structure of hadronization effects and the required renormalon subtractions. We derive new universality relations that relate hadronization effects and renormalon subtractions in the  $\tau_1$ and $\tau_{1a}$ DIS event shapes to those appearing in the DIS thrust $\tau_{1b}$ event shape. These universality relations can be exploited to constrain hadronization effects and improve strong coupling constant extractions~\cite{Becher:2008cf, Abbate:2010xh,Abbate:2012jh,Bell:2023dqs,Benitez:2024nav,Benitez:2025vsp} through global analyses of DIS event shapes.

We first review the form of the factorization formula in the resummation region, which is given by~\cite{Kang:2012zr,Kang:2013wca,Kang:2013nha} 
\bea
\label{eq:factorization_resum}
 d\sigma_{\rm resum} \left [\xi,\{{\cal O}_i\} \right ] &=&\sigma_0  \>H(Q_H, \mu; \mu_H)  \int ds_J \int dt_B \> {\cal S}\left(\xi - \frac{t_B}{Q_B}-\frac{s_J}{Q_J}, \mu;\mu_S\right) 
\nn \\
\times J(s_J, \mu;\mu_J)&&\hspace*{-0.7cm} \left [ \sum_{q} L_{q}\> B_{q}(t_B, x_*,\mu;\mu_B)  + \sum_{\bar{q}} L_{\bar{q}}\> B_{\bar{q}}(t_B, x_*,\mu;\mu_B) \right ] ,
\eea
where $\sigma_0 =\sigma_0 (\{{\cal O}_i\} )$ is the single photon exchange Born cross section, and its exact form depends on the chosen set of kinematic observables $\{{\cal O}_i\} $. The $L_q$ and $L_{\bar{q}}$ coefficients contain the necessary modifications to include the Born-level Z-boson contributions for NC DIS or to give the born cross section for CC DIS. Field theoretic definitions of the hard ($H$), jet ($J$), beam ($B_{q,\bar{q}}$), and soft (${\cal S}$) functions can be found in Ref.~\cite{Kang:2013wca}. The quark or anti-quark beam functions ($B_{q,\bar{q}}$) are matched~\cite{Stewart:2009yx} onto the PDFs as
\bea
\label{eq:beam}
B_{q,\bar{q}}(t_B, x,\mu;\mu_B) &=&\sum_{i} \int_x^1 \frac{dz}{z} {\cal I}_{({q,\bar{q}})i}\left(t_B, \frac{x}{z},  \mu;\mu_B\right) f_{i/p}(z,\mu_B),
\eea
where the ${\cal I}_{q i}$ and ${\cal I}_{\bar{q}i}$ are perturbatively calculable matching coefficients. The index $i$ runs over the  initial parton species, including the different flavors of quarks and anti-quarks, and the gluon.  The hard, beam, jet, and soft functions in Eq.~(\ref{eq:factorization_resum}) include their renormalization group evolution from their natural scales $\mu_H, \mu_B, \mu_J$, and $\mu_S$, respectively, to the common scale $\mu$. These natural scales are of typical size $\mu_H\sim Q_H$, $\mu_B\sim\mu_J\sim \sqrt{\xi\> Q_H}$ , and $\mu_S\sim \xi$, corresponding to scalings that minimize large logarithms in their perturbative expansions.  In order to smoothly match the resummation and fixed order regions of the 1-Jettiness spectrum using Eq.~(\ref{eq:spectrum_xsec}), the beam, jet, and soft scales must smoothly merge with the hard scale in the fixed order region. This is achieved through the use of profile functions~\cite{Berger:2010xi,Stewart:2011cf}. In our analysis we use the same profile functions as in Ref.~\cite{Cao:2024ota},  based on those introduced in Ref.~\cite{Kang:2013nha}.

The renormalization group equations have the general form
\bea
\label{eq:RGevolmom}
H(Q_H, \mu; \mu_H) &=& U_H (Q_H,\mu, \mu_H)\>H(Q_H, \mu_H), \nn \\
{\cal I}_{({q,\bar{q}})i}\left(t, x,  \mu;\mu_B\right) &=& \int dt' \>U_B(t-t',\mu,\mu_B)\>{\cal I}_{({q,\bar{q}})i}\left(t', x,\mu_B\right) , \nn \\
J(s, \mu;\mu_J) &=& \int ds' \>U_J(s-s',\mu,\mu_J)\> J(s', \mu_J) , \nn \\
{\cal S}(k, \mu;\mu_J) &=& \int dk' \>U_S(k-k',\mu,\mu_J)\> {\cal S}(k', \mu_J) , 
\eea
where  $U_H (Q_H,\mu, \mu_H), U_B(t,\mu,\mu_B), U_J(s,\mu,\mu_J),$ and $U_S(k,\mu,\mu_S)$ are the evolution factors.
The beam, jet, and soft function evolution equations have a  convolution structure. The PDFs in Eq.~(\ref{eq:beam}) are evaluated at the beam function scale, $\mu_B$, using standard DGLAP evolution. The quark jet function $J$ is the same for all light quark and anti-quark flavors, a consequence of charge conjugation and light quark flavor symmetry of QCD, so that we can write $J_q(s_J, \mu;\mu_J) = J_{\bar{q}}(s_J, \mu;\mu_J)\equiv J(s_J, \mu;\mu_J) $. This allows us to  factor out the jet function from the sum over quark and anti-quark flavors, as shown in Eq.~(\ref{eq:factorization_resum}).  The factorization formula in Eq.~(\ref{eq:factorization_resum}) can be expressed in terms of the position space beam, jet, and soft functions as
\bea
\label{eq:fac_resum_pos}
d\sigma_{\rm resum} \left [\xi,\{ {\cal O}_i \} \right ] &=&\sigma_0\>  H(Q_H, \mu ; \mu_H)\>   \int \frac{dy}{2\pi}   e^{iy \xi} \>J(\frac{y}{Q_J}, \mu; \mu_J)\>{\cal S}\left(y,\mu ;\mu_S\right)\nn \\
&&\times \Big [ \sum_{q}\sum_{i}  L_q \int_{x_*}^1 \frac{dz}{z} \> {\cal I}_{qi}\left(\frac{x_*}{z}, \frac{y}{Q_B}, \mu; \mu_B\right) f_{i/p}(z,\mu_B) \nn \\
&&+  \sum_{\bar{q}}\sum_{i}  L_{\bar{q}} \int_{x_*}^1 \frac{dz}{z} \> {\cal I}_{\bar{q}i}\left( \frac{x_*}{z},  \frac{y}{Q_B}, \mu ; \mu_B\right) f_{i/p}(z,\mu_B) \Big ],
\eea
where the momentum and position space functions are are related to each other via the Fourier transforms
\bea
\label{eq:FT_BJS}
{\cal I}_{(q,\bar{q})i}(t_B, x,\mu_B) &=& \int \frac{dy}{2\pi}\> e^{iy t_B} {\cal I}_{(q,\bar{q})i}(y, x,\mu_B),\nn \\
J(s_J,\mu_J) &=& \int \frac{dy}{2\pi} \> e^{iys_J} J(y,\mu_J), \\
{\cal S}(\tau_1 , \mu_S)&=&\int \frac{dy}{2\pi} \> e^{iy\tau_1} \>{\cal S}(y, \mu_S). \nn
\eea
We note that in position space, the renormalization group evolution equations for the beam, jet, and soft functions become multiplicative:
\bea
\label{eq:BJSevolPos}
{\cal I}_{(q,\bar{q})i}(y, x,\mu; \mu_B) &=& U_B(y,\mu,\mu_B) \> {\cal I}_{(q,\bar{q})i}(y,x,\mu_B), \nn \\
J(y,\mu;\mu_J) &=& U_J(y,\mu,\mu_J) \> J(y,\mu_J),  \\
{\cal S}(y, \mu;\mu_S) &=& U_S(y,\mu,\mu_S) {\cal S}(y, \mu_S),\nn
\eea
where $U_B(y_B,\mu,\mu_B), U_J(y_J,\mu,\mu_J),$ and $U_S(y_S,\mu,\mu_S)$ are the position space evolution factors.

\subsection{Universality of Hadronization Effects}
\label{sec:UnivHad}

The soft functions that appear in the factorization and resummation formulae for  the jet-based $\xi$ global event shapes are given by
\bea
\label{eq:Stau1tau1a}
{\cal S}(\xi,\mu) &=& \int dk_B \int dk_J \>\delta(\xi-\lambda_B k_B-\lambda_J k_J) \>{\cal S}_{\xi} (k_B,k_J,\mu), 
\eea
 where ${\cal S}_{\xi}(k_B,k_J,\mu)$   denotes the generalized~\cite{Jouttenus:2011wh} hemisphere soft function,  explicitly defined  in appendix~\ref{appexSoftFunc} for the $\tau_1$ and $\tau_{1a}$ event shapes in Eqs.~(\ref{eq:Stau1}) and (\ref{eq:Stau1a}), respectively.  For the $\tau_{1a}$ event shape, we note the useful relation $Q {\cal S}(Q\tau_{1a},\mu) = {\cal S}(\tau_{1a},\mu)$ which easily follows from Eq.~(\ref{eq:Stau1tau1a}) with $\xi=Q\tau_{1a}$. The arguments $k_B=\sum_{i\in B}n_B\cdot k_i$ and $k_J=\sum_{i\in J}n_J\cdot k_i$ denote the total momentum components of the soft particles in the beam and jet regions along $n_B^\mu$ and $n_J^\mu$, respectively. From Eq.~(\ref{eq:tau1tau1a2}), we see that the $\lambda_{B,J}$ factors are given by
 \bea
\label{eq:lambdaBJgen}
\lambda_{B}=\frac{\omega_{B}}{Q_B}, \>\>\> \lambda_{J}=\frac{\omega_{J}}{Q_J},
\eea
where $\omega_B$ and $\omega_J$ are given in Eq.~(\ref{eq:wBwJ}). 
Using the choices of $Q_{B}$ and $Q_J$ in Eq.~(\ref{eq:beam_ref_choices}) for  $\tau_1$ and $\tau_{1a}$,  these factors become
\bea
\label{eq:lambdaBJ}
&&\xi=\tau_1: \qquad \>\>\>\>\>\lambda_B=1, \qquad  \lambda_J = 1, \nn \\
&&\xi=Q\tau_{1a}: \qquad \lambda_B=\frac{\omega_B}{Q}, \qquad \lambda_J = \frac{\omega_J}{Q}. 
\eea

The kinematic dependence of the non-perturbative hadronization effects can be made explicit by relating the  generalized hemisphere soft functions ${\cal S}_{\xi}(k_B,k_J,\mu)$  to the standard DIS hemisphere soft function,  ${\cal S}_{\rm hemi.}(k_1,k_2,\mu)$, that appears in the factorization formula of the DIS thrust event shape, $\tau_{1b}$,~\cite{Kang:2013nha,Ee:2025scz} in the Breit frame.  The arguments $k_1$ and $k_2$ are given by $k_1=\sum_{k\in 1}n_1\cdot p_k$ and $k_2=\sum_{k\in 2}n_2\cdot p_k$,  corresponding to contributions from the soft particles of momenta $p_k$ in the two back-to-back hemisphere regions defined by the null reference vectors  
 \bea
 \label{eq:n1n2}
n_1=(1,\hat{n}), \qquad n_2=(1,-\hat{n}). 
 \eea
Unlike the soft functions ${\cal S}_{\xi}(k_B,k_J,\mu)$ that can depend on the dynamically varying scalar product, $n_B\cdot n_J$,   ${\cal S}_{\rm hemi.}(k_1,k_2,\mu)$  depends on $n_1\cdot n_2$ which takes on a fixed value for every event:
\bea
 n_1\cdot n_2=2.
 \eea
It was shown in Ref.~\cite{Kang:2013nha}, and reviewed in Appendix~\ref{appexSoftFunc}, that the soft functions ${\cal S}_{\xi}(k_B,k_J,\mu)$   are related to ${\cal S}_{\rm hemi.}(k_1,k_2,\mu)$ by the relation
\bea
\label{eq:Sgenhemi}
{\cal S}_{\xi}(k_B,k_J,\mu) =\frac{1}{R_BR_J} {\cal S}_{\rm hemi.}\left(\frac{k_B}{R_B},\frac{k_J}{R_J},\mu\right).
\eea
$R_B$ and $R_J$ are rescaling factors used to define a new set of transformed reference null vectors $n_B'$ and $n_J'$ as
\bea
\label{eq:nBRBnJRJ}
n_B' = n_B/R_B, \qquad n_J'= n_J/R_J,
\eea
such that they have the same scalar product as $n_1$ and $n_2$:
 \bea
 \label{eq:nBnJ2}
 n_B'\cdot n_J'=n_1\cdot n_2=2.
 \eea
As explained in appendix~\ref{appexSoftFunc}, the rescaling factors $R_B$ and $R_J$ for $\tau_1$~\cite{Cao:2024ota} and $\tau_{1a}$~\cite{Kang:2013nha} , are given by
\bea
\label{eq:RBRJtau1}
\tau_1:&& \qquad R_B = R_J = \sqrt{\frac{n_B\cdot n_J}{2}}\equiv r_S, \nn \\
\tau_{1a}:&& \qquad R_B = \sqrt{\frac{\omega_J \>n_B\cdot n_J}{2\omega_B}}, \qquad R_J=\sqrt{\frac{\omega_B \>n_B\cdot n_J}{2\omega_J}} .
\eea

The fact that  ${\cal S}(\xi,\mu)$ can be expressed in terms of the standard hemisphere soft function, ${\cal S}_{\rm hemi.}(k_1,k_2,\mu)$, through (\ref{eq:Sgenhemi}), corresponds to a statement of universality: the 1-Jettiness soft functions ${\cal S}(\xi,\mu)$ will inherit  their properties directly from ${\cal S}_{\rm hemi.}(k_1,k_2,\mu)$ through Eqs.~(\ref{eq:Stau1tau1a}) and (\ref{eq:Sgenhemi}). In particular, there will be a direct relationship between the hadronization models and renormalon subtractions implemented for the standard DIS hemisphere soft function and the 1-Jettiness  soft functions.  Using Eq.~(\ref{eq:Sgenhemi}), and a simple change of integration variables in Eq.~(\ref{eq:Stau1tau1a}),  ${\cal S}(\xi,\mu)$  can be written in terms of the standard hemisphere soft function as
\bea
\label{eq:Stauxi}
{\cal S}(\xi,\mu) = \int dk_B \int dk_J \>\delta(\xi - {\cal R}_B k_B -{\cal R}_J k_J) \>{\cal S}_{\rm hemi.}(k_B,k_J,\mu) ,
\eea
where we have defined ${\cal R}_{B,J}$ as
\bea
{\cal R}_B = \lambda_B R_B, \qquad {\cal R}_J = \lambda_J R_J. 
\eea
For $\xi=\tau_1$, we see from Eq.~(\ref{eq:lambdaBJ}) that ${\cal R}_B=R_B$ and ${\cal R}_J=R_J$. Likewise for $\xi=Q\tau_{1a}$, we have  ${\cal R}_B^2={\cal R}_J^2=\omega_B\omega_J n_B\cdot n_J/(2Q^2)=2q_B\cdot q_J/Q^2= 2 x P\cdot q/Q^2=1$, where we used $2q_B\cdot q_J=2xP\cdot (xP+q)$ in the $\xi \ll Q_H$ limit. Summarizing these results, we have
 \bea
\label{eq:scriptRBJ}
&&\xi=\tau_1: \qquad \>\>\>\>\>{\cal R}_B=R_B, \>\>\> {\cal R}_J = R_J, \nn \\
&&\xi=Q\tau_{1a}: \qquad {\cal R}_B=1, \>\>\> {\cal R}_J=1.
\eea
For $\xi=Q\tau_{1a}$, since ${\cal R}_B={\cal R}_J=1$, the kinematic dependence cancels exactly between the $\lambda_{B,J}$ and $R_{B,J}$ factors. This implies that ${\cal S}(Q\tau_{1a},\mu)$ is  related to ${\cal S}_{\rm hemi.}$ in exactly the same manner as the DIS thrust, $\tau_{1b}$, soft function is related to ${\cal S}_{\rm hemi.}$, as was first shown in Ref.~\cite{Kang:2013nha}.  By contrast,  for $\xi=\tau_1$ there is a non-trivial hard kinematic dependence in the relation of ${\cal S}(\tau_1,\mu)$ to ${\cal S}_{\rm hemi.}$  since  ${\cal R}_B=R_B$ and ${\cal R}_J=R_J$. Thus,  the soft functions for DIS thrust, $\tau_{1a}$, $\tau_1$, and other possible definitions of $\xi$, simply correspond to different projections of ${\cal S}_{\rm hemi.}$, determined by the corresponding values of ${\cal R}_B$ and ${\cal R}_J$ in Eq.~(\ref{eq:Stauxi}).

This result implies that the hadronization effects, including hadron mass effects~\cite{Mateu:2012nk},  for DIS thrust and the jet-based 1-Jettiness class of event shapes $\xi$ are all determined by hadronization effects  in  ${\cal S}_{\rm hemi.}$. Furthermore, the structure of renormalons and the required implementation of renormalon subtractions is also determined by ${\cal S}_{\rm hemi.}$.  The interplay of any hard kinematic dependence through ${\cal R}_B$ and ${\cal R}_J$ with hadronization effects in ${\cal S}_{\rm hemi.}$ leads to a non-trivial kinematic dependence of hadronization effects in ${\cal S}(\xi,\mu)$, as in the case of $\xi=\tau_1$. 
As emphasized in Ref.~\cite{Boughezal:2026dvu},  
this non-trivial kinematic dependence provides an independent kinematic lever arm for constraining hadronization effects in the universal ${\cal S}_{\rm hemi.}$ soft function.

\subsection{Renormalon Subtracted DIS  Hemisphere Soft Function}
\label{RenormSubShemi}

In this section, we review the implementation of hadronization effects and renormalon subtractions in ${\cal S}_{\rm hemi.}(k_1,k_2,\mu)$, the DIS hemisphere soft function in the Breit frame. The corresponding results for the 1-Jettiness soft functions ${\cal S}(\xi,\mu)$ then immediately follow from Eq.~(\ref{eq:Stauxi}).  We follow the analysis of Refs.~\cite{Hoang:2007vb,Hoang:2008fs} for the hemisphere soft function~\cite{Korchemsky:1998ev,Korchemsky:1999kt,Bauer:2002aj,Lee:2006nr,Hoang:2007vb,Fleming:2007qr,Fleming:2007xt,Schwartz:2007ib,Becher:2008cf,Hoang:2008fs,Jain:2008gb,Hoang:2008yj,Abbate:2010xh,Kelley:2011ng,Hornig:2011iu,Monni:2011gb,Bachu:2020nqn,Hoang:2025uaa} 
 for $e^+e^-\to $ hadrons. While the DIS hemisphere function  is generally different from the hemisphere soft function for  $e^+e^-\to $ hadrons due to a difference in the soft  Wilson lines paths, the same procedure can still be applied. It was shown in Ref.~\cite{Kang:2015moa} that the standard hemisphere soft function for DIS and $e^+e^-\to $ hadrons are the same at least up to ${\cal O}(\alpha_s^2)$ in perturbation theory. We will use this result  later for renormalon subtractions. However, the hemisphere soft functions for DIS and  $e^+e^-\to $ hadrons are not generally the same at the non-perturbative level.

Non-perturbative hadronization effects are incorporated in ${\cal S}_{\rm hemi.}(k_1,k_2,\mu)$  through a convolution with a  shape function~\cite{Hoang:2007vb} 
\bea
\label{eq:hemiconv}
 {\cal S}_{\rm hemi.}(k_1,k_2,\mu) = \int dk_1' \int dk_2' \>  {\cal S}_{\rm hemi.}^{\rm part.}(k_1-k_1',k_2-k_2',\mu)\>  {\cal S}_{\rm hemi.}^{\rm mod.}(k_1',k_2') ,
\eea
where $ {\cal S}_{\rm hemi.}^{\rm part.}$ and ${\cal S}_{\rm hemi.}^{\rm mod.}$ denote the partonic  and non-perturbative shape function, respectively. The shape function satisfies the normalization condition
\bea
\label{eq:norm}
\int dk_1' \int dk_2'\> {\cal S}_{\rm hemi.}^{\rm mod.}(k_1',k_2') =1,
\eea
and is chosen to have non-zero support only for $k_{1,2}'\geq 0$, with an exponential fall off for $k_{1,2}'\gg \Lambda_{\rm QCD}$. Since the partonic soft function ${\cal S}_{\rm hemi.}^{\rm part.}(k_1-k_1',k_2-k_2',\mu)$ in Eq.~(\ref{eq:hemiconv}) only has support for $k_{1,2}\geq k_{1,2}'$, the full soft function $ {\cal S}_{\rm hemi.}(k_1,k_2,\mu)$ has non-zero support only for $k_{1,2}\geq 0$. This is consistent with the expectation of the zero-momentum partonic threshold of the partonic soft function.

However, the QCD mass gap  requires a minimum hadronic energy deposit, $\bar{\Delta}\sim \Lambda_{\rm QCD}$, indicating that  $ {\cal S}_{\rm hemi.}(k_1,k_2,\mu) $ should have non-zero support only for $k_{1,2} \geq \bar{\Delta}$. Through Eq.~(\ref{eq:hemiconv}), this can be achieved by introducing the gap parameter, $\bar{\Delta}$, in ${\cal S}_{\rm hemi.}^{\rm mod.}$ so that
\bea
\label{eq:hemiconvgap}
 {\cal S}_{\rm hemi.}(k_1,k_2,\bar{\Delta},\mu) = \int dk_1' \int dk_2' \>  {\cal S}_{\rm hemi.}^{\rm part.}(k_1-k_1',k_2-k_2',\mu)\>  {\cal S}_{\rm hemi.}^{\rm mod.}(k_1'-\bar{\Delta},k_2'-\bar{\Delta}). 
\eea
It is well-known~\cite{Gardi:2000yh,Hoang:2007vb,Hoang:2008fs} that ${\cal S}_{\rm hemi.}^{\rm part.}$, computed in the $\overline{\rm MS}$ scheme, has a leading infrared renormalon, corresponding to a pole at $u=1/2$ in the Borel plane.  This renormalon leads to an ${\cal O}(\Lambda_{\rm QCD})$ ambiguity in the  partonic threshold of ${\cal S}_{\rm hemi.}^{\rm part.}$. Through Eq.~(\ref{eq:hemiconvgap}), this ambiguity translates into an ${\cal O}(\Lambda_{\rm QCD})$ ambiguity in the gap parameter $\bar{\Delta}$ itself. Thus, the renormalon in the partonic soft function can be subtracted by  writing~\cite{Hoang:2008fs,Abbate:2010xh,Jain:2008gb,Hoang:2008yj} the gap parameter as
\bea
\label{eq:gap}
\bar{\Delta} =\Delta(R,\mu) + \delta(R,\mu) ,
\eea
where $\delta(R,\mu)$ is a perturbative series that contains the same renormalon as ${\cal S}_{\rm hemi.}^{\rm part.}$ and  has the form
\bea
\label{eq:deltaexp}
\delta(R,\mu) = R e^{\gamma_E}\sum_{i=1}^{\infty} \left [ \frac{\alpha_s(\mu)}{4\pi}\right ]^i \delta_i(R,\mu) .
\eea 
The scale $R$ in $\delta(R,\mu)$ defines the scheme choice for the renormalon subtraction. $\Delta(R,\mu)$ then gives a renormalon-free definition of the gap parameter.  Since the $\bar{\Delta}$ gap parameter in Eq.~(\ref{eq:gap}) is independent of both $R$ and $\mu$
\bea
\mu \frac{d}{d\mu} \bar{\Delta} =0, \qquad \qquad R \frac{d}{dR} \bar{\Delta} =0,
\eea
the non-perturbative renormalon-free gap parameter $\Delta(R,\mu)$ inherits an anomalous dimension. The corresponding evolution equations in the two-dimensional $(R,\mu)$-space  
\bea
\label{eq:Rmuevol}
\mu \frac{d}{d\mu}\Delta(R,\mu) = -\mu \frac{d}{d\mu}\delta(R,\mu), \qquad \qquad R \frac{d}{dR} \Delta(R,\mu) = -R \frac{d}{dR} \delta(R,\mu) ,
\eea
are determined by the perturbative series for $\delta(R,\mu)$ in Eq.~(\ref{eq:deltaexp}). 

The appropriate scheme choice parameter is of size $R\sim 1 $ GeV, corresponding to the non-perturbative scale and a shift of size of order $\Lambda_{\rm QCD}$ between the original gap parameter, $\bar{\Delta}$, and the renormalon-free gap parameter, $\Delta(R,\mu)$. This also ensures that the peak in the shape function $S_{\rm mod .} (k_1,k_2)$ remains near $k_1,k_2\sim \Lambda_{\rm QCD}$ after the renormalon subtraction.  On the other hand, the renormalization scale $\mu$ that appears in the hemisphere soft function in Eq.~(\ref{eq:hemiconvgap})  is of natural size $\mu_S \sim M_i^2/Q_H$, where $M_i$ denotes the jet mass of the $i$-th hemisphere, corresponding to the scale that minimizes logarithms in the soft function. As a result, in the tail region $\mu_S\sim M_i^2/Q_H\gg \Lambda_{\rm QCD}$, large logarithms of $\mu_S/R\gg 1$ appear in the perturbative series for $\delta(R,\mu_S)$. This requires  resummation, achieved by solving the $R$- and $\mu$-evolution~\cite{Hoang:2008fs,Hoang:2008yj} equations in Eq.~(\ref{eq:Rmuevol}). A closed form expression for the solution to these evolution equations that  evolve the renormalon-free gap parameter in $R$-$\mu$ space from  $\Delta(R_\Delta,\mu_\Delta)$ to  $\Delta(R,\mu)$, where $R_\Delta,\mu_\Delta$ are the input  scales, is given in Refs.~\cite{Benitez:2024nav,Ee:2025scz}.

Using Eq.~(\ref{eq:gap}) in Eq.~(\ref{eq:hemiconvgap}), followed by the change of integration variables, $k_{1,2}'\to k_{1,2}'+\delta$, the renormalon-subtracted hemisphere soft function can be brought into the form
\bea
\label{eq:ShemiSub}
 {\cal S}_{\rm hemi.}(k_1,k_2,\delta,\mu) &=& \int dk_1' \int dk_2' \>  {\cal S}_{\rm hemi.}^{\rm part.}(k_1-k_1' -\delta,k_2-k_2' -\delta,\mu) \nn \\
 &\times& {\cal S}_{\rm hemi.}^{\rm mod.}(k_1'-\Delta,k_2'-\Delta), 
\eea
where $ {\cal S}_{\rm hemi.}(k_1,k_2,\delta,\mu)$ is now expressed in terms of the renormalon-free gap parameter, $\Delta=\Delta(R,\mu)$, and the presence of the $\delta=\delta(R,\mu)$ shift in the arguments of ${\cal S}_{\rm hemi.}^{\rm part.}$ subtracts the renormalon in the perturbative series for $ {\cal S}_{\rm hemi.}^{\rm part.}$. It is important to expand the partonic soft function, including the renormalon subtraction terms involving $\delta(R,\mu)$, consistently at each order in $\alpha_s$ for a proper implementation of the renormalon subtraction.

A renormalon-free definition of the gap parameter, $\Delta(R,\mu)$, for $ {\cal S}_{\rm hemi.}(k_1,k_2,\delta,\mu) $ was given in Ref.~\cite{Hoang:2008fs}. It can be defined in terms of the position space hemisphere soft function
\bea
{\cal S}_{\rm hemi.}(y_1,y_2,\delta,\mu) = \int dk_1\>\int dk_2\>e^{-iy_1k_1}\>e^{-iy_2k_2}\> {\cal S}_{\rm hemi.}(k_1,k_2,\delta,\mu), 
\eea
which can be brought into the form
\bea
\label{eq:ShemiPos}
{\cal S}_{\rm hemi.}(y_1,y_2,\delta,\mu) &=& \int dk_1\>\int dk_2\>\int dk_1' \int dk_2' \> e^{-iy_1k_1}\>e^{-iy_2k_2} \nn \\
&\times& {\cal S}_{\rm hemi.}^{\rm part.}(k_1-k_1' -\delta,k_2-k_2' -\delta,\mu) 
  \>{\cal S}_{\rm hemi.}^{\rm mod.}(k_1'-\Delta,k_2'-\Delta), 
\eea
using Eq.~(\ref{eq:ShemiSub}). Expressing ${\cal S}_{\rm hemi.}^{\rm part.}$ and ${\cal S}_{\rm hemi.}^{\rm mod.}$ in terms of their corresponding position space functions and performing all  integrations, Eq.~(\ref{eq:ShemiPos}) can be brought into the simple multiplicative form
\bea
{\cal S}_{\rm hemi.}(y_1,y_2,\delta,\mu) &=& {\cal S}_{\rm hemi.}^{\rm part.}(y_1,y_2,\delta,\mu)\> {\cal S}_{\rm hemi.}^{\rm mod.}(y_1,y_2)\>e^{-i\Delta (y_1+y_2)},
\eea
where the renormalon-subtracted  position space partonic soft function, ${\cal S}_{\rm hemi.}^{\rm part.}(y_1,y_2,\delta,\mu)$, is given by
\bea
\label{eq:ShemiPosRenSub}
 {\cal S}_{\rm hemi.}^{\rm part.}(y_1,y_2,\delta,\mu) &=&  {\cal S}_{\rm hemi.}^{\rm part.}(y_1,y_2,\mu)\>e^{-i\delta (y_1+y_2)} ,
\eea
and ${\cal S}_{\rm hemi.}^{\rm part.}(y_1,y_2,\mu)$ and ${\cal S}_{\rm hemi.}^{\rm mod.}(y_1,y_2)$ are the position space functions
\bea
{\cal S}_{\rm hemi.}^{\rm part.}(y_1,y_2,\mu) &=& \int dk_1 \>\int dk_2\> e^{-iy_1k_1}\> e^{-iy_2k_2}\>{\cal S}_{\rm hemi.}^{\rm part.}(k_1,k_2,\mu), \nn \\
{\cal S}_{\rm hemi.}^{\rm mod.}(y_1,y_2) &=& \int dk_1 \>\int dk_2\> e^{-iy_1k_1}\> e^{-iy_2k_2}\>{\cal S}_{\rm hemi.}^{\rm mod.}(k_1,k_2). 
\eea
Since ${\cal S}^{\rm part.}_{\rm hemi. }(y_1,y_2,\delta,\mu)$ satisfies the exponentiation property~\cite{Hoang:2008fs} to all orders in perturbation theory, according to the non-abelian exponentiation theorem~\cite{Gatheral:1983cz,Frenkel:1984pz}, a renormalon subtraction scheme can be conveniently defined by the condition~\cite{Hoang:2008fs} 
\bea
\frac{R}{2}e^{\gamma_E}\frac{d}{d\ln(i y_1)}\ln \left [{\cal S}^{\rm part.}_{\rm hemi.}(y_1,y_2,\delta,\mu) \right ] \Big |_{y_1=y_2=(iRe^{\gamma_E})^{-1}}&=& 0 .
\eea
Using Eq.~(\ref{eq:ShemiPosRenSub}) in the above condition yields an explicit expression for  $\delta(R,\mu)$ given by
\bea
\label{eq:deltaR}
\delta(R,\mu)=\frac{R}{2}e^{\gamma_E}\frac{d}{d\ln(i y_1)}\ln \left [{\cal S}^{\rm part.}_{\rm hemi.}(y_1,y_2,\mu)  \right ] \Big |_{y_1=y_2=(iRe^{\gamma_E})^{-1}} ,
\eea
which can be used to compute the coefficients $\delta_i(R,\mu)$ in the perturbative series in Eq.~(\ref{eq:deltaexp}).  Using the fact~\cite{Kang:2015moa} that  the  hemisphere soft functions for DIS and $e^+e^-\to $ hadrons are equal up to ${\cal O}(\alpha_s^2)$ in perturbation theory, the coefficients $\delta_1$ and $\delta_2$ in Eq.~(\ref{eq:deltaexp}) can be directly lifted from the analysis~\cite{Hoang:2008fs}  of $e^+e^-\to$ hadrons. Closed form expressions for the ${\cal O}(\alpha_s)$ and  ${\cal O}(\alpha_s^2)$ coefficients, $\delta_1(R,\mu)$ and $\delta_2(R,\mu)$,  needed for N$^2$LL and N$^3$LL resummation, respectively, can be found in Refs.~\cite{Hoang:2008fs,Abbate:2010xh,Ee:2025scz}.

\subsection{Renormalon Subtracted DIS 1-Jettiness Soft Functions}

Using Eq.~(\ref{eq:ShemiSub}) in Eq.~(\ref{eq:Stauxi}),  we can write the 1-Jettiness soft functions ${\cal S}(\xi,\mu)$  as
\bea
{\cal S}(\xi,\delta, \mu) &=& \int dk_B \int dk_J  \int dk_B' \int dk_J'\>\delta(\xi-{\cal R}_B k_B-{\cal R}_J k_J)\nn \\
&\times& {\cal S}^{\rm part.}_{\rm hemi.}(k_B-k_B'-\delta,k_J-k_J'-\delta,\mu)\>   {\cal S}_{\rm hemi.}^{\rm mod.}(k_B'-\Delta,k_J'-\Delta). 
\eea
Performing another change of integration variables, $k_B\to u_B=k_B-k_B'-\delta$ and $k_J\to u_J=k_J-k_J'-\delta$, followed by relabeling the integration variables, $k_{B,J}'\to k_{B,J}$, ${\cal S}(\xi,\delta,\mu) $ can be brought into the form
\bea
\label{eq:Stau1sub}
{\cal S}(\xi,\delta,\mu) &=& \int dk_B \int dk_J\>{\cal S}^{\rm part.}\Big (\xi-{\cal R}_B k_B-{\cal R}_J k_J - \left({\cal R}_B+{\cal R}_J\right)\delta,\mu\Big) \nn \\
&\times& {\cal S}_{\rm hemi.}^{\rm mod.}(k_B-\Delta,k_J-\Delta),
\eea
where the  partonic soft function, ${\cal S}^{\rm part.}(k,\mu)$, is related to the partonic hemisphere soft function as
\bea
{\cal S}^{\rm part.}(k,\mu) = \int du_B \int du_J\> \delta (k-{\cal R}_B u_B - {\cal R}_J u_J) \>{\cal S}^{\rm part.}_{\rm hemi.}(u_B,u_J,\mu). 
\eea
Performing the change of integration variables, $u={\cal R}_Bk_B+{\cal R}_Jk_J$ and $\zeta={\cal R}_Bk_B-{\cal R}_Jk_J$, in Eq.~(\ref{eq:Stau1sub}), the final form of the renormalon-subtracted 1-jettiness soft function is given by
\bea
\label{eq:SxiRS}
{\cal S}(\xi,\delta,\mu) &=& \int_0^\infty du\>{\cal S}^{\rm part.}\left (\xi-u - \left({\cal R}_B+{\cal R}_J\right)\delta,\mu\right) \> F^{\rm mod.}(u,{\cal R}_B,{\cal R}_J,\Delta),
\eea
where the shape function $F^{\rm mod.}$ is related to  the hemisphere shape function ${\cal S}_{\rm hemi.}^{\rm mod.}$ as
\bea
\label{eq:FmodRS}
F^{\rm mod.}(u,{\cal R}_B,{\cal R}_J,\Delta) = \int_{-u+2{\cal R}_B\Delta}^{u-2{\cal R}_J\Delta} \frac{d\zeta}{2{\cal R}_B {\cal R}_J} \>{\cal S}_{\rm hemi.}^{\rm mod.}(\frac{u+\zeta}{2{\cal R}_B}-\Delta,\frac{u-\zeta}{2{\cal R}_J}-\Delta) ,
\eea
and the normalization condition in Eq.~(\ref{eq:norm}) becomes equivalent to the condition
\bea
\label{eq:FmodRSnorm}
\int du\> F^{\rm mod.}(u,{\cal R}_B,{\cal R}_J,\Delta) =1.
\eea
Eqs.~(\ref{eq:SxiRS}) and (\ref{eq:FmodRS}), provide the general formula for the renormalon-subtracted 1-jettiness soft functions, ${\cal S}(\xi,\delta,\mu)$. We note that these general formulae apply for the entire class of 1-Jettiness observables defined in Eq.~(\ref{tau1andtau1a}) for any choice of the $Q_B$ and $Q_J$ constants that define the 1-Jettiness observable.

Using the expressions for the ${\cal R}_B$ and ${\cal R}_J$ kinematic factors given in Eq.~(\ref{eq:scriptRBJ}), we can give the explicit expressions for ${\cal S}(\tau_1,\delta,\mu)$ and ${\cal S}(Q\tau_{1a},\delta,\mu)$, the soft functions for the $\tau_1$ and $\tau_{1a}$ event shapes.
The expression for ${\cal S}(\tau_1,\delta,\mu)$ is obtained by setting $R_B=R_J=r_S$ as
\bea
\label{eq:Stau1RSrs}
{\cal S}(\tau_1,\delta,\mu) &=& \int_0^\infty du\>{\cal S}^{\rm part.}\left (\tau_1-u - 2r_S \delta,\mu\right) \> F^{\rm mod.}(u-2r_S\Delta, r_S),
\eea
where the shape function, $F^{\rm mod.}(u,r_S)$, given in terms of the DIS thrust hemisphere shape function as
\bea
\label{eq:Fmodtau1urS}
F^{\rm mod.}(u,r_S) = \frac{1}{2r_S^2}\int_{-u}^{u} \frac{d\zeta}{2} \> {\cal S}_{\rm hemi.}^{\rm mod.}\left(\frac{u+\zeta}{2r_S},\frac{u-\zeta}{2r_S}\right).
\eea
The corresponding expressions for $\xi=Q\tau_{1a}$ can be obtained by setting $r_S=1$ in Eqs.~(\ref{eq:Stau1RSrs}) and (\ref{eq:Fmodtau1urS}).

\subsection{Factorization and Resummation  with Renormalon Subtractions}
\label{fac_resum_renorm}

The factorization and resummation formula in Eq.~(\ref{eq:factorization_resum}) for 1-Jettiness, $\xi$,  is given without renormalon subtractions. Implementing the renormalon subtraction just corresponds to replacing the soft function ${\cal S}(\xi,\mu) $, defined in the $\overline{{\rm MS}}$ scheme, with the renormalon subtracted soft function  ${\cal S}(\xi,\delta, \mu) $ given in Eqs.~(\ref{eq:SxiRS}) and (\ref{eq:FmodRS}). It becomes convenient to work with the factorization formula expressed in terms of the beam, jet, and soft functions defined in position space, where their renormalization group equations take on the multiplicative form in Eq.~(\ref{eq:BJSevolPos}). The momentum and position space factorization formulae, without renormalon subtractions, are related to each other by
\bea
\label{eq:facresum}
d\sigma_{\rm resum}  \left [\xi,\{{\cal O}_i \}\right ] &=& \int \frac{dy}{2\pi} e^{iy \xi} \>d\sigma_{\rm resum}\left [y,\{{\cal O}_i \}\right ],
\eea
where $d\sigma_{\rm resum}\left [y,\{{\cal O}_i \}\right ]$ denotes cross section in position space, without the renormalon subtraction. $d\sigma_{\rm resum}\left [y,\{{\cal O}_i \}\right ]$ involves the position space soft function  ${\cal S}(y,\mu) $, as seen in Eq.~(\ref{eq:fac_resum_pos}). Thus, implementing the renormalon subtraction  corresponds to replacing  ${\cal S}(y,\mu) $ with
${\cal S}(y,\delta, \mu) $, the Fourier transform of the renormalon-subtracted momentum space soft function
\bea
\label{eq:Sydelta}
{\cal S}(y,\delta,\mu) =  \int d\xi\>e^{-iy\xi} \>{\cal S}(\xi,\delta,\mu). 
\eea
Using Eq.~(\ref{eq:SxiRS}) in Eq.~(\ref{eq:Sydelta}), followed by the change of integration variable $\xi\to \xi - ({\cal R}_B+{\cal R}_J)\delta$, ${\cal S}(y,\delta,\mu)$ takes the form
\bea
\label{eq:SyRS}
{\cal S}(y,\delta,\mu) &=&  e^{-i y ({\cal R}_B \delta +{\cal R}_J\delta) }\>{\cal S}^{\rm part.}(y,\mu)\>\int du\>e^{-iyu} F^{\rm mod.}(u, R_B,R_J,\Delta) , 
\eea
where ${\cal S}^{\rm part.}(y,\mu)$  denotes the position space partonic soft function  without renormalon subtraction
\bea
{\cal S}^{\rm part.}(y,\mu) =  \int d\xi\>e^{-iy\xi} \>{\cal S}^{\rm part.}(\xi,\mu). 
\eea
 The renormalon subtractions are now entirely encoded in the pre-factor $e^{-i y ({\cal R}_B \delta +{\cal R}_J\delta) }$ in Eq.~(\ref{eq:SyRS}). This form of the soft function implies that the renormalon-subtracted position space cross section, $d\sigma_{\rm resum}\left [y,\delta, \{{\cal O}_i \}\right ]$, can be written in the form
\bea
d\sigma_{\rm resum} \left [y,\delta, \{{\cal O}_i\} \right ] = e^{-i y ({\cal R}_B \delta +{\cal R}_J\delta) } \>d\sigma_{\rm resum}^{\rm part.} \left [y,\{{\cal O}_i\} \right ] \>\int du\>e^{-iyu} F^{\rm mod.}(u, {\cal R}_B,{\cal R}_J,\Delta),
\eea
where $d\sigma_{\rm resum}^{\rm part.} \left [y,\{{\cal O}_i\} \right ]$ is the position space partonic cross section without renormalon subtractions.
The corresponding momentum space  cross section is given by the Fourier transform
\bea
\label{eq:fac_intermediate}
d\sigma_{\rm resum}  \left [\xi,\delta, \{{\cal O}_i \}\right ] = \int du \int \frac{dy}{2\pi} e^{iy (\xi- u - {\cal R}_B \delta - {\cal R}_J\delta)} d\sigma_{\rm resum}^{\rm part.} \left [y,\{{\cal O}_i\} \right ] \>F^{\rm mod.}(u, {\cal R}_B,{
\cal R}_J,\Delta) ,
\eea
which can be brought into the form
\bea
\label{eq:fac_intermediate_1}
d\sigma_{\rm resum}  \left [\xi,\delta, \{{\cal O}_i \}\right ] = \int du \>d\sigma_{\rm resum}^{\rm part.} \left [\xi- u - {\cal R}_B \delta - {\cal R}_J\delta,\{{\cal O}_i\} \right ] \>F^{\rm mod.}(u, {\cal R}_B,{\cal R}_J,\Delta) .
\eea
Performing a Taylor series expansion of $d\sigma_{\rm resum}^{\rm part.}$  in powers of ${\cal R}_B\delta + {\cal R}_J\delta$, we can write
\bea
d\sigma_{\rm resum} \left [\xi,\delta, \{{\cal O}_i\} \right ] &=& \sum_{n=0}^\infty \frac{(-{\cal R}_B\delta-{
\cal R}_J\delta)^n}{n!}\int du  \>\frac{d^n}{d\xi^n}d\sigma_{\rm resum}^{\rm part.} \left [\xi-u,\{{\cal O}_i\} \right ] \nn \\
&\times& F^{\rm mod.}(u, {\cal R}_B,{\cal R}_J,\Delta) ,
\eea
which can be expressed more compactly as
\bea
\label{eq:fac_had_renormalon}
d\sigma_{\rm resum} \left [\xi,\delta, \{{\cal O}_i\} \right ] = \int du  \>d\sigma_{\rm resum}^{\rm part.} \left [\xi-u,\delta, \{{\cal O}_i\} \right ] F^{\rm mod.}(u, {\cal R}_B,{\cal R}_J,\Delta) ,
\eea
where the renormalon-subtracted partonic cross section, $d\sigma_{\rm resum}^{\rm part.} \left [\xi,\delta, \{{\cal O}_i\} \right ]$, is related to the partonic cross section without renormalon subtraction, $d\sigma_{\rm resum}^{\rm part.} \left [\xi, \{{\cal O}_i\} \right ]$, as
\bea
\label{eq:xsec_part_renormalon}
d\sigma_{\rm resum}^{\rm part.} \left [\xi,\delta, \{{\cal O}_i\} \right ] = e^{-({\cal R}_B\delta+{\cal R}_J\delta ) \frac{d}{d\xi}}\>d\sigma_{\rm resum}^{\rm part.} \left [\xi,\{{\cal O}_i\} \right ]. 
\eea
Eqs.~(\ref{eq:fac_had_renormalon}) and (\ref{eq:xsec_part_renormalon}) are  key results that show the universality of the leading hadronization effects and renormalon subtractions for 1-jettiness DIS event shapes. First, the soft hadronization effects are captured by the shape function $F^{\rm mod.}$ which is related through Eq.~(\ref{eq:FmodRS}) to the universal DIS hemisphere shape function ${\cal S}_{\rm hemi.}^{\rm mod.}$. Furthermore, the renormalon subtractions are determined by $\delta$ which is a perturbative series that contains the same renormalon as in ${\cal S}_{\rm hemi.}^{\rm part.}$, and is determined by Eq.~(\ref{eq:deltaR}). Finally, any additional hard kinematic dependence of hadronization effects is accounted for by the factors of ${\cal R}_B$ and ${\cal R}_J$ in both $F^{\rm mod.}$  and in the renormalon subtractions as seen in Eqs.~(\ref{eq:fac_had_renormalon}) and (\ref{eq:xsec_part_renormalon}).

Using Eqs.~(\ref{eq:Stau1RSrs}) and (\ref{eq:Fmodtau1urS}), the form of Eqs.~(\ref{eq:fac_had_renormalon}) and (\ref{eq:xsec_part_renormalon}) can be simplified for the $\tau_1$ event shape to give
\bea
\label{eq:sigtau1ren}
d\sigma_{\rm resum} \left [\tau_1,\delta, \{{\cal O}_i\} \right ] &=& \int du  \>d\sigma_{\rm resum}^{\rm part.} \left [\tau_1-u,\delta, \{{\cal O}_i\} \right ] \>F^{\rm mod.}(u-2r_S\Delta, r_S)  ,
\eea
where the renormalon-subtracted partonic cross section, $d\sigma_{\rm resum}^{\rm part.} \left [\tau_1,\delta, \{{\cal O}_i\} \right ] $, is related to the partonic cross section without the renormalon subtraction, $d\sigma_{\rm resum}^{\rm part.} \left [\tau_1,\{{\cal O}_i\} \right ]$ by
\bea
\label{eq:xsec_part_renormalon_tau1}
d\sigma_{\rm resum}^{\rm part.} \left [\tau_1,\delta, \{{\cal O}_i\} \right ] = e^{-2r_S\delta \frac{d}{d\xi}}\>d\sigma_{\rm resum}^{\rm part.} \left [\tau_1,\{{\cal O}_i\} \right ] .
\eea
We note that for the purposes of numerical implementation, it is more convenient to rewrite Eq.~(\ref{eq:fac_had_renormalon}) in the form
\bea
\label{eq:fac_had_renormalon_num}
d\sigma_{\rm resum} \left [\xi,\delta, \Delta, \{{\cal O}_i\} \right ] = \int du  \>d\sigma_{\rm resum}^{\rm part.} \left [\xi-u,\{{\cal O}_i\} \right ] \>e^{-(R_B\delta+R_J\delta ) \frac{d}{du}}F^{\rm mod.}(u, R_B,R_J,\Delta) ,
\eea 
which for the $\tau_1$ event shape takes the simplified form
\bea
\label{eq:sigtau1ren2}
d\sigma_{\rm resum} \left [\tau_1,\delta, \Delta, \{{\cal O}_i\} \right ] &=& \int du  \>d\sigma_{\rm resum}^{\rm part.} \left [\tau_1-u,\{{\cal O}_i\} \right ] \>e^{-2r_S\delta \frac{d}{du}}F^{\rm mod.}(u-2r_S\Delta, r_S)  .
\eea
This form can be understood by starting with Eq.~(\ref{eq:fac_intermediate}), and instead of carrying out the Fourier transform integral over the partonic cross section to arrive at Eq.~(\ref{eq:fac_intermediate_1}), performing a change of the integration variable $u\to u+R_B\delta + R_J\delta$ and then Taylor expanding the resulting $F^{\rm mod.}(u-R_B\delta -R_J\delta, R_B,R_J,\Delta) $ in powers of $(R_B\delta + R_J\delta)$. The corresponding formulae for $\xi=Q\tau_{1a}$ can be obtained by setting $r_S=1$ in Eqs.~(\ref{eq:sigtau1ren}), (\ref{eq:xsec_part_renormalon_tau1}), and (\ref{eq:sigtau1ren2}).

\subsection{Universal Non-perturbative Shift in Tail Region}

In the tail region  $\Lambda_{\rm QCD} \ll \xi \ll Q_H$, since $F^{\rm mod.}(u, {\cal R}_B,{\cal R}_J,\Delta)$ peaks near $u\sim \Lambda_{\rm QCD}$, the 1-Jettiness spectrum can be predicted using an OPE of the resummed cross section   in Eq.~(\ref{eq:fac_had_renormalon}) to get
\bea
\label{eq:OPE}
d\sigma_{\rm resum}\left [\xi,\delta,\{{\cal O}_i\} \right ]  = d\sigma_{\rm resum}^{\rm part.}\left [\xi, \delta, \{{\cal O}_i\} \right ]- \frac{d\sigma_{\rm resum}^{\rm part.}\left [\xi, \delta, \{{\cal O}_i\} \right ]}{d\xi} 2 \Omega_1
+\cdots ,
\eea
where the renormalon-subtracted first moment $2\Omega_1=2\Omega_1(R,\mu_S,{\cal R}_B,{\cal R}_J) $ is given by
\bea
\label{eq:moment}
2\Omega_1(R,\mu_S,{\cal R}_B,{\cal R}_J) = \int du\> u \>F^{\rm mod.}(u, {\cal R}_B,{\cal R}_J,\Delta) .
\eea
The dependence of $\Omega_1$ on the renormalon subtraction scheme parameter $R$ and the soft scale $\mu_S$ arises from the renormalon-free gap parameter $\Delta=\Delta(R,\mu_S)$.
We note that $\Omega_1$ can contain an additional hard kinematic dependence through the factors of ${\cal R}_B$ and ${\cal R}_J$. 
The first moment, $2\bar{\Omega}_1$, defined in the $\overline{{\rm MS}}$ scheme without the renormalon subtraction,  is given by
\bea
\label{eq:OmegaRBRJ}
2\bar{\Omega}_1({\cal R}_B,{\cal R}_J)=\int du\> u \>F^{\rm mod.}(u, {\cal R}_B,{\cal R}_J,\bar{\Delta}), 
\eea
where the original gap parameter, $\bar{\Delta}$, without renormalon subtraction is used. Using the relationship $\bar{\Delta}=\Delta(R,\mu_S)+\delta(R,\mu_S)$, $2\Omega_1$ and $2\bar{\Omega}_1$ are related to each other as
\bea
2\Omega_1(R,\mu_S,{\cal R}_B,{\cal R}_J) = 2\bar{\Omega}_1({\cal R}_B,{\cal R}_J) - ({\cal R}_B+{\cal R}_J)\>\delta(R,\mu_S).
\eea
As seen in Eq.~(\ref{eq:OPE}), the leading power correction in the tail region simply corresponds to the well-known~\cite{Korchemsky:1994is,Dokshitzer:1995zt,Dokshitzer:1995qm,Dokshitzer:1997ew,Korchemsky:1999kt,Berger:2003pk,Salam:2001bd,Belitsky:2001ij,Berger:2004xf,Bauer:2003di,Lee:2006fn,Lee:2006nr} non-perturbative shift in the partonic result determined by the first moment of the shape function
\bea
\label{eq:NPshift}
d\sigma_{\rm resum}\left [\xi,\delta, \{{\cal O}_i\} \right ]  = d\sigma_{\rm resum}^{\rm part.}\left [\xi- 2 \Omega_1, \delta, \{{\cal O}_i\} \right ] , \qquad \Lambda_{\rm QCD} \ll \xi \ll Q_H.
\eea
However, we now see that for the 1-Jettiness global event shapes $\xi$, this universal shift can have an additional  dependence on the hard kinematics through ${\cal R}_B$ and ${\cal R}_J$. Eqs.~(\ref{eq:moment}) and (\ref{eq:NPshift}) correspond to the more detailed version of Eqs.~(\ref{eq:OmegaOi}) and (\ref{eq:OPEshift}), respectively.

For the $\tau_1$ event shape the first moments $2\Omega_1=2\Omega_1(R,\mu_S,r_S)$ and $2\bar{\Omega}_1$ take on the simplified form
\bea
2\Omega_1(R,\mu_S,r_S)&=&\int du\> u \>F^{\rm mod.}(u-2r_S\Delta,r_S), \nn \\
 2\bar{\Omega}_1(r_S)&=&\int du\> u \>F^{\rm mod.}(u-2r_S\bar{\Delta},r_S), 
\eea
and are related to each other by
\bea
2\Omega_1(R,\mu_S,r_S) = 2\bar{\Omega}_1(r_S) - 2r_S\>\delta(R,\mu_S).
\eea
Once again the corresponding expressions for $2\Omega_1$ for $\xi=Q\tau_{1a}$ are obtained by setting $r_S=1$.

\subsection{Hadronization: Shape Function Model}
\label{sec:shape}

We have seen that hadronization effects for 1-Jettiness observables are captured by the $F^{\rm mod.}(u, {\cal R}_B,{\cal R}_J,\Delta)$ shape function in Eq.~(\ref{eq:FmodRS}).  For $\xi=\tau_1$, it simplifies to the form in Eq.~(\ref{eq:Fmodtau1urS}), and for $\xi=Q\tau_{1a}$ and DIS thrust it simplifies to Eq.~(\ref{eq:Fmodtau1urS}) with $r_S=1$. We see that  $F^{\rm mod.}$ is entirely determined by the DIS hemisphere shape function $S^{\rm mod.}_{\rm hemi.}(k_1,k_2)$.  This implies that the dominant hadronization effects for  the $\tau_1$, $\tau_{1a}$,   DIS thrust , and other definitions of the $\xi$ event shape arise through  $S^{\rm mod.}_{\rm hemi.}(k_1,k_2)$  for both the NC and CC DIS processes. In particular, a single choice for the shape function model for $S^{\rm mod.}_{\rm hemi.}(k_1,k_2)$ determines the dominant hadronization effects across all of these observables. For the purposes of generating numerical results, we use the shape function model~\cite{Korchemsky:2000kp}
\bea
\label{eq:shapehemi}
S^{\rm mod.}_{\rm hemi.}(k_1,k_2) &=& \theta(k_1)\theta(k_2)\frac{{\cal N}(a,b,\Lambda)}{\Lambda^2}\left ( \frac{k_1 k_2}{\Lambda^2}\right )^{a-1}{\rm exp}\left (\frac{-k_1^2 -k_2^2 - 2b k_1 k_2}{\Lambda^2} \right ),
\eea
where ${\cal N}(a,b,\Lambda)$ is a normalization factor to ensure the normalization condition in Eq.~(\ref{eq:norm}), and $a,b,$ and $\Lambda$ are free parameters that determine the model. We note that one can also express the two-dimensional shape function $S^{\rm mod.}_{\rm hemi.}(k_1,k_2) $ using a basis of products of one-dimensional functions~\cite{Hoang:2025uaa}.

\subsection{1-Jettiness Spectrum: Combining the Peak, Tail, and Fixed-Order Regions}

Eq.~(\ref{eq:spectrum_xsec}) gives the schematic formula that smoothly connects the peak ($\xi \sim \Lambda_{\rm QCD}$), tail ($\Lambda_{\rm QCD} \ll \xi \ll Q_H$), and fixed order ($\xi \sim Q_H$) regions of the 1-Jettiness spectrum. Here we elaborate on this schematic formula which has the more detailed form 
\bea
\label{eq:spec_formula}
d\sigma \left [\xi,\{{\cal O}_i\} \right ] &=& d\sigma_{\rm resum}\left [\xi, \delta, \{{\cal O}_i\} \right ]  - \>d\sigma_{\rm resum}^{\rm FO} \left [\xi, \{{\cal O}_i\} \right ]   +\>  d\sigma^{\rm FO} \left [\xi, \{{\cal O}_i\}\right ] .
\eea
$d\sigma_{\rm resum}\left [\xi,\delta, \{{\cal O}_i\} \right ]$, given by Eqs.~(\ref{eq:fac_had_renormalon}) and (\ref{eq:xsec_part_renormalon}), denotes the renormalon-subtracted cross section predicted by the SCET factorization and resummation formula in the peak and tail regions. Making its scale dependence explicit, it takes the form
\bea
\label{eq:dsigresum_exp_scale}
d\sigma_{\rm resum}\left [\xi, \delta, \{ {\cal O}_i\}\right ]=d\sigma_{\rm resum}\left [\xi,\delta,  \{ {\cal O}_i\},\mu;\mu_H,\mu_B,\mu_J,\mu_S, R \right ] .
\eea 
It is evaluated at an arbitrary reference scale $\mu$ to which the hard, beam, jet, and soft functions are evolved from their natural scales $\mu_H,\mu_B,\mu_J,$ and $\mu_S$, respectively. We choose the reference scale to be near the hard scale $\mu=\mu_H\sim Q_H$.  This factorized and resummed cross section also depends on the gap subtraction scale, $R\sim \mu_S$, which arises when switching the soft function from the $\overline{\rm MS}$-scheme to the renormalon-free R-gap scheme.  The $d\sigma_{\rm resum}^{\rm FO}$ contribution in Eq.~(\ref{eq:spec_formula}), with its scale dependence made explicit, is given by 
 \bea
 \label{eq:dsigresumFO}
 d\sigma_{\rm resum}^{\rm FO}\left [\xi, \{ {\cal O}_i\}\right ] = d\sigma_{\rm resum}^{\rm part.}\left [\xi, \{ {\cal O}_i\},\mu=\mu_H=\mu_B=\mu_J=\mu_S\right ],
 \eea
 corresponding to turning off resummation in the partonic SCET factorization and resummation formula where the soft function is computed in the $\overline{\rm MS}$-scheme. It gives the most singular terms in the  $\xi\to0$ limit in the partonic cross section in fixed order perturbation theory. Finally, the  $d\sigma^{\rm FO}$ contribution, with its scale dependence made explicit, is given by
 \bea
 \label{eq:dsigFO}
 d\sigma^{\rm FO} \left [\xi,\{{\cal O}_i\}\right ]=d\sigma^{\rm FO} \left [\xi,\{{\cal O}_i\},\mu=\mu_H\right ],
 \eea
and gives the full fixed-order perturbation theory result for the partonic cross section in the $\overline{{\rm MS}}$-scheme, evaluated at the scale, $\mu=\mu_H\sim Q_H$. It contains both the singular and non-singular terms of partonic cross section at fixed order in perturbation theory. It differs from  $d\sigma_{\rm resum}^{\rm FO}$  only by the additional non-singular terms.

In the peak and tail regions, $\xi \ll Q_H$, where the singular terms in fixed-order perturbation theory dominate, there is a cancellation between $d\sigma_{\rm resum}^{\rm FO}$ and $ \>d\sigma^{\rm FO}$ in Eq.~(\ref{eq:spec_formula}) up to the suppressed non-singular terms in $d\sigma^{\rm FO}$. Thus, as expected, in this region the cross section is dominated by the renormalon-subtracted resummed SCET factorization formula, $d\sigma_{\rm resum}\left [\xi, \delta, \{{\cal O}_i\} \right ] $,  the first term in Eq.~(\ref{eq:spec_formula}).  

\begin{table}[]
    \centering
    \begin{tabular}{|c|c|c|c|}
    \hline
      1-Jettiness & EIC-1 & EIC-2 & HERA\\
       Observable     & $\sqrt{s}=90.0$ GeV & $\sqrt{s}=140.0$ GeV  & $\sqrt{s}=319.0$ GeV\\
      \hline
       $d\sigma[\>\tau_{1(a)},P_{J_T}, y_J\>]$ & $P_{J_T}=[20,30]$ GeV  &  $P_{J_T}> 30$ GeV & $P_{J_T}>70$ GeV \\
       & $|y_J|<2.5$&$|y_J|<2.5$ & $|y_J|<2.5$\\
       \hline
       $d\sigma[\>\tau_{1(a)},Q^2, x\>]$ & $\> Q^2> 800$ GeV$^2$  & $\> Q^2> 800$ GeV$^2$ & $\> Q^2> 5000$ GeV$^2$ \\
          & $x>0.1$ & $x>0.01$ & $x> 0.001$\\
              \hline
       $d\sigma[\>\tau_{1(a)},Q^2, y\>]$ & $\> Q^2> 800$ GeV$^2$  & $\> Q^2> 800$ GeV$^2$ & $Q^2> 5000$ GeV$^2$ \\
       &$y=[0.2,0.6]$ & $y=[0.2,0.6]$ & $y=[0.2,0.6]$\\
       \hline
       $d\sigma[\>\tau_{1(a)},p_{T_e}, y_e\>]$ & $p_{T_e}=[20,30]$ GeV & $p_{T_e}> 30$ GeV & $p_{T_e}= [50,80]$ GeV \\
              & $|y_e|< 2.5$&$|y_e|< 2.5$& $|y_e|< 2.5$\\
       \hline
    \end{tabular}
    \caption{Kinematic settings chosen for various observables 1-Jettiness observables. All these observables can be applied to NC and CC DIS, expect for the one in the last row which only applies for NC DIS as it is differential in  $p_{T_e}, y_e$, requiring measuring the momentum of the scattered lepton. CC DIS differential in $Q^2,x,$ or $y$, requires reconstructing the missing energy corresponding to the scattered neutrino from the hadronic state using the Jacquet-Blondel~\cite{Jacquet:1979,Akhundov:1993wvk}  method. }
    \label{tab:kin_set}
\end{table}

In the tail and fixed order regions, $\xi\gg \Lambda_{\rm QCD}$, one can  perform an OPE of $d\sigma_{\rm resum}\left [\xi, \delta, \{{\cal O}_i\} \right ]$, as seen in Eq.~(\ref{eq:OPE}), which is equivalent to the universal shift of the partonic resummation formula as seen in Eq.~(\ref{eq:NPshift}).  In the fixed-order region, $\xi\sim Q_H$, where there are no large logarithms, the leading term, $d\sigma_{\rm resum}^{\rm part.}$, in the OPE  in Eq.~(\ref{eq:OPE}) cancels with $d\sigma_{\rm resum}^{\rm FO}$, up to higher-order terms suppressed in perturbation theory. Correspondingly, the full result is given by just the fixed-order contribution $d\sigma^{\rm FO}$. As seen from Eqs.~(\ref{eq:dsigresumFO}) and (\ref{eq:dsigFO}), in order for this cancellation to occur it is important that the scales $\mu_H,\mu_B,\mu_J,$ and $\mu_S$ smoothly merge into $\mu=\mu_H\sim Q_H$ in the fixed-order region. This is achieved through the use of profile functions~\cite{Berger:2010xi,Stewart:2011cf} which make the scales $\mu_B,\mu_J,$ and $\mu_S$ functions of $\xi$. Furthermore, since we must have $R\sim \mu_S$ in order to avoid large logarithms in the gap subtraction, $\delta(R,\mu_S)$, the renormalon subtraction scale, $R$, also becomes a function of $\xi$ through the profile function for $\mu_S$.  

Note that the cancellation between  $d\sigma_{\rm resum}^{\rm part.}\left [\xi, \delta, \{{\cal O}_i\} \right ]$ and $d\sigma_{\rm resum}^{\rm FO}\left [\xi, \{ {\cal O}_i\}\right ] $ in the fixed-order region is not exact, but leaves over the renormalon subtraction terms in $d\sigma_{\rm resum}^{\rm part.}\left [\xi, \delta, \{{\cal O}_i\} \right ]$ since $d\sigma_{\rm resum}^{\rm FO}$ is computed in  the $\overline{{\rm MS}}$-scheme without the renormalon subtraction. These left over renormalon subtraction terms then combine with $d\sigma^{\rm FO}\left [\xi, \{ {\cal O}_i\}\right ]$, also computed in the $\overline{\rm MS}$-scheme without renormalon subtractions, to give the full partonic fixed-order result in perturbation theory with the appropriate renormalon subtractions. We also note that the power corrections of the OPE in Eq.~(\ref{eq:OPE}) remain in the fixed order region, and correspondingly the non-perturbative matrix elements, such as the first moment $2\Omega_{1}(R,\mu_S, {\cal R}_B,{\cal R}_J)$ defined in Eq.~(\ref{eq:moment}), are interpreted in the same R-gap scheme.

We note that in Refs.~\cite{Abbate:2010xh,Abbate:2012jh,Benitez:2024nav,Ee:2025scz} a modified prescription is used in place of
Eq.~(\ref{eq:spec_formula}) to connect the peak, tail, and fixed-order regions. In particular, they implement a convolution with the shape function and the renormalon subtractions globally to all three terms in Eq.~(\ref{eq:spec_formula}).  It is argued in Refs.~\cite{Abbate:2010xh,Abbate:2012jh,Benitez:2024nav} that while the convolution  with the shape function can only be rigorously proven for the SCET factorization and resummation formula, the first term in Eq.~(\ref{eq:spec_formula}), the modified prescription treating all three terms in Eq.~(\ref{eq:spec_formula}) on an equal footing has better numerical stability and behavior  in the fixed-order region. This modified prescription is expected to give results that formally differ only at the power suppressed level. In our analysis, we restrict ourselves to working with Eq.~(\ref{eq:spec_formula}) without any  modifications.

\section{Numerical Results}
\label{sec:NR}

\begin{figure}
    \includegraphics[scale=0.52]{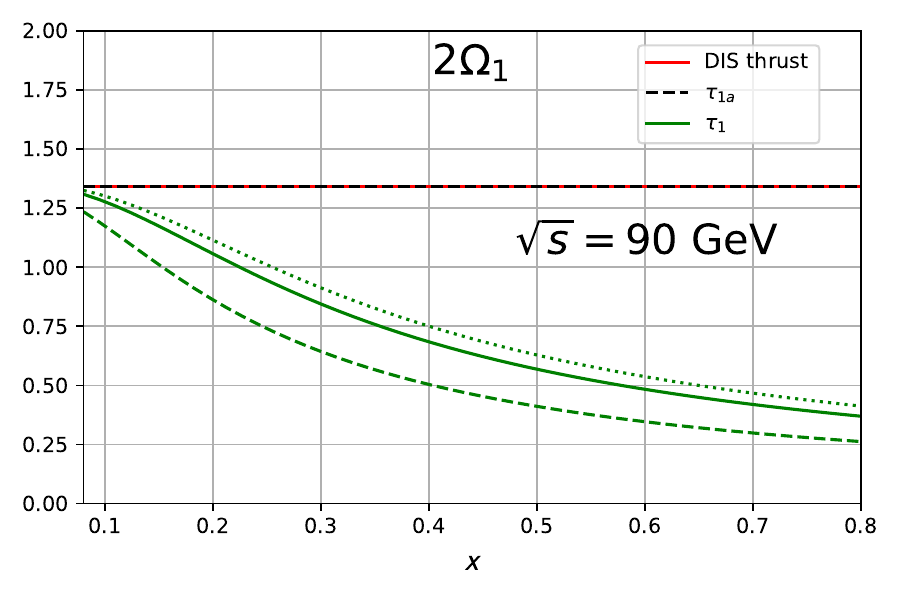}
    \includegraphics[scale=0.52]{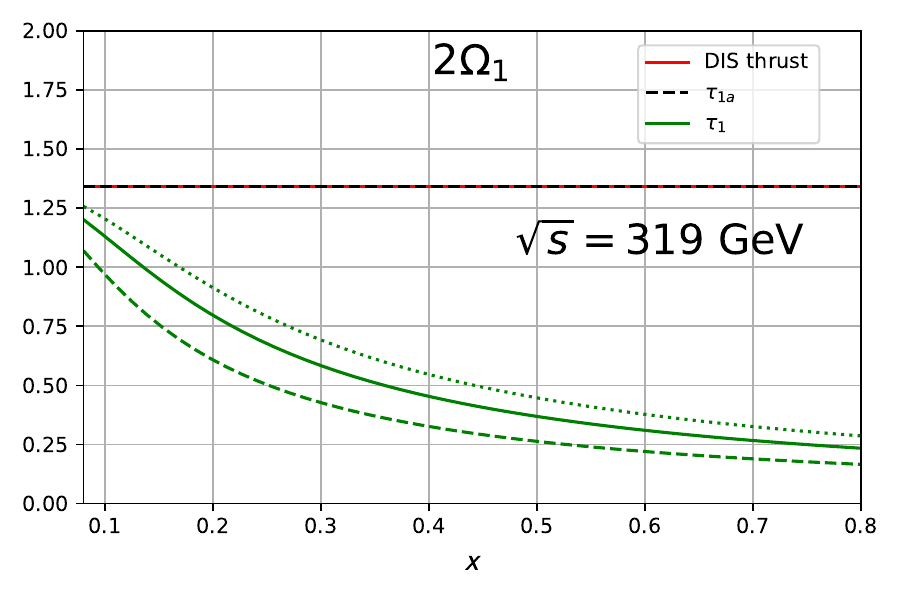}
    \caption{Kinematic dependence of the first moment $2\Omega_1$ for DIS thrust (red), $\tau_{1a}$ (black), and $\tau_1$ (green) for typical EIC (left) and HERA (right) kinematics. For the $\sqrt{s}=90$ GeV (EIC) we show the $x$-dependence of $2\Omega_1$ for $\tau_1$ for $Q^2$ values of 200 GeV$^2$ (dashed), 400 GeV$^2$ (solid) and 500 GeV$^2$ (dotted). For $\sqrt{s}=319$ GeV (HERA) we show curves for $\tau_1$ for 1000 GeV$^2$ (dashed), 2000 GeV$^2$ (solid) and 3000 GeV$^2$ (dotted). For $\tau_{1a}$ and DIS thrust, the $2\Omega_1$ values are identical and have no kinematic dependence.}
 \label{fig:2omega1}   
\end{figure} 

In this section, we provide numerical results up to the N$^3$LL$+{\cal O}(\alpha_s^2)$ level of accuracy for single photon exchange contributions to NC DIS for the $\tau_1$ and $\tau_{1a}$ 1-Jettiness global event shapes. We also give N$^2$LL$+{\cal O}(\alpha_s)$ results for NC and CC DIS when including the contributions of the $Z$- and $W^\pm$-exchange in NC and CC DIS, respectively. This extends previous results that were restricted to NC DIS, and only included contributions from single photon exchange. We present these results for a wide range of  kinematics relevant to  the EIC and HERA. We also provide results for a variety of differential cross sections $d\sigma\left [\tau_1,\{{\cal O}_i\} \right ] $ and $d\sigma \left [\tau_{1a},\{{\cal O}_i\} \right ] $ for different choices of the set of kinematic observables $\{{\cal O}_i\} $ given in Eq.~(\ref{eq:Oi}). Table~\ref{tab:kin_set} summarizes the different observables and kinematic settings used for generating theoretical predictions. We compare these  predictions against PYTHIA8.312~\cite{Sj_strand_2015,bierlich2022comprehensive} simulation results. The general form of the 1-jettiness differential cross section for $\xi=\tau_1$ or $\xi=Q\tau_{1a}$, across the entire spectrum, is given by Eq.~(\ref{eq:spec_formula}). It smoothly connects the peak, tail, and fixed order regions of the entire 1-Jettiness spectrum. 

\begin{figure}
\centering
    \includegraphics[width=0.45\textwidth]{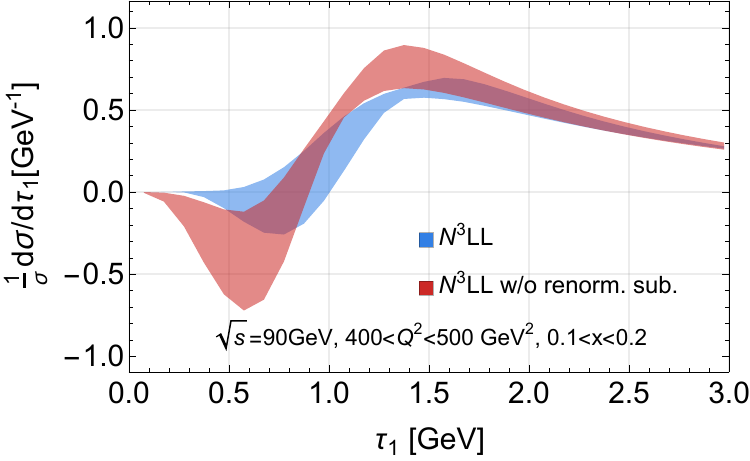}
    \includegraphics[width=0.45\textwidth]{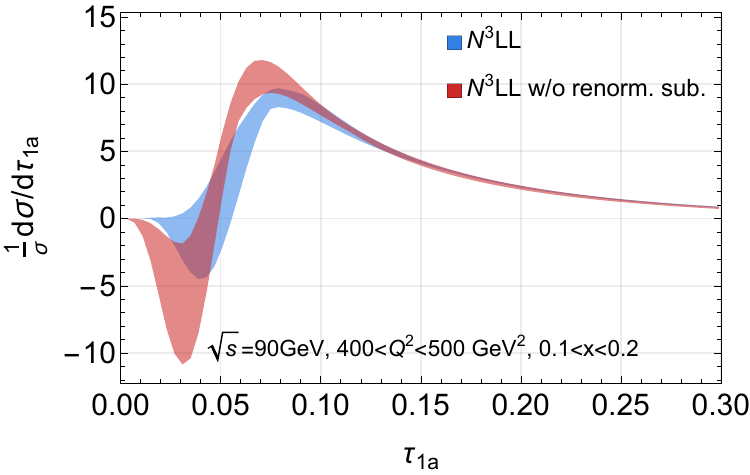}
    \includegraphics[width=0.45\textwidth]{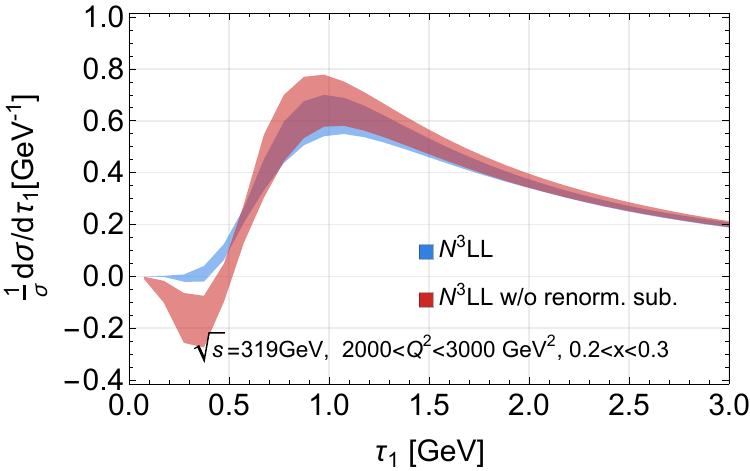}
    \includegraphics[width=0.45\textwidth]{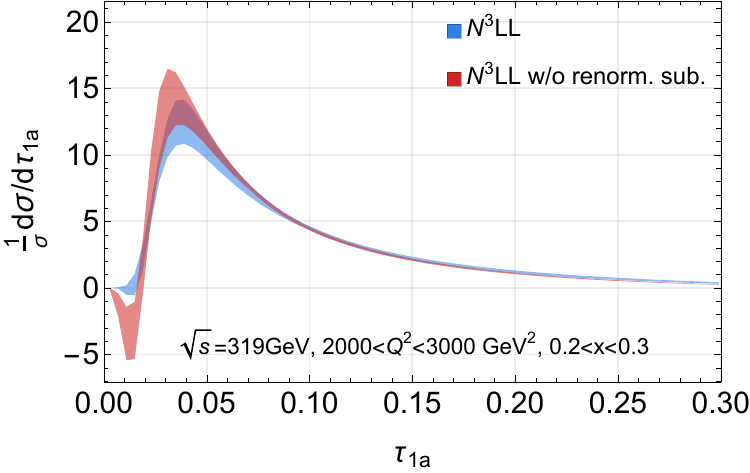}
    \caption{N$^3$LL distributions for $\tau_1$ (left panels) and $\tau_{1a}$ (right panels) with (blue) and without (pink) renormalon subtractions.}
 \label{fig:renor}   
\end{figure} 

\begin{figure}
\centering
    \includegraphics[width=0.45\textwidth]{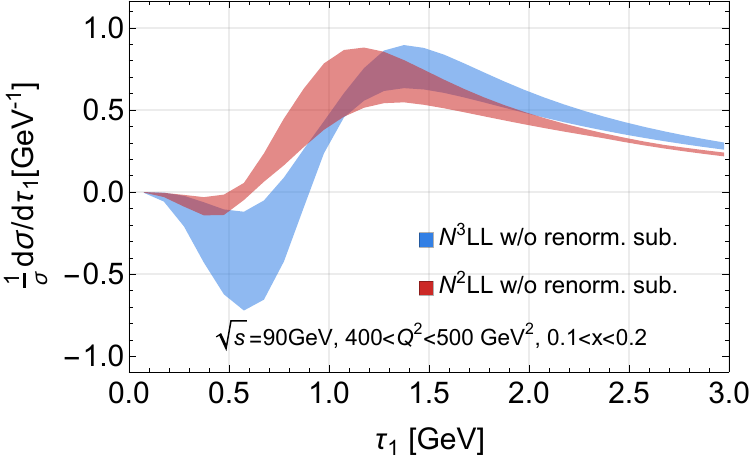}
    \includegraphics[width=0.45\textwidth]{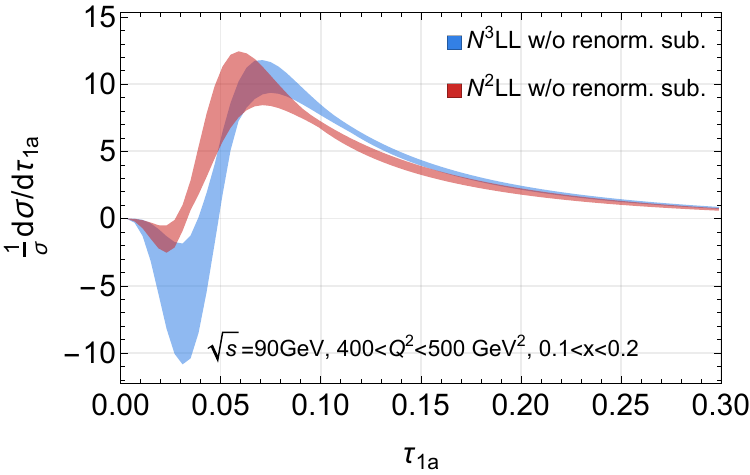}
    \includegraphics[width=0.45\textwidth]{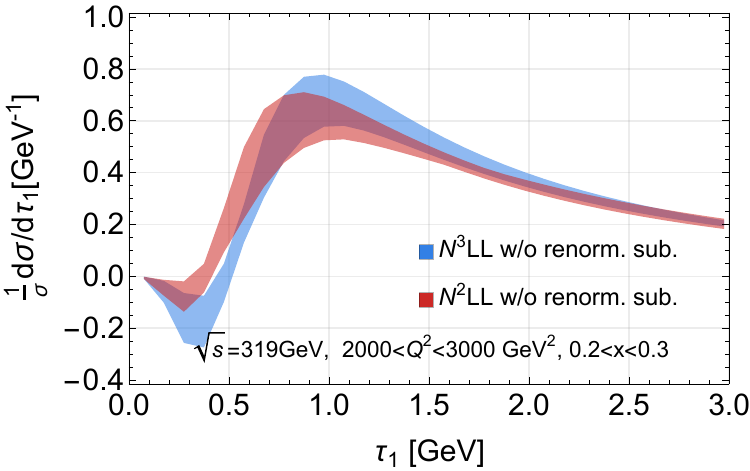}
    \includegraphics[width=0.45\textwidth]{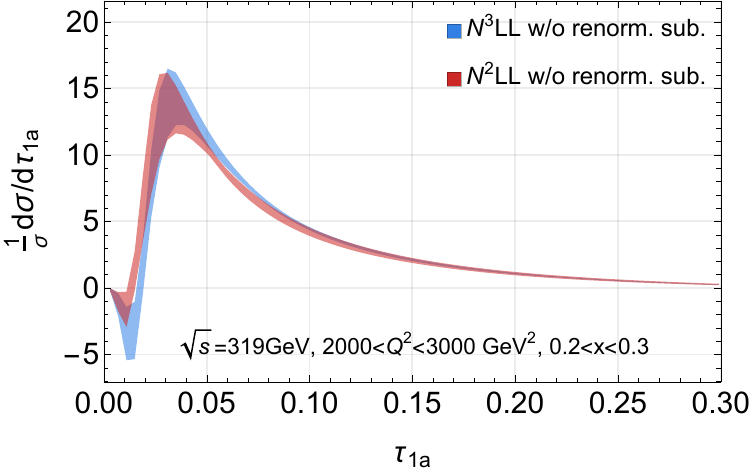}
    \caption{Distributions for $\tau_1$ (left panels) and $\tau_{1a}$ (right panels) at N$^3$LL (blue) and N$^2$LL (pink) with renormalon subtractions turned off.}
 \label{fig:gammanorenor}   
\end{figure}

The renormalon-subtracted SCET factorized and resummed cross section, $d\sigma_{\rm resum}\left [\xi, \delta, \{{\cal O}_i\} \right ]$, corresponding to the first term in Eq.~(\ref{eq:spec_formula}), depends on the hard, beam, jet, soft, and renormalon subtraction scales $\mu_H,\mu_B,\mu_J,\mu_S, R\sim \mu_S$, respectively, the shape function $F^{\rm mod.}(u, R_B,R_J,\Delta)$, the renormalon-free gap parameter $\Delta=\Delta(R,\mu_S)$, and the gap subtraction series $\delta=\delta(R,\mu_S)$. The  gap parameter $\Delta=\Delta(R,\mu_S)$ is obtained 
after $R$ and $\mu$ evolution from the input value $\Delta(R_\Delta,\mu_\Delta)$ at the reference scales $R=R_\Delta, \mu=\mu_\Delta$. We follow Ref.~\cite{Ee:2025scz}, which gives the explicit solution to the  $R$ and $\mu$ evolution equations, and use their reference input scales:
\bea
\label{eq:gapparam}
 R_\Delta=\mu_\Delta=2 \>{\rm GeV}, \qquad  \Delta(R_\Delta,\mu_\Delta)=0.05\> {\rm GeV}. 
 \eea
For the scales $\mu_B,\mu_J,$ and $\mu_S$ we use the profile functions introduced for $\tau_{1a}$ in Ref.~\cite{Kang:2013nha}, and adapted for $\tau_1$ in Ref.~\cite{Cao:2024ota}. Following Ref.~\cite{Ee:2025scz}, in order to keep $\delta_1(R,\mu_S)=-8 C_F {\rm log}(\mu_S/R)$ non-zero, we deviate from setting the profile functions of  $R$ and $\mu_S$ exactly equal to each other, $R(\xi)=\mu_S(\xi)$, and instead use $R(\xi)=\mu_S(\xi)|_{\mu_0\to R_0}$ with $R_0=0.85 \>\mu_0$, where $\mu_0$  is a parameter in the $\mu_S$ profile function, as shown in Refs.~\cite{Kang:2013nha,Cao:2024ota}, and is set to $\mu_0=1$ GeV. We also use the same scale variations given in Ref.~\cite{Cao:2024ota} in order to determine the uncertainty bands for the theoretical predictions of the $\tau_1$ and $\tau_{1a}$ distributions.

\begin{figure}
\centering
    \includegraphics[width=0.45\textwidth]{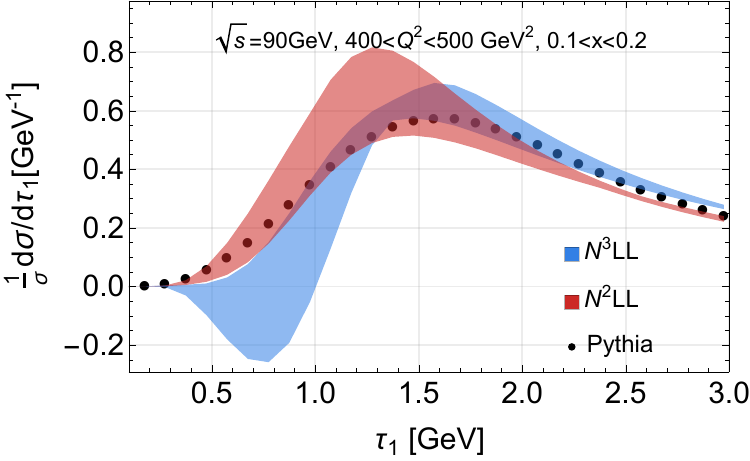}
    \includegraphics[width=0.45\textwidth]{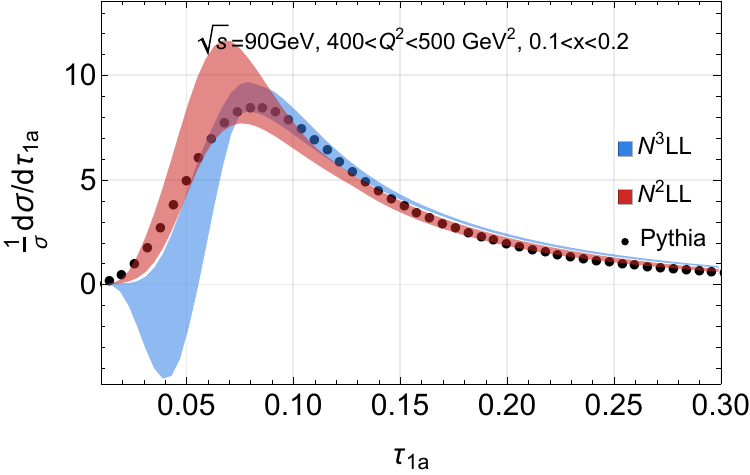}
    \includegraphics[width=0.45\textwidth]{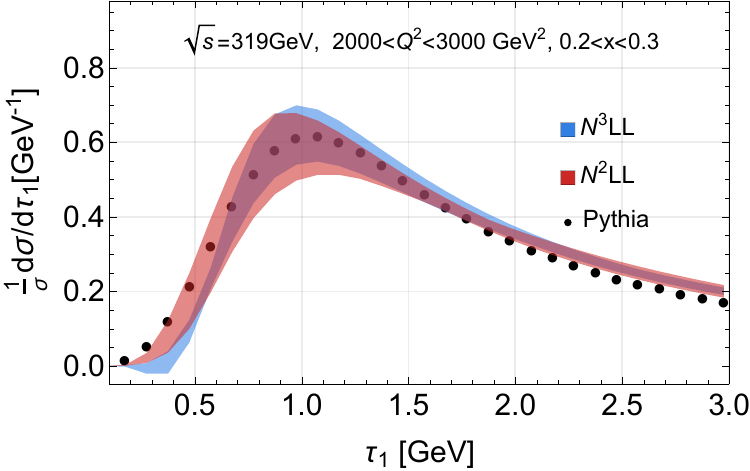}
    \includegraphics[width=0.45\textwidth]{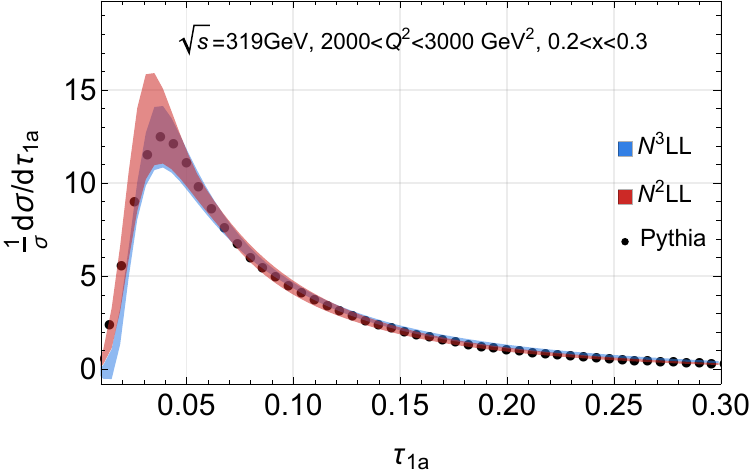}
    \caption{Distributions for $\tau_1$ (left panel) and $\tau_{1a}$ (right panel) at N$^3$LL (blue) and N$^2$LL (pink) with renormalon subtractions included.}
 \label{fig:gammannll}   
\end{figure} 

\begin{figure}
\centering
    \includegraphics[width=0.45\textwidth]{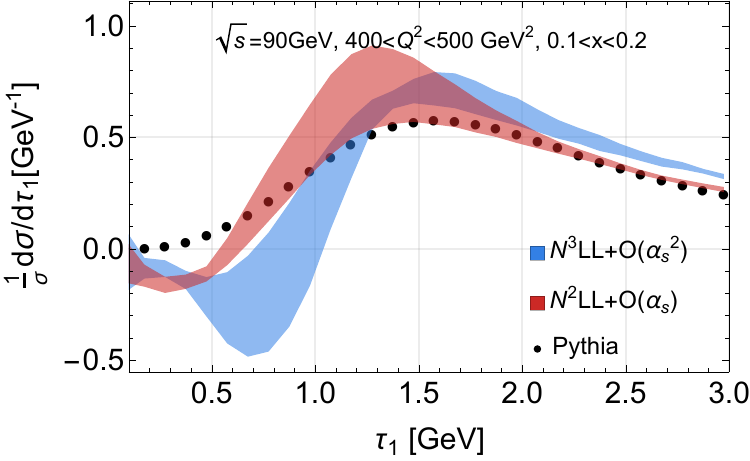}
    \includegraphics[width=0.45\textwidth]{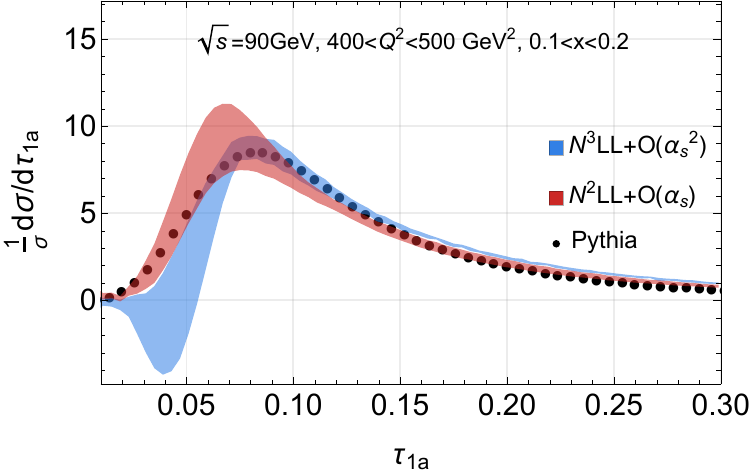}
    \includegraphics[width=0.45\textwidth]{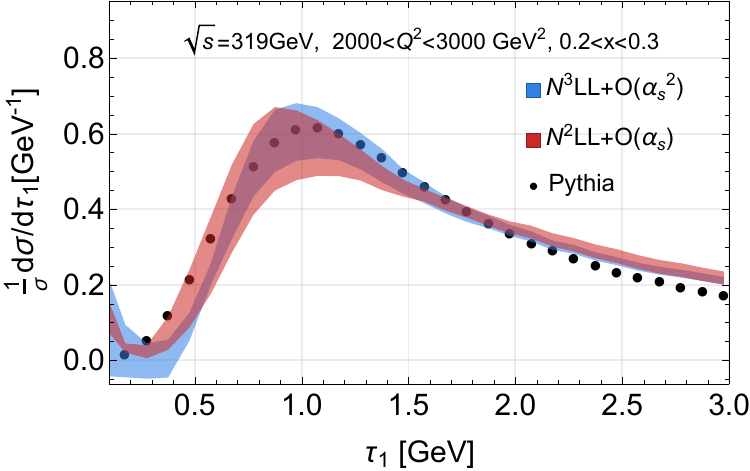}
    \includegraphics[width=0.45\textwidth]{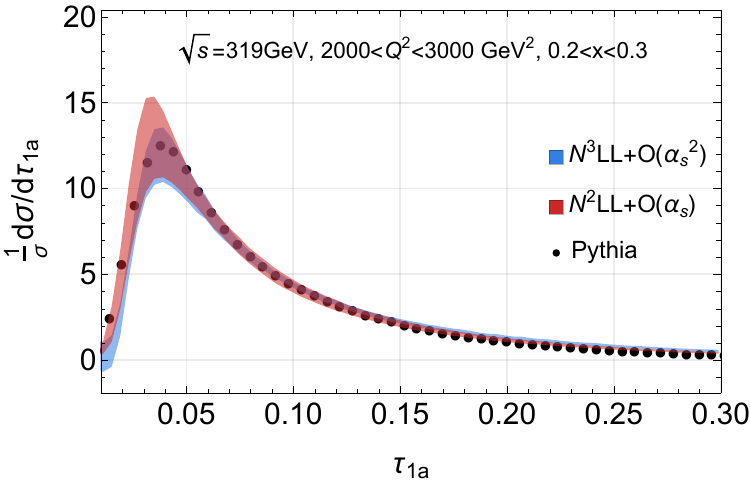}
    \caption{Distributions for $\tau_1$ (left panel) and $\tau_{1a}$ (right panel) at N$^3$LL+$O(\alpha_s^2)$ (blue) and N$^2$LL+$O(\alpha_s)$ (pink) with renormalon subtractions included.}
 \label{fig:gammannlo}   
\end{figure}

The shape function $F^{\rm mod.}(u, R_B,R_J,\Delta)$  is given in terms of the  DIS hemisphere shape function, $S^{\rm mod.}_{\rm hemi.}(k_1,k_2)$. This universality also extends to the renormalon subtractions since $\Delta(R,\mu_S)$ and $\delta(R,\mu_S)$ are the same renormalon-free gap parameter and gap subtraction series that appear in the DIS hemisphere soft function. We demonstrate this universality of hadronization and the required renormalon subtractions  by comparing theoretical predictions and simulation results for the wide range of the observables and kinematics, using the shape function model in Eqs.~(\ref{eq:SxiRS}) and (\ref{eq:FmodRS}) for a single choice of the parameters $a,b,\Lambda$ in Eq.~(\ref{eq:shapehemi}). In particular, we choose:
\bea
\label{eq:shapeparam}
a=2.0, \qquad b=0.0, \qquad \Lambda= 0.4 \>{\rm GeV}.
\eea 
We show that this single choice for the hemisphere shape function model can be used to describe both the $\tau_1$ and $\tau_{1a}$ observables for a wide range of kinematics at HERA and the EIC, shown in Table~\ref{tab:kin_set}.  We note that all the parameter choices made in this section are illustrative only. Ultimately, they must be determined through a more careful analysis with simulation or real data.

\begin{figure}
    \centering
    \includegraphics[width=0.45\textwidth]{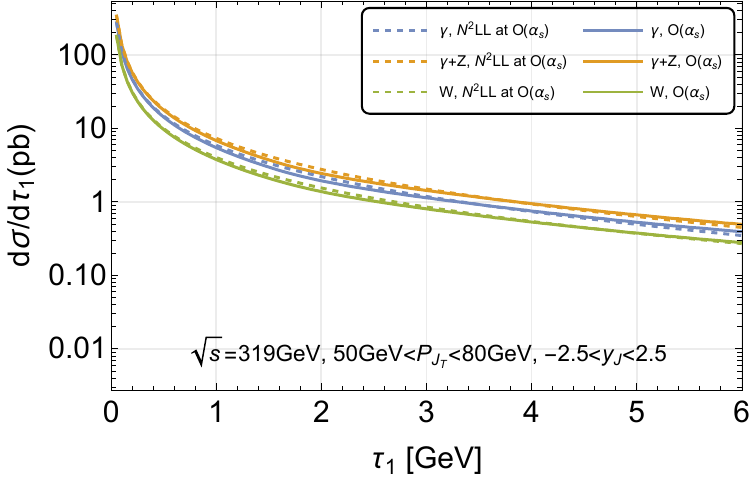} 
      \includegraphics[width=0.45\textwidth]{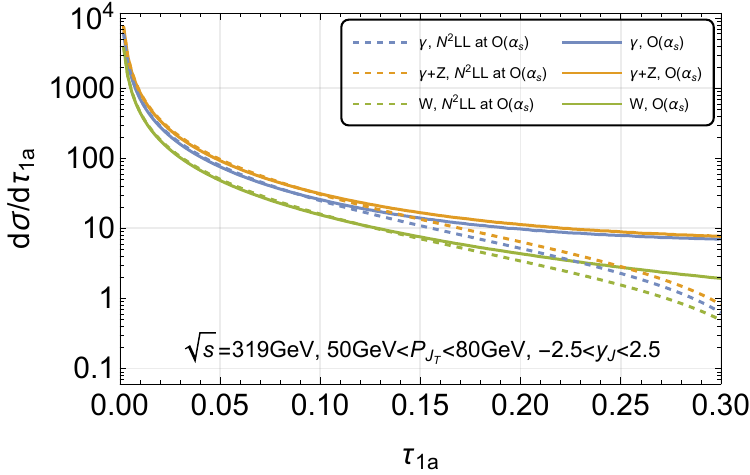}
            \includegraphics[width=0.45\textwidth]{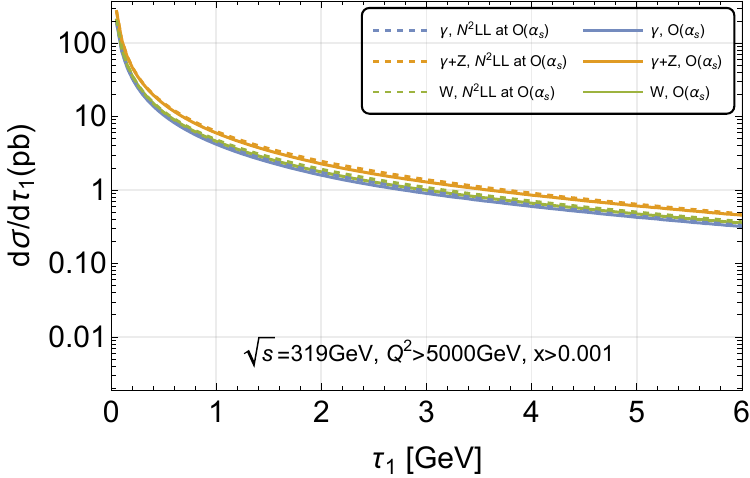}
    \includegraphics[width=0.45\textwidth]{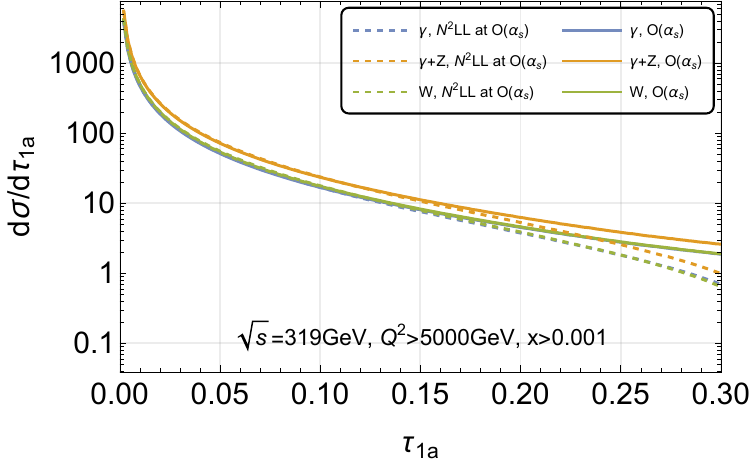}
    \caption{Comparison of the perturbative N$^2$LL resummed result expanded to ${\cal O}(\alpha_s)$ (dashed line) and the full fixed-order ${\cal O}(\alpha_s)$ result from DISTRESS (solid line). We show predictions for $\tau_1$ (left panel) and $\tau_{1a}$ (right panel) for the HERA center of mass energy, $\sqrt{s}=319$ GeV, in different kinematic bins for NC DIS with only the single photon exchange contribution (blue) and with both the single photon and $Z$-boson exchange contribution (orange) and CC (green) DIS.}
    \label{fig:check-HERA}
\end{figure} 

\begin{figure}
    \centering
    \includegraphics[width=0.45\textwidth]{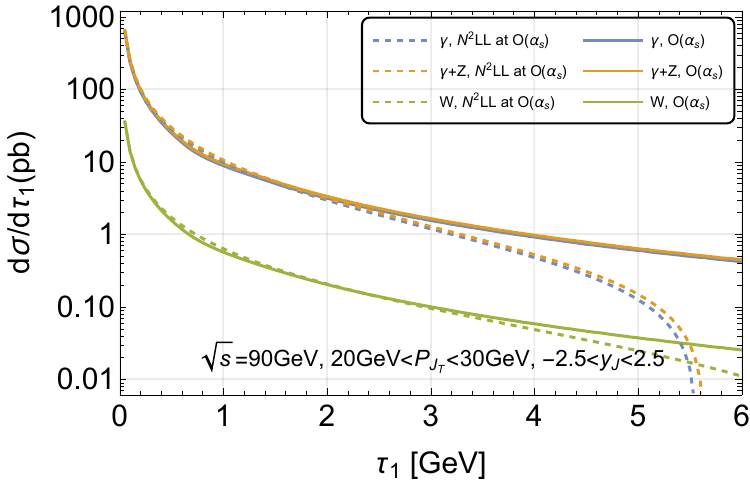}
    \includegraphics[width=0.45\textwidth]{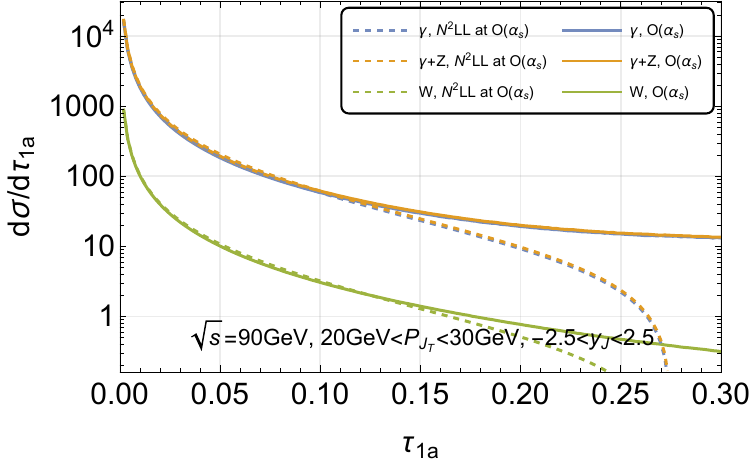}
    \includegraphics[width=0.45\textwidth]{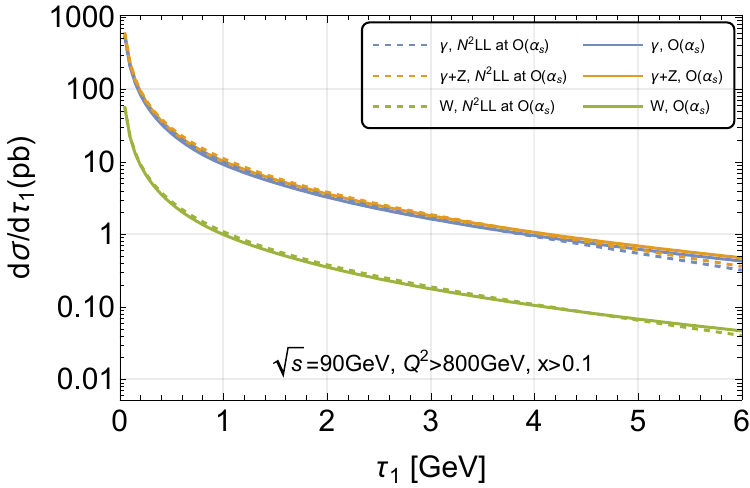}
    \includegraphics[width=0.45\textwidth]{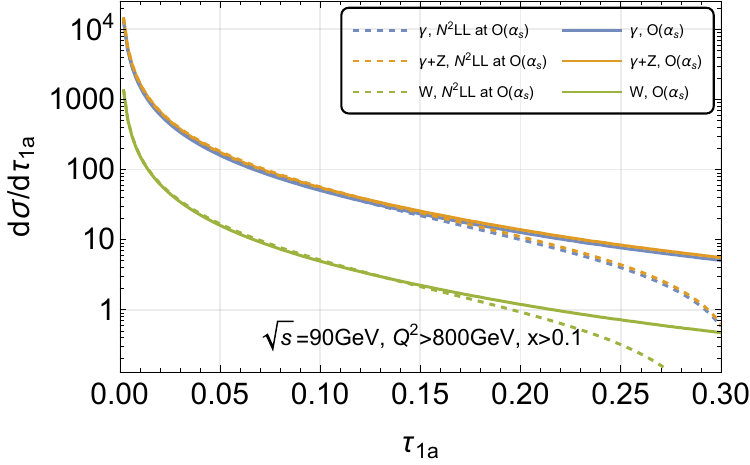}
    \caption{Comparison of the perturbative N$^2$LL resummed result expanded expanded to ${\cal O}(\alpha_s)$ (dashed line) and the full fixed-order ${\cal O}(\alpha_s)$ result from DISTRESS (solid line). We show predictions for $\tau_1$ (left panel) and $\tau_{1a}$ (right panel) for the EIC center of mass energy, $\sqrt{s}=90$ GeV, in different kinematic bins for NC DIS with only the single photon exchange contribution (blue) and with both the single photon and Z-boson exchange contribution (orange) and CC (green) DIS.}
    \label{fig:check-EIC1}
\end{figure} 

In Fig.~\ref{fig:2omega1}, we show the kinematic dependence of the first moment $2\Omega_1$ of the shape function $F^{\rm mod.}(u, R_B,R_J,\Delta)$, given in  Eq.~(\ref{eq:moment}), for $\tau_1$, $\tau_{1a}$, and DIS thrust. The shape function for  $\tau_{1a}$ and DIS thrust are identical, corresponding to ${\cal R}_B={\cal R}_J=1$, so that it has no hard kinematic dependence. As a result, $2\Omega_1$ takes on a constant value across all $(x,Q^2)$ kinematic bins, as seen in Fig.~\ref{fig:2omega1}. By contrast, for the $\tau_1$ event shape, ${\cal R}_B={\cal R}_J=R_B=R_J=\sqrt{(n_B\cdot n_J)/2}$, as seen in Eq.~(\ref{eq:RBRJtau1}). Using Eqs.~(\ref{eq:qBqJ}), (\ref{eq:ref_vectors}), and (\ref{eq:wBwJ}) in the $\xi\ll Q_H$ limit, we get $n_B\cdot n_J=1-\tanh y_J$, where $y_J=1/2\ln [Q^2(1-y)/(y^2s)]$. Using $Q^2=xys$, we see that ${\cal R}_{B,J}$ are functions of the hard kinematic variables $(x,Q^2)$. Correspondingly, $F^{\rm mod.}(u, R_B,R_J,\Delta)$ and its first moment $2\Omega_1$ for $\tau_1$ have a non-trivial dependence on $(x,Q^2)$. Fig.~\ref{fig:2omega1} shows this kinematic dependence of $2\Omega_1$ for $\tau_1$ as a function of $x$ for different $Q^2$ values for typical EIC and HERA kinematics. For a given choice of the hemisphere shape function model $S_{\rm hemi.}^{\rm mod.}(k_1,k_2)$, the kinematic dependence of $F^{\rm mod.}(u, R_B,R_J,\Delta)$ is exactly calculable through Eq.~(\ref{eq:FmodRS}). Thus, this kinematic dependence for $\tau_1$ can serve as an independent lever arm for constraining the universal shape function $S_{\rm hemi.}^{\rm mod.}(k_1,k_2)$ that determines hadronization effects for $\tau_1$, $\tau_{1a}$, DIS thrust, and other possible definitions of $\xi$. Given the wide range of $(x,Q^2)$ kinematic bins available at the EIC and HERA, the universality of hadronization effects combined with the kinematic lever arm could allow for including the typically ignored peak region, where hadronization effects are most severe, for precision extractions of the strong coupling constant.

In Fig.~\ref{fig:renor}, we show the hadron-level N$^3$LL distributions for $\tau_1$ and $\tau_{1a}$ for typical EIC and HERA kinematics, with (blue) and without (pink) renormalon subtractions. We see that the renormalon subtractions lead to better behavior in the small $\xi$ region, taming the severity of the unphysical negative dips as shown in the analysis of Ref.~\cite{Hoang:2007vb}. The renormalon subtractions also lead to better convergence and stability of the peak position in going from the N$^2$LL to N$^3$LL distributions, as seen by comparing Figs.~\ref{fig:gammanorenor} and \ref{fig:gammannll} that show the distributions without and with renormalon subtractions. In Fig.~\ref{fig:gammannll}, we also show Pythia simulation which is in good agreement with the theoretical predictions across the $\tau_1$ and $\tau_{1a}$ observables in widely separated kinematic bins. In particular, it shows that $F^{\rm mod.}(u, {\cal R}_B,{\cal R}_J,\Delta)$ for $\tau_1$ and $\tau_{1a}$  is correctly capturing the different kinematic dependence of hadronization effects in $\tau_1$ and $\tau_{1a}$, as emphasized in Ref.~\cite{Boughezal:2026dvu}.  

In Fig.~\ref{fig:gammannlo}, we show the complete N$^3$LL+${\cal O}(\alpha_s^2)$ and  N$^2$LL+${\cal O}(\alpha_s)$   results by including the fixed-order contributions, and combining it with the  resummation result as in Eq.~(\ref{eq:spec_formula}). We have used the same shape function parameters in Eq.~(\ref{eq:shapeparam}), chosen to give good agreement with just the resummation results in Fig.~\ref{fig:gammannll}.  We still see good agreement using these same parameters between the N$^3$LL+${\cal O}(\alpha_s^2)$ and N$^2$LL+${\cal O}(\alpha_s)$ distributions and Pythia simulation. Compared to the N$^3$LL and N$^2$LL distributions in Fig.~\ref{fig:gammannll}, we see slight numerical effects from the leftover non-singular terms in the combination $d\sigma^{\rm FO} \left [\xi, \{{\cal O}_i\}\right ]-d\sigma_{\rm resum}^{\rm FO} \left [\xi, \{{\cal O}_i\} \right ]  $ in the small $\xi$ region. 

Next we provide results that include the contributions from the exchange of the massive $Z$ and $W$ bosons for NC and CC DIS, respectively,  at the N$^2$LL+${\cal O}(\alpha_s)$ level of accuracy.  We provide cross checks of the ${\cal O}(\alpha_s)$ corrections to the $\gamma$, $Z$, and $W$ boson exchange contributions  implemented in the DISTRESS code by comparing  them with the corresponding 
N$^2$LL partonic resummed result expanded to ${\cal O}(\alpha_s)$. Figs.~\ref{fig:check-HERA} and \ref{fig:check-EIC1} show this comparison for the contribution to NC DIS from only $\gamma$-exchange (blue) and $\gamma+Z$-exchange (orange) and to CC DIS from $W$-exchange (green). Figs.~\ref{fig:check-HERA} and \ref{fig:check-EIC1}
 correspond to some of the kinematic settings of HERA and  EIC-1 in Table~\ref{tab:kin_set}, respectively. In these figures, the left and right panels correspond to $\tau_1$ and $\tau_{1a}$, respectively, and each row of panels is for the kinematic variables in some of the rows in Table~\ref{tab:kin_set}.  We see that the partonic cross sections $d\sigma_{\rm resum}^{\rm FO} \left [\xi, \{{\cal O}_i\} \right ]$   and   $d\sigma^{\rm FO} \left [\xi, \{{\cal O}_i\}\right ]$ at  ${\cal O}(\alpha_s)$ merge in the resummation region $\xi\ll Q_H$, as expected, providing an important cross check.   
 
\begin{figure}
    \centering
           \includegraphics[width=0.45\textwidth]{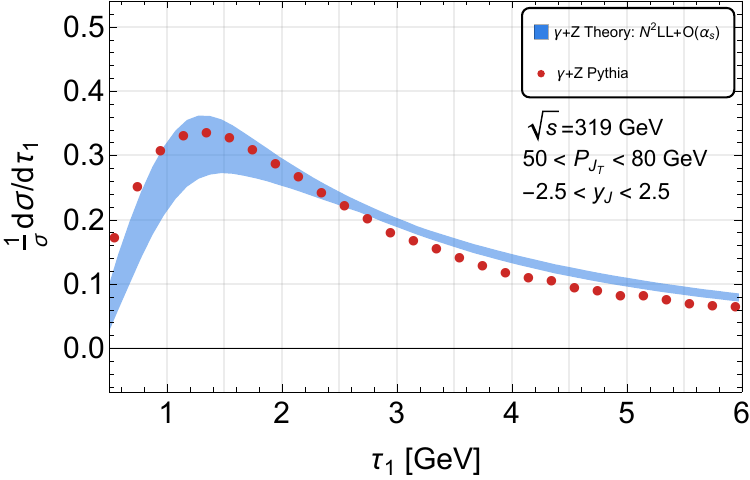}
           \includegraphics[width=0.45\textwidth]{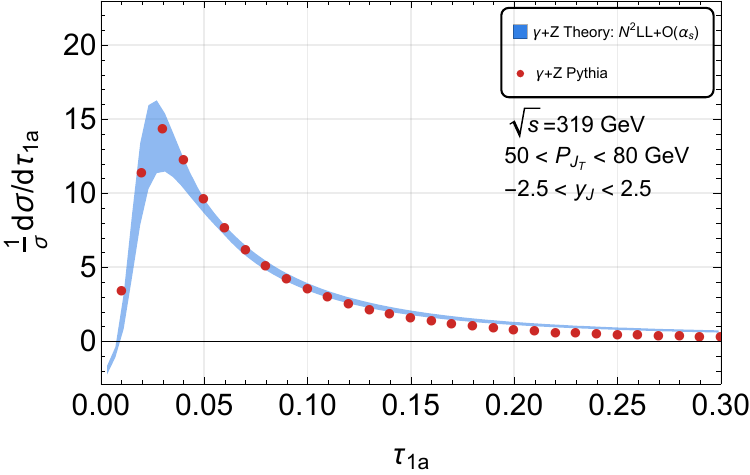}
           \includegraphics[width=0.45\textwidth]{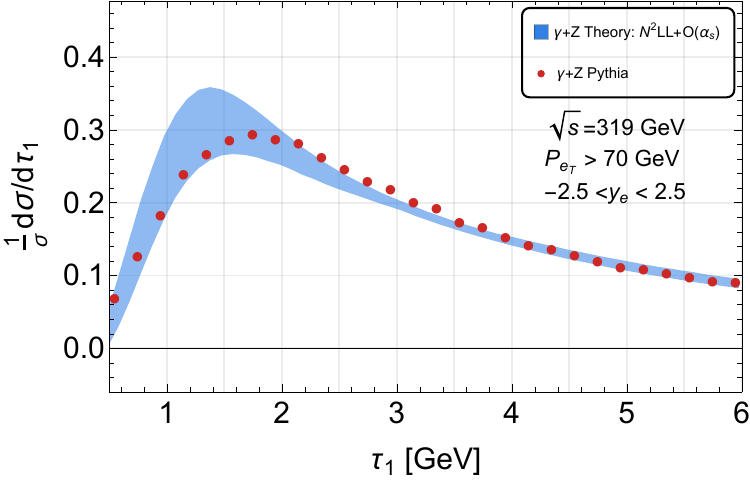}
           \includegraphics[width=0.45\textwidth]{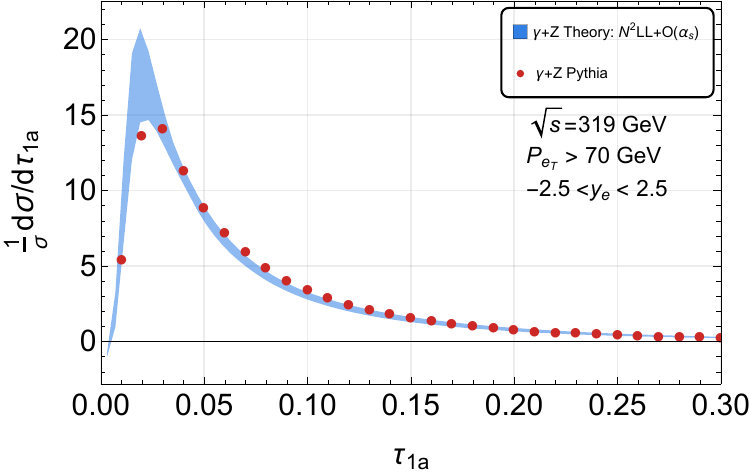}
           \includegraphics[width=0.45\textwidth]{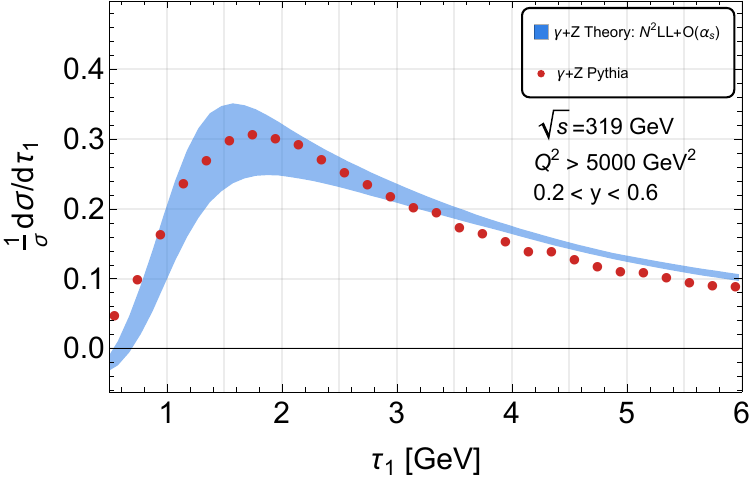}
           \includegraphics[width=0.45\textwidth]{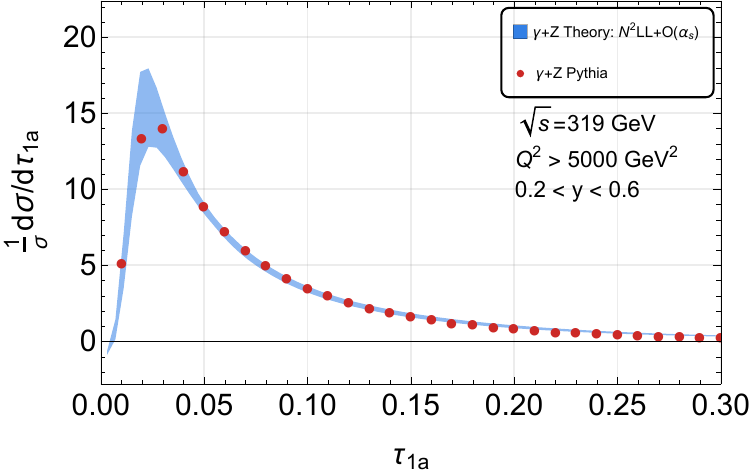}
           \includegraphics[width=0.45\textwidth]{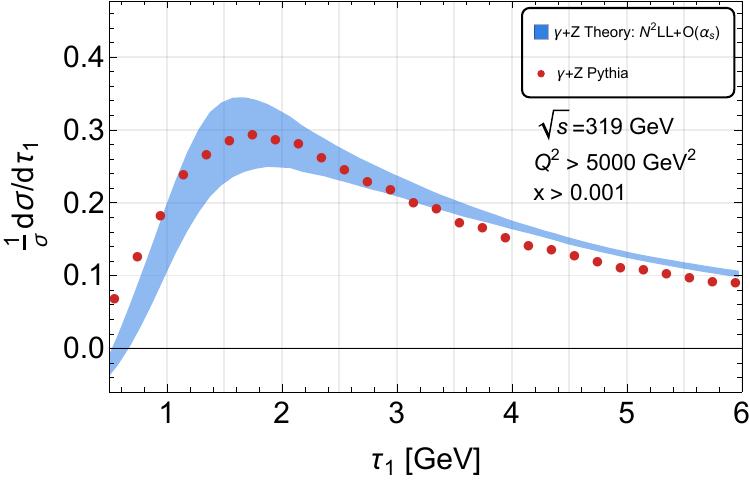}
           \includegraphics[width=0.45\textwidth]{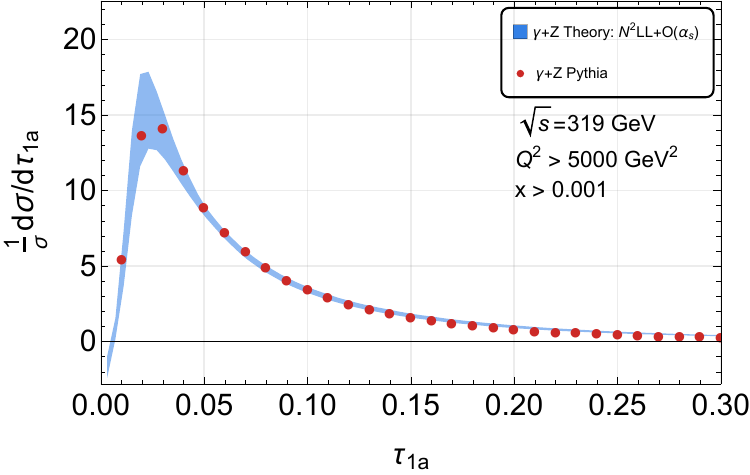}
        \caption{Comparison of the N$^2$LL+${\cal O}(\alpha_s)$ theory prediction (blue band) for NC DIS with the hadronic PYTHIA 8.312 simulation results (red dots). We show predictions for $\tau_1$ (left panel) and $\tau_{1a}$ (right panel) for the HERA center of mass energy, $\sqrt{s}=319$ GeV, in different kinematic bins. The theory predictions use the shape function model parameters in Eq.~(\ref{eq:shapeparam}) and the gap parameter in Eq.~(\ref{eq:gapparam}).}
    \label{fig:TheoryvsPythiaheraz}
\end{figure}

\begin{figure}
    \centering
           \includegraphics[width=0.45\textwidth]{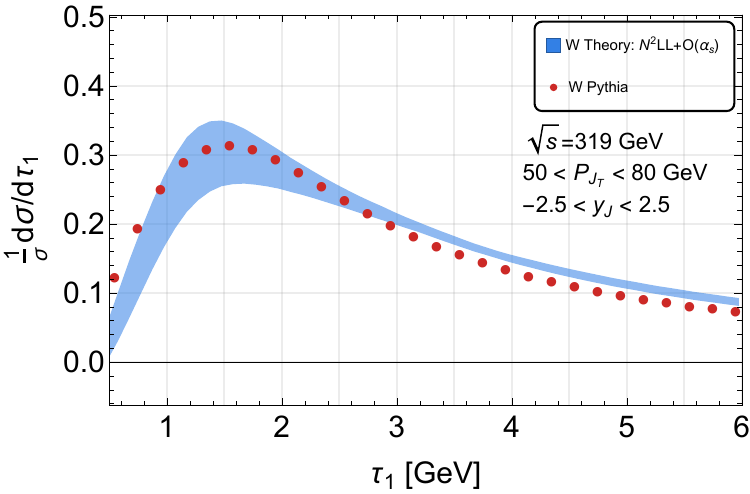}
           \includegraphics[width=0.45\textwidth]{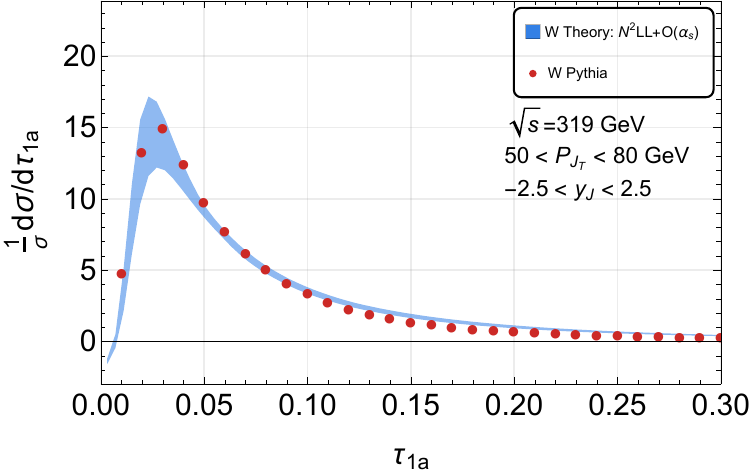}
           \includegraphics[width=0.45\textwidth]{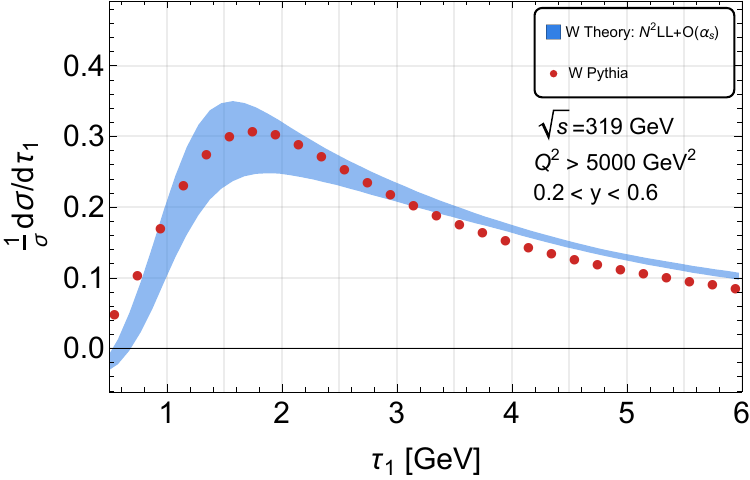}
           \includegraphics[width=0.45\textwidth]{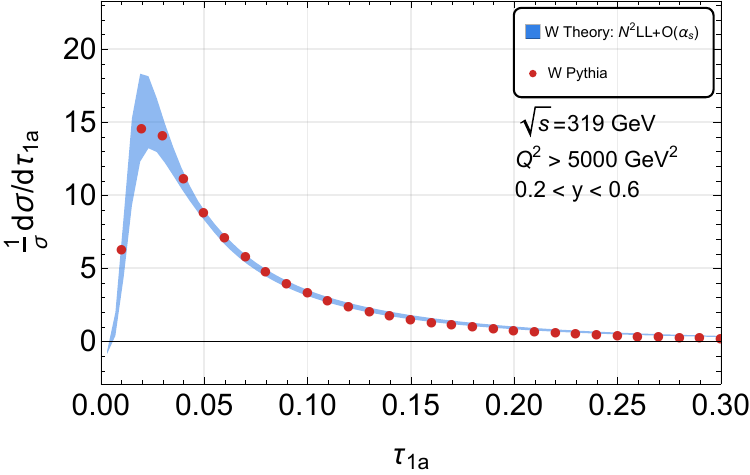}
           \includegraphics[width=0.45\textwidth]{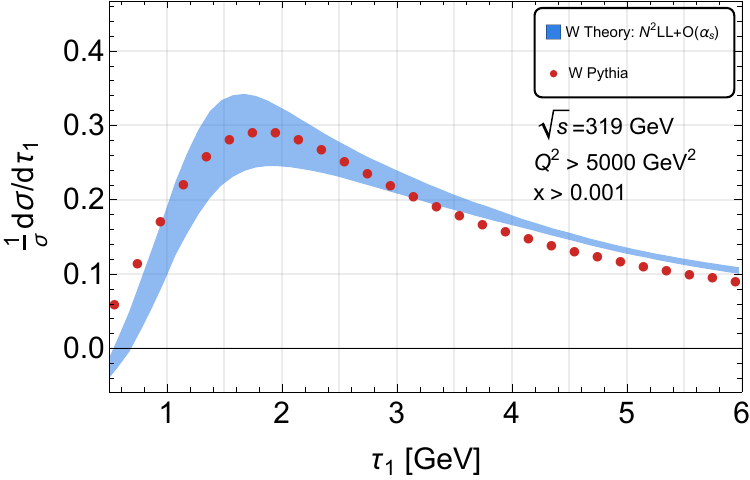}
           \includegraphics[width=0.45\textwidth]{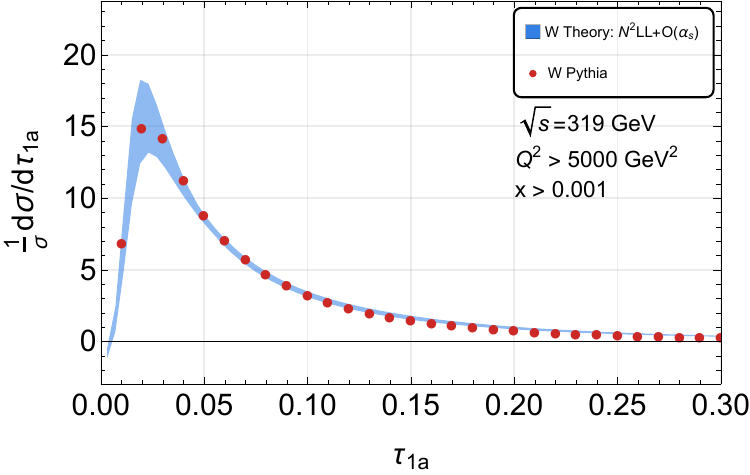}
        \caption{Comparison of the N$^2$LL+${\cal O}(\alpha_s)$ theory prediction (blue band) for CC DIS with the hadronic PYTHIA 8.312 simulation results (red dots). We show predictions for $\tau_1$ (left panel) and $\tau_{1a}$ (right panel) for the HERA center of mass energy, $\sqrt{s}=319$ GeV, in different kinematic bins. The theory predictions use the shape function model parameters in Eq.~(\ref{eq:shapeparam}) and the gap parameter in Eq.~(\ref{eq:gapparam}).}
    \label{fig:TheoryvsPythiaheraw}
\end{figure}

\begin{figure}
    \centering
           \includegraphics[width=0.45\textwidth]{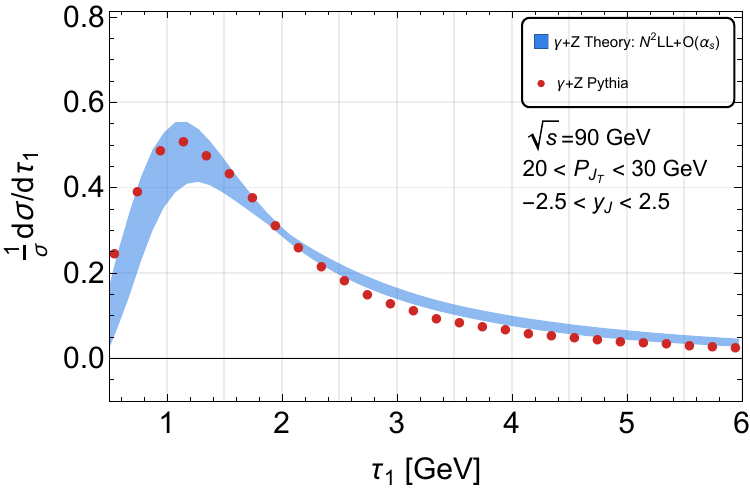}
           \includegraphics[width=0.45\textwidth]{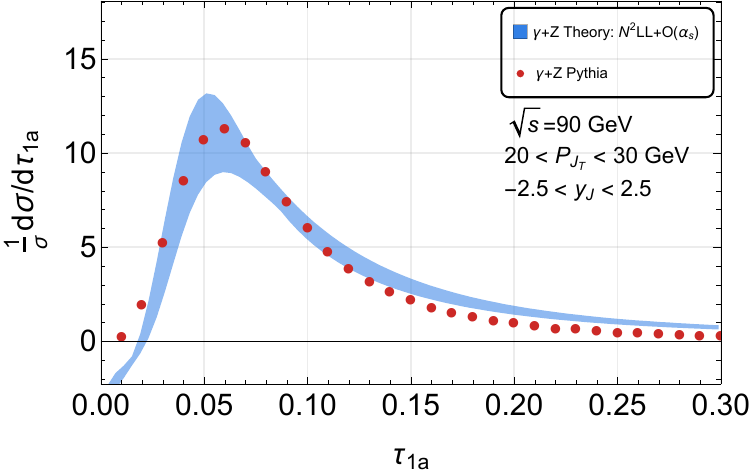}
           \includegraphics[width=0.45\textwidth]{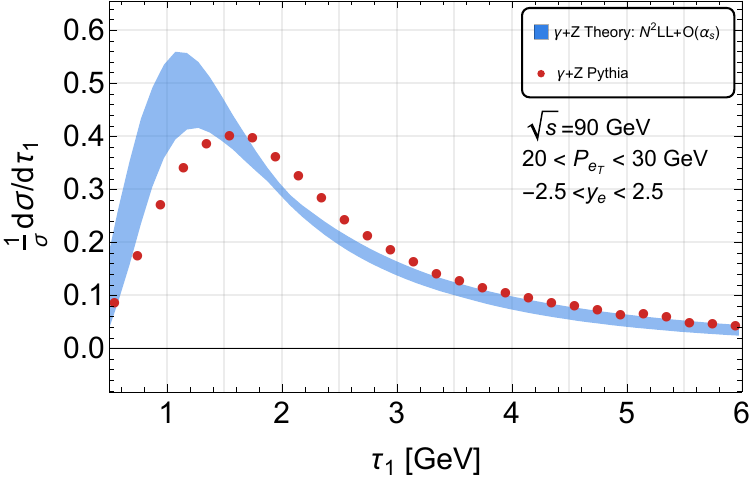}
           \includegraphics[width=0.45\textwidth]{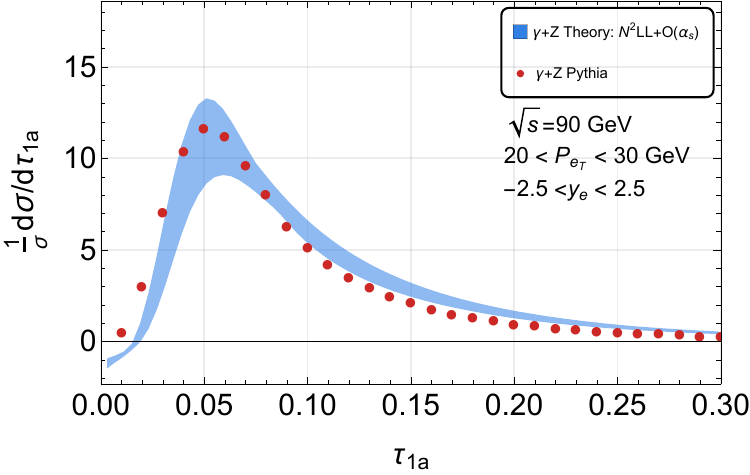}
           \includegraphics[width=0.45\textwidth]{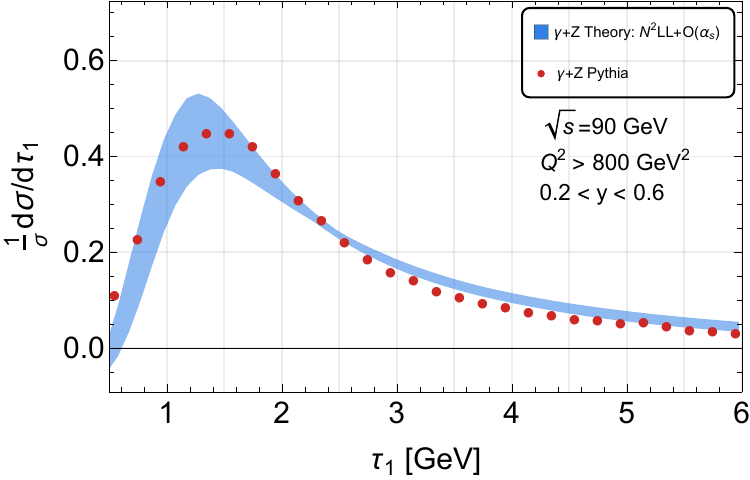}
           \includegraphics[width=0.45\textwidth]{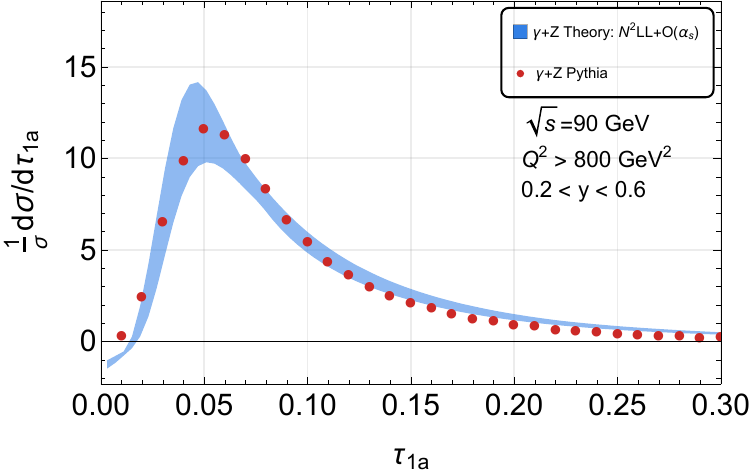}
           \includegraphics[width=0.45\textwidth]{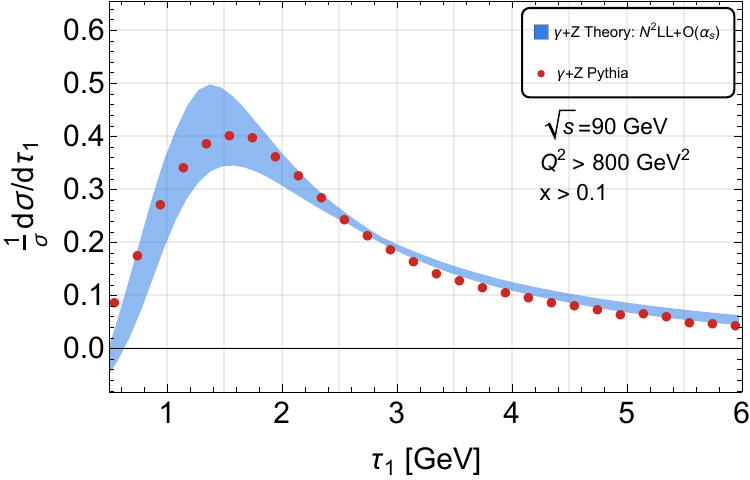}
           \includegraphics[width=0.45\textwidth]{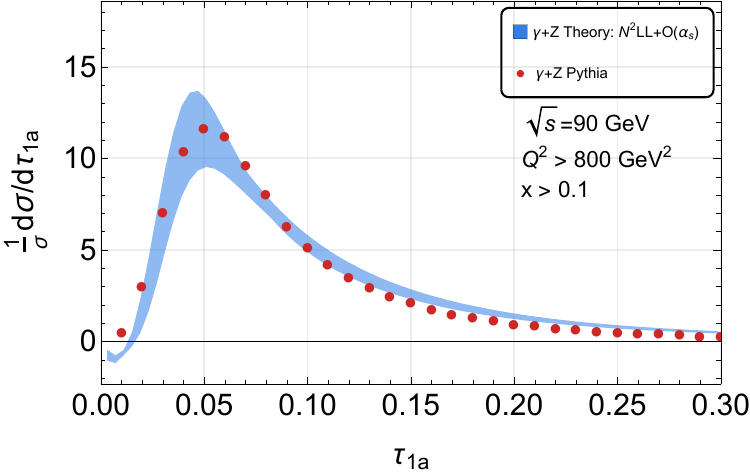}
        \caption{Comparison of the N$^2$LL+${\cal O}(\alpha_s)$ theory prediction (blue band) for NC DIS with the hadronic PYTHIA 8.312 simulation results (red dots). We show predictions for $\tau_1$ (left panel) and $\tau_{1a}$ (right panel) for the EIC center of mass energy, $\sqrt{s}=90$ GeV, in different kinematic bins. The theory predictions use the shape function model parameters in Eq.~(\ref{eq:shapeparam}) and the gap parameter in Eq.~(\ref{eq:gapparam}).}
    \label{fig:TheoryvsPythiaEIC1z}
\end{figure}

\begin{figure}
    \centering
           \includegraphics[width=0.45\textwidth]{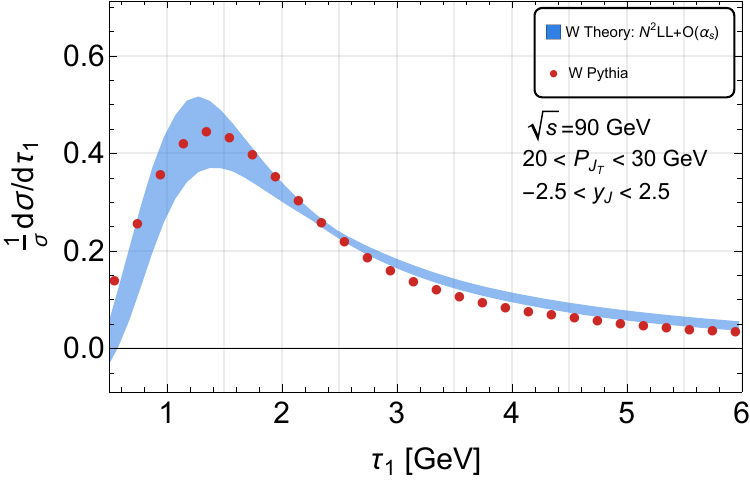}
           \includegraphics[width=0.45\textwidth]{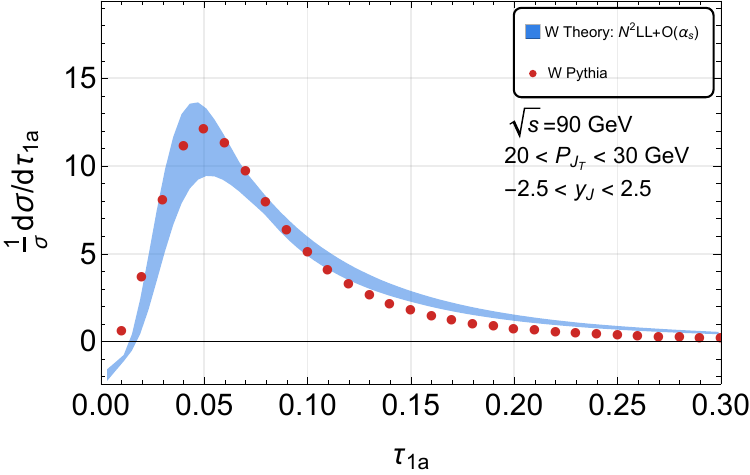}
           \includegraphics[width=0.45\textwidth]{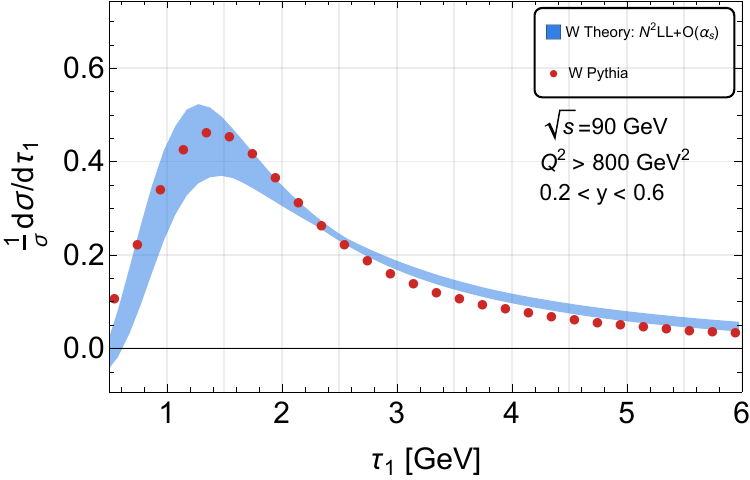}
           \includegraphics[width=0.45\textwidth]{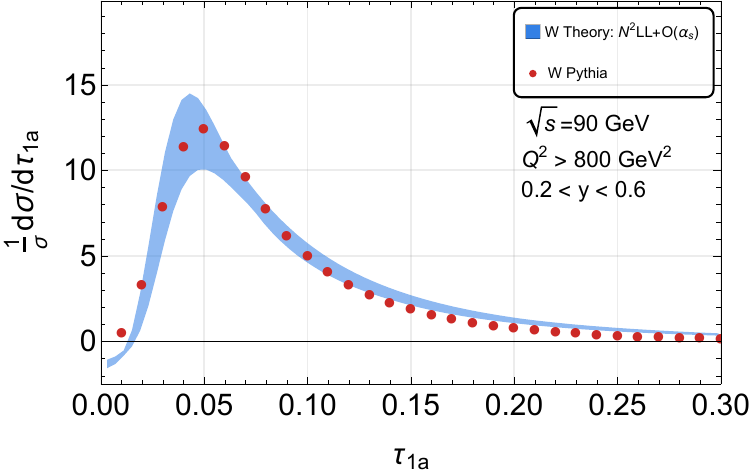}
           \includegraphics[width=0.45\textwidth]{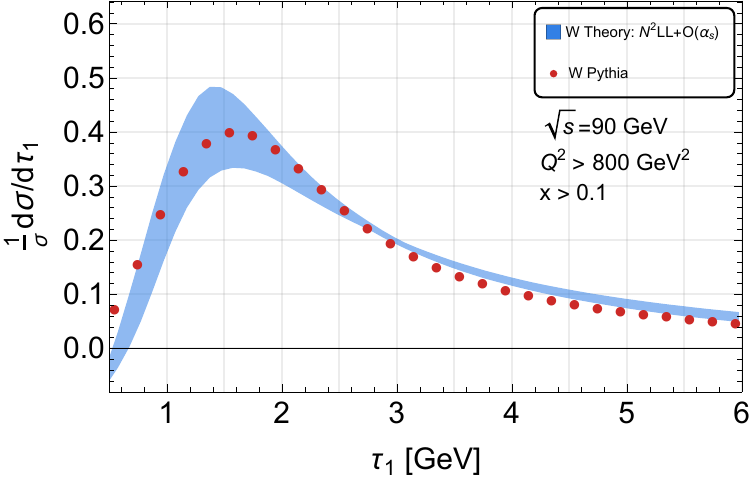}
           \includegraphics[width=0.45\textwidth]{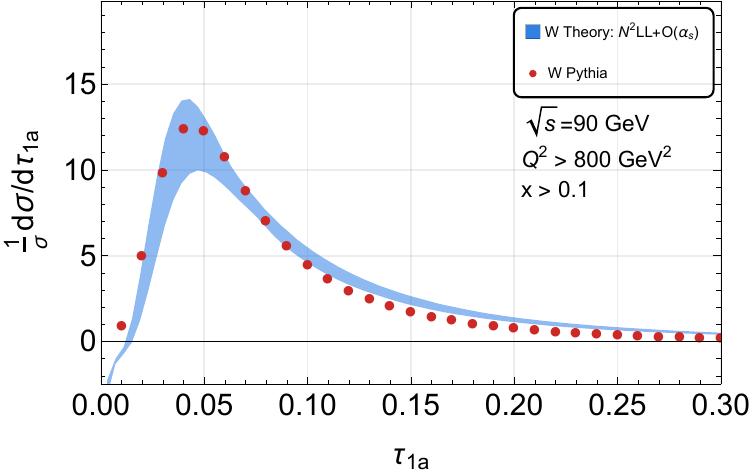}
        \caption{Comparison of the N$^2$LL+${\cal O}(\alpha_s)$ theory prediction (blue band) for CC DIS with the hadronic PYTHIA 8.312 simulation results (red dots). We show predictions for $\tau_1$ (left panel) and $\tau_{1a}$ (right panel) for the EIC center of mass energy, $\sqrt{s}=90$ GeV, in different kinematic bins. The theory predictions use the shape function model parameters in Eq.~(\ref{eq:shapeparam}) and the gap parameter in Eq.~(\ref{eq:gapparam}).}
    \label{fig:TheoryvsPythiaEIC1w}
\end{figure}

\begin{figure}
    \centering
           \includegraphics[width=0.45\textwidth]{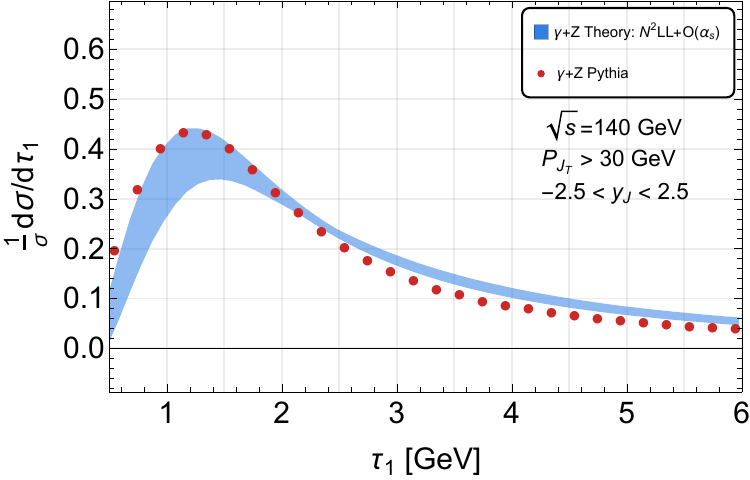}
           \includegraphics[width=0.45\textwidth]{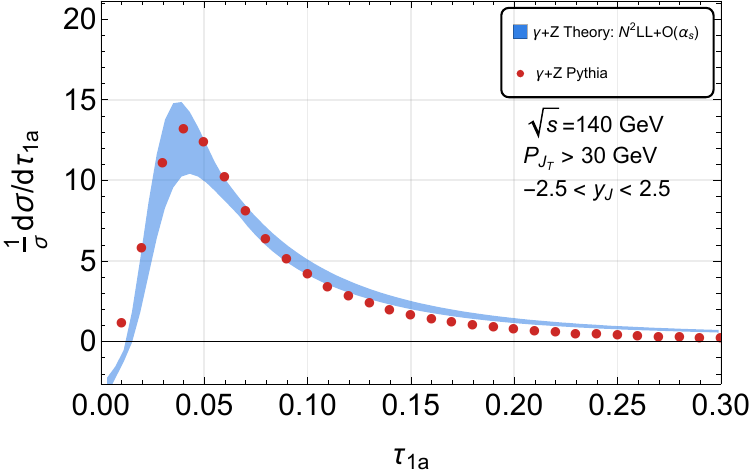}
           \includegraphics[width=0.45\textwidth]{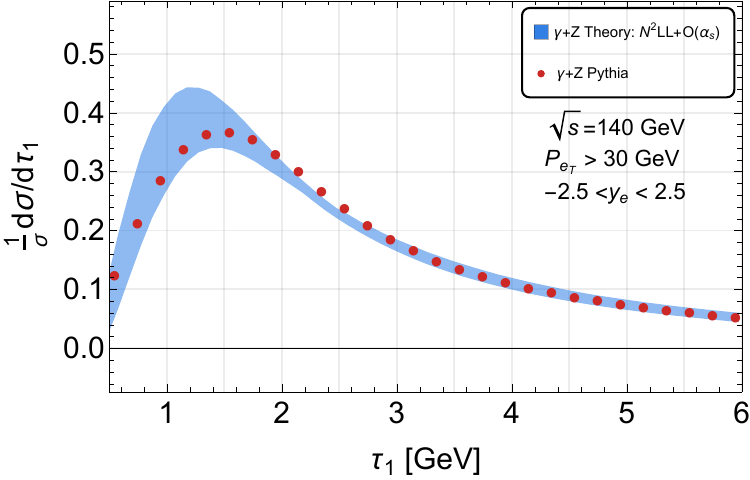}
           \includegraphics[width=0.45\textwidth]{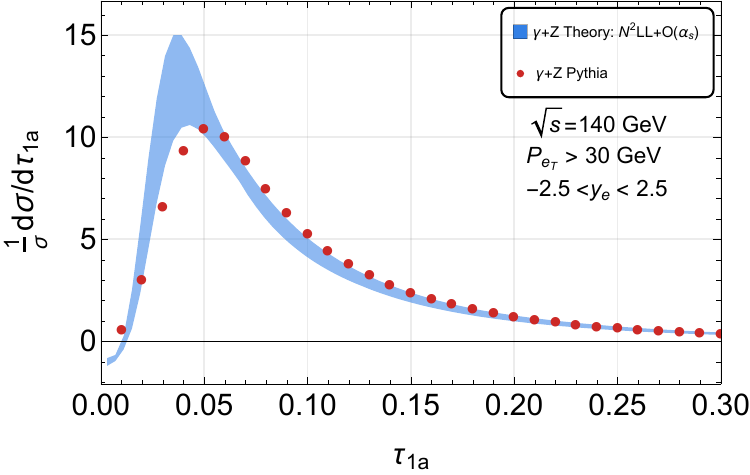}
           \includegraphics[width=0.45\textwidth]{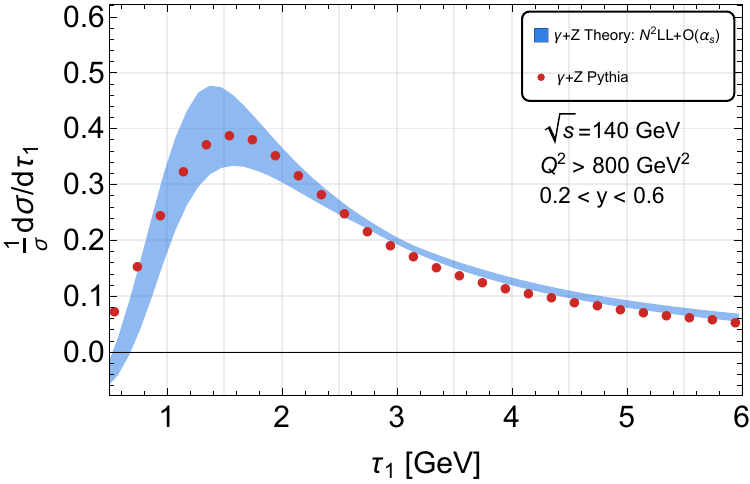}
           \includegraphics[width=0.45\textwidth]{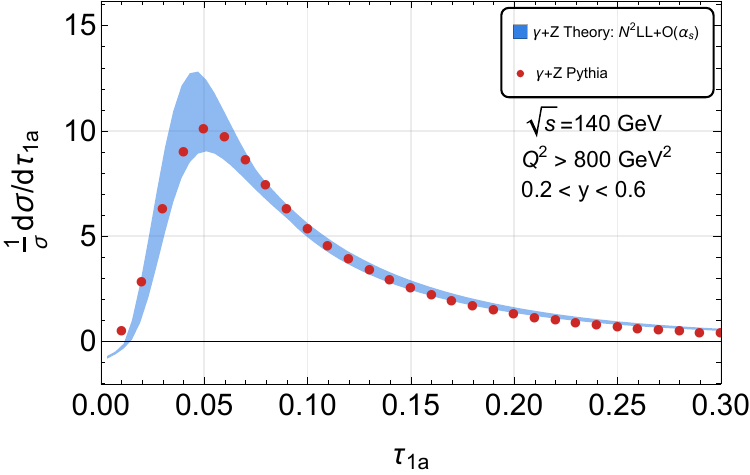}
           \includegraphics[width=0.45\textwidth]{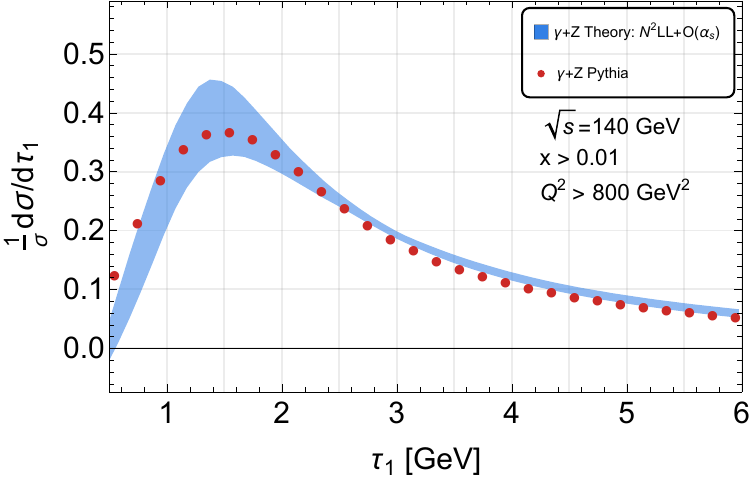}
           \includegraphics[width=0.45\textwidth]{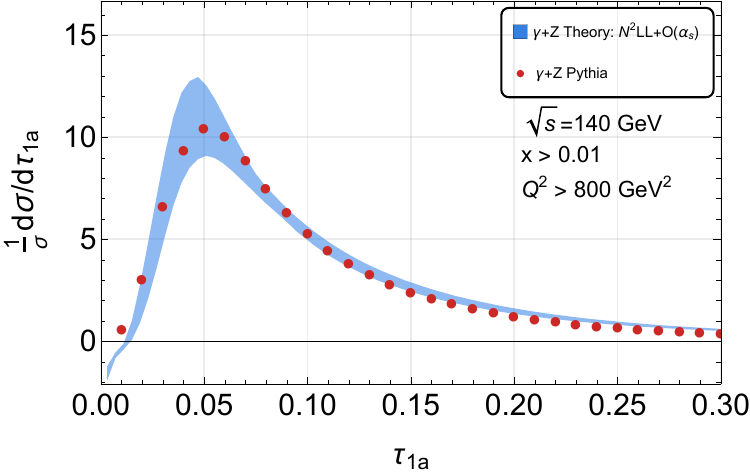}
        \caption{Comparison of the N$^2$LL+${\cal O}(\alpha_s)$ theory prediction (blue band) for NC DIS with the hadronic PYTHIA 8.312 simulation results (red dots). We show predictions for $\tau_1$ (left panel) and $\tau_{1a}$ (right panel) for the EIC center of mass energy, $\sqrt{s}=140$ GeV, in different kinematic bins. The theory predictions use the shape function model parameters in Eq.~(\ref{eq:shapeparam}) and the gap parameter in Eq.~(\ref{eq:gapparam}).}
    \label{fig:TheoryvsPythiaEIC2z}
\end{figure}

\begin{figure}
    \centering
           \includegraphics[width=0.45\textwidth]{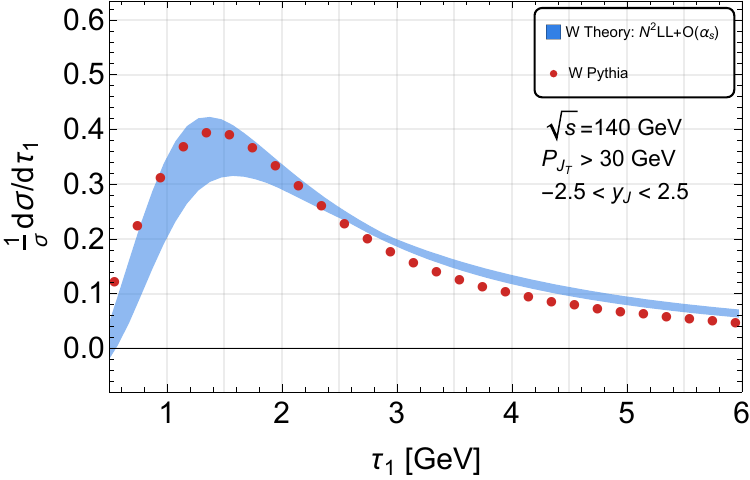}
           \includegraphics[width=0.45\textwidth]{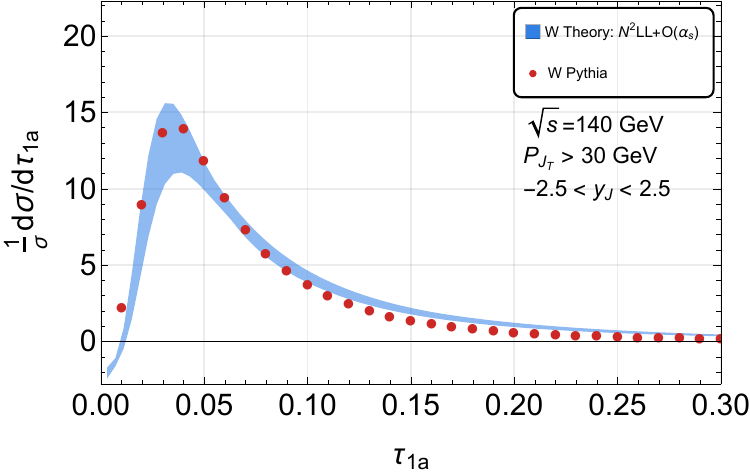}
           \includegraphics[width=0.45\textwidth]{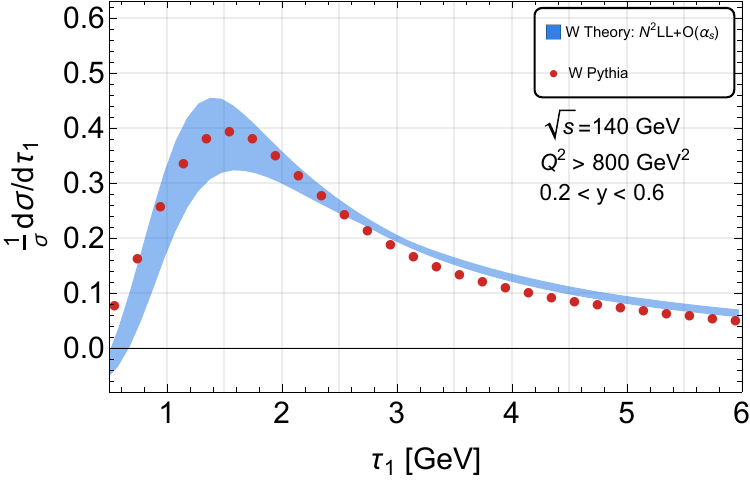}
           \includegraphics[width=0.45\textwidth]{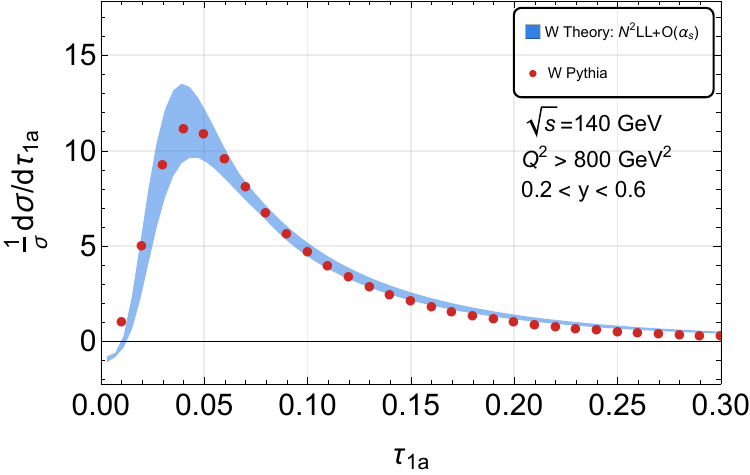}
           \includegraphics[width=0.45\textwidth]{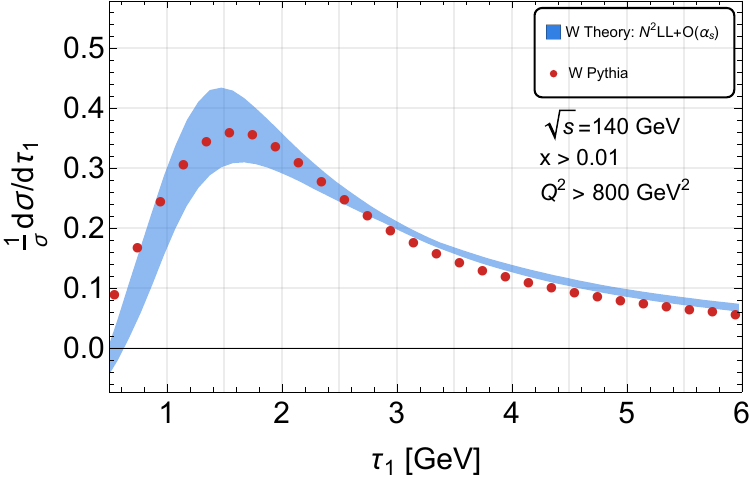}
           \includegraphics[width=0.45\textwidth]{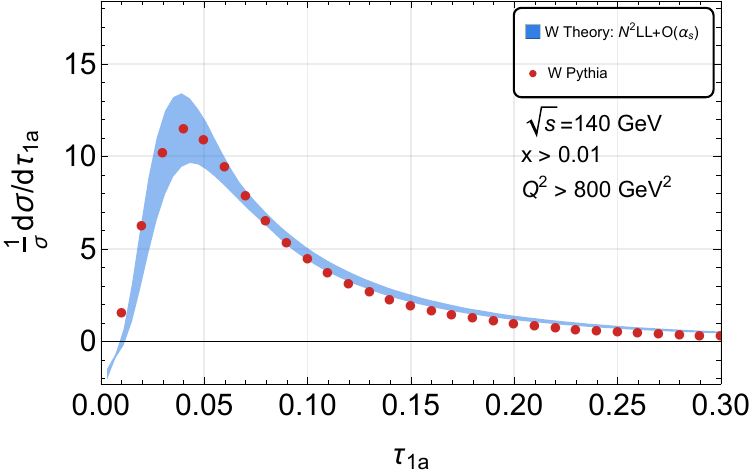}
        \caption{Comparison of the N$^2$LL+${\cal O}(\alpha_s)$ theory prediction (blue band) for CC DIS with the hadronic PYTHIA 8.312 simulation results (red dots). We show predictions for $\tau_1$ (left panel) and $\tau_{1a}$ (right panel) for the EIC center of mass energy, $\sqrt{s}=140$ GeV, in different kinematic bins. The theory predictions use the shape function model parameters in Eq.~(\ref{eq:shapeparam}) and the gap parameter in Eq.~(\ref{eq:gapparam}).}
    \label{fig:TheoryvsPythiaEIC2w}
\end{figure}

 In Figs.~\ref{fig:TheoryvsPythiaheraz}--\ref{fig:TheoryvsPythiaEIC2w},
 we show the  comparison of the N$^2$LL+${\cal O}(\alpha_s)$ theory predictions (blue band) with Pythia simulation (red dots) results, including the contributions of the massive $Z$ and $W$ boson exchange. We see that there is good agreement between the theory predictions and Pythia simulation results for a wide range of 1-Jettiness observables and kinematics using the same underlying shape function model parameters in Eq.~(\ref{eq:shapeparam}). These results demonstrate that the universality of hadronization effects, and their non-trivial but calculable kinematic dependence, extend across NC and CC DIS for a wide range of kinematics for  the $\tau_1$ and $\tau_{1a}$ event shapes.

\section{Conclusions}
\label{sec:Conc}

We have studied the structure of leading hadronization effects and the contributions of massive electroweak gauge bosons  in the 1-Jettiness class of DIS global event shapes. We have shown that the leading hadronization effects for DIS thrust, $\tau_1$, $\tau_{1a}$, and other possible definitions of 1-Jettiness, are described by the same underlying universal shape function. For a subclass of jet-based event shapes such as $\tau_1$, the leading hadronization effects acquire a non-trivial dependence on the underlying hard scattering kinematics. This hard kinematic dependence is explicitly calculable and serves as an independent lever arm for constraining the underlying universal shape function through measurements across a wide range of kinematic bins available at HERA and the EIC. We also include the required renormalon subtractions for the 1-Jettiness event shapes and show that they are related to those for DIS thrust up to calculable kinematic factors. We provide general expressions for the shape functions for the 1-Jettiness class of event shapes in terms of the underlying universal shape function with the calculable hard kinematic dependence made explicit. We presented results up to the N$^3$LL+${\cal O}(\alpha_s^2)$ level of accuracy for NC DIS when including only the single photon exchange contribution.

We have also presented results for the $\tau_1$ and $\tau_{1a}$ 1-Jettiness DIS global event shapes at the N$^2$LL+${\cal O}(\alpha_s)$ level of accuracy for neutral current (NC) and charged current (CC) DIS, extending previous results by  including the ${\cal O}(\alpha_s)$ QCD corrections to  contributions mediated by the exchange of the massive $Z$ and $W$ electroweak gauge bosons. The ${\cal O}(\alpha_s)$ results were implemented by modifying the DISTRESS code. This is the first time that 1-Jettiness has been applied to CC DIS which probes different flavor combinations of the initial state PDFs.  

We have presented a wide range of theoretical predictions and comparisons to Pythia simulation for the NC and CC DIS event shapes $\tau_1$ and $\tau_{1a}$ across a wide range of kinematics. These results demonstrate the universality of hadronization effects and the potential to use the non-trivial but calculable kinematic dependence in $\tau_1$ as a independent lever arm to constrain universal hadronization effects. Given the wide kinematic range available at HERA and the EIC, a global analysis can be used to constrain hadronization effects, and could allow for including the typically ignored peak region, where hadronization effects are most severe, in precision extractions of the strong coupling. Such a global analysis can also include Centauric 1-Jettiness~\cite{Dotson:2026ttc} with adjustable beam and jet regions and the  1-Jettiness jet charge~\cite{Chien:2025rbp} which includes jet charge measurements for  improved quark flavor separation of proton structure and hadronization dynamics.

\begin{center}
\textbf{Acknowledgements} \\
\end{center}

H. C. is partially supported by a CFNS Joint Postdoctoral Fellowship. Z.~K. is supported by the National Science Foundation under grant No.~PHY-2515057. H. C. and F. P. are supported by the U.S. Department of Energy, Office of High Energy Physics, under contract No. DE-SC0010143. X. L. is supported by the National Natural Science Foundation of China under Grant No. 12547109 and Fundamental Research Funds for the
Central Universities, Beijing Normal University.
This research was supported in part through the computational resources and staff contributions provided for the Quest high performance computing facility at Northwestern University which is jointly supported by the Office of the Provost, the Office for Research, and Northwestern University Information Technology. Z.~K. and  X.~L. would like to thank the Erwin-Schr\"odinger International Institute for Mathematics and Physics at the University of Vienna for partial support during the Programme `New Paradigms for Harnessing Quantum Field Theory at Colliders', July 27 – August 28, 2026. S.M. thanks Jefferson Lab for their hospitality and support through the Visiting Faculty Program (VFP) during which part of this work was carried out.

\appendix

\section{Relationship Between the Generalized and Hemisphere Soft function}
\label{appexSoftFunc}

 It was shown in Ref.~\cite{Kang:2013nha} that the  generalized hemisphere soft function for the $\tau_{1a}$ event shape is related to the standard DIS thrust hemisphere soft function. This formalism was then applied to show a  similar result for the $\tau_1$ event shape in Ref.~\cite{Cao:2024ota} for $\tau_1$. These results are summarized in Eqs.~(\ref{eq:Sgenhemi}) and (\ref{eq:RBRJtau1}). Below we revisit this derivation in order to point out the differences in these results between the $\tau_1$ and $\tau_{1a}$ event shapes. These  differences will also lead to differences in the implementation of the required renormalon subtractions of in the soft functions for $\tau_1$ and $\tau_{1a}$.

The DIS hemisphere soft function is defined as 
\bea
\label{eq:stanhemi}
{\cal S}(k_1, k_2, \mu) &=& \frac{1}{N_c} \>{\rm tr}\>\sum_{X_s} \Big |\langle X_s | [Y_{n_2}^\dagger Y_{n_1}] (0) |0\rangle \Big |^2 \nn \\
&& \delta \left [k_1- \sum_{i \in X_s} \theta \left ( n_2\cdot k_i - n_1\cdot k_i \right ) n_1 \cdot k_i\right ] \nn \\
&& \delta \left [k_2 - \sum_{i \in X_s} \theta \left ( n_1\cdot k_i - n_2\cdot k_i \right ) n_2 \cdot k_i\right ] ,
\eea
where the $n_1$ and $n_2$ reference vectors are defined in Eq.~(\ref{eq:n1n2}). Based on the definitions of $\tau_{1}$ and $\tau_{1a}$ in Eq.~(\ref{tau1andtau1a}), 
the corresponding generalized hemisphere soft functions are given by
\bea
\label{eq:Stau1}
\tau_1: \>\>{\cal S}_{\tau_1}(k_B, k_J, \mu) &=& \frac{1}{N_c} \>{\rm tr}\>\sum_{X_s} \Big |\langle X_s | [Y_{n_J}^\dagger Y_{n_B}] (0) |0\rangle \Big |^2 \nn \\
&& \delta \left [k_B - \sum_{i \in X_s} \theta \left ( \frac{2q_J\cdot k_i}{Q_J} - \frac{2q_B\cdot k_i}{Q_B} \right ) n_B \cdot k_i\right ] \nn \\
&&\delta \left [k_J - \sum_{i \in X_s} \theta \left ( \frac{2q_B\cdot k_i}{Q_B}- \frac{2q_J\cdot k_i}{Q_J} \right ) n_J \cdot k_i\right ],
\eea
and
\bea
\label{eq:Stau1a}
\tau_{1a}: \>\>{\cal S}_{\tau_{1a}}(k_B, k_J, \mu) &=& \frac{1}{N_c} \>{\rm tr}\>\sum_{X_s} \Big |\langle X_s | [Y_{n_J}^\dagger Y_{n_B}] (0) |0\rangle \Big |^2 \nn \\
&& \delta \left [k_B - \sum_{i \in X_s} \theta \left (  \frac{2q_J\cdot k_i}{Q^2}  - \frac{2q_B\cdot k_i}{Q^2} \right ) n_B \cdot k_i\right ] \nn \\
&&\delta \left [k_J - \sum_{i \in X_s} \theta \left (\frac{2q_B\cdot k_i}{Q^2}-  \frac{2q_J\cdot k_i}{Q^2}  \right ) n_J \cdot k_i\right ],
\eea
respectively. The beam and jet reference vectors are $q_B^\mu=\omega_B n_B^\mu/2$ and $q_J^\mu=\omega_J n_J^\mu/2$, where $\omega_B$ and $\omega_J$ are defined in Eq.~(\ref{eq:wBwJ}). Furthermore, for $\tau_1$ we have $Q_B=\omega_B$ and $Q_J=\omega_J$. We can then bring the generalized soft functions for $\tau_1$ and $\tau_{1a}$ into the form
\bea
\label{eq:tau1S}
\tau_1: \>\>{\cal S}_{\tau_1}(k_B, k_J, \mu) &=& \frac{1}{N_c} \>{\rm tr}\>\sum_{X_s} \Big |\langle X_s | [Y_{n_J}^\dagger Y_{n_B}] (0) |0\rangle \Big |^2 \nn \\
&& \delta \left [k_B - \sum_{i \in X_s} \theta \left ( n_J\cdot k_i- n_B\cdot k_i \right ) n_B \cdot k_i\right ] \nn \\
&&\delta \left [k_J - \sum_{i \in X_s} \theta \left ( n_B\cdot k_i- n_J\cdot k_i \right ) n_J \cdot k_i\right ],
\eea
and
\bea
\label{eq:tau1aS}
\tau_{1a}: \>\>{\cal S}_{\tau_{1a}}(k_B, k_J, \mu) &=& \frac{1}{N_c} \>{\rm tr}\>\sum_{X_s} \Big |\langle X_s | [Y_{n_J}^\dagger Y_{n_B}] (0) |0\rangle \Big |^2 \nn \\
&& \delta \left [k_B - \sum_{i \in X_s} \theta \left (  \omega_J \>n_J\cdot k_i- \omega_B \>n_B \cdot k_i \right ) n_B \cdot k_i\right ] \nn \\
&&\delta \left [k_J - \sum_{i \in X_s} \theta \left (\omega_B \>n_B\cdot k_i -  \omega_J \> n_J\cdot k_i  \right ) n_J \cdot k_i\right ],
\eea
respectively. One can relate these soft functions to the DIS hemisphere soft function via a rescaling of the beam and jet reference vectors, $n_B^\mu$ and $n_J^\mu$, as shown in Eq.~(\ref{eq:nBRBnJRJ}). These rescalings
are chosen such that they satisfy the same condition as the reference vectors  for the standard DIS thrust hemisphere soft function as shown in 
 Eq.~(\ref{eq:nBnJ2}). Note that the Wilson lines are invariant~\cite{Kang:2013nha} under these rescalings, corresponding to the boost invariance of the soft Wilson lines. Thus, with the appropriate choice of these rescaling  factors,  $R_B$ and $R_J$, the generalized soft functions can be related to the standard hemisphere soft function as
\bea
{\cal S}_{\tau_1,\tau_{1a}}(k_B, k_J, \mu) &=& \frac{1}{N_c} \>{\rm tr}\>\sum_{X_s} \Big |\langle X_s | [Y_{n_J'}^\dagger Y_{n_B'}] (0) |0\rangle \Big |^2 \nn \\
&& \frac{1}{R_B}\delta \left [\frac{k_B}{R_B} - \sum_{i \in X_s} \theta \left ( n_J'\cdot k_i - n_B'\cdot k_i \right ) n_B' \cdot k_i\right ] \nn \\
&& \frac{1}{R_J}\delta \left [\frac{k_J}{R_J}  - \sum_{i \in X_s} \theta \left ( n_B'\cdot k_i - n_J'\cdot k_i \right ) n_J' \cdot k_i\right ] , \nn \\
&=& \frac{1}{R_BR_J}{\cal S}_{\rm hemi.}(\frac{k_B}{R_B}, \frac{k_J}{R_J}, \mu) ,
\eea
where the last equality follows by comparison with the form of the hemisphere soft function in Eq.~(\ref{eq:stanhemi}).
From Eqs.~(\ref{eq:tau1S}) and (\ref{eq:tau1aS}), we see that this can be done by choosing $R_B$ and $R_J$ for $\tau_1$ and $\tau_{1a}$ as shown in Eq.~(\ref{eq:RBRJtau1}).

\bibliographystyle{h-physrev3.bst}
\bibliography{disjettiness}
\end{document}